\pdfoutput=1
\documentclass[a4paper,11pt]{article}
\usepackage{jheppub}
\usepackage{tikz-cd}
\usepackage{tikz}
\usetikzlibrary{arrows.meta,calc,positioning,shapes.geometric}
\usepackage{amsmath}
\usepackage{array,booktabs}
\usepackage{amsthm}
\newtheorem{theorem}{Theorem}
\newtheorem{lemma}[theorem]{Lemma}
\newtheorem{corollary}[theorem]{Corollary}
\newtheorem{proposition}[theorem]{Proposition}
\theoremstyle{definition}
\newtheorem{definition}[theorem]{Definition}
\newtheorem{example}[theorem]{Example}
\newtheorem{remark}[theorem]{Remark}
\usepackage{enumerate}
\usepackage[utf8]{inputenc}
\usepackage{amssymb,amsbsy,mathtools}
\usepackage{braket}
\usepackage{graphicx}
\usepackage{xcolor}
\usepackage{natbib}
\usepackage{bm}
\usepackage{soul}

\newcommand{\dd}{\,\mathrm{d}}
\newcommand{\cH}{\mathcal{H}}

\newcommand{\Mop}{\mathsf{M}}

\usepackage[nottoc]{tocbibind}

\usepackage{dsfont}
\let\mathbbm\mathds

\newcommand{\Tr}{\text{Tr}}

\usepackage{caption}
\newcommand{\mA}{\mathcal{A}}

\newcommand{\mP}{\mathcal{P}}
\newcommand{\mR}{\mathcal{R}}
\newcommand{\nn}{\nonumber}
\newcommand{\mN}{\mathcal{N}}
\newcommand{\mI}{\mathcal{I}}

\newcommand{\mH}{\mathcal{H}}
\usepackage{subcaption}
\usepackage{hyperref}

\newcommand{\mM}{\mathcal{M}}

\newcommand{\Ad}{\operatorname{Ad}}
\newcommand{\Dom}{\operatorname{Dom}}
\newcommand{\mE}{\mathcal{E}}

\newcommand{\cL}{\mathcal{L}}
\begin{document}

\title{Subsystem Flow: Quantum Error Correction,\\ Petz Recovery, and Ergodic Inclusions}

\author{Yidong Chen$^1$,}
\affiliation{$^1$Department of Mathematics, University of Illinois at Urbana-Champaign, Illinois, IL 61801, USA}
\author{Nima Lashkari$^2$,}
\author{Kwing Lam Leung$^2$}
\affiliation{$^2$Department of Physics and Astronomy, Purdue University, West Lafayette, IN 47907, USA}
\emailAdd{yidongc2@illinois.edu}
\emailAdd{nima@purdue.edu}
\emailAdd{leung60@purdue.edu}

\abstract{In quantum error correction (QEC), the goal is to engineer an encoding that protects logical information against the erasure of a physical subsystem. In local many-body quantum systems, by contrast, nature provides a subsystem inclusion $\mN\subset \mM$ in a fixed reference state $\ket{\Omega}$. Asking whether the inclusion admits a reference-preserving code that protects subsystem $\mN$ against erasure of the complement is equivalent to asking whether $\mN$ is preserved under the modular dynamics of $\mM$ in this state (modular autonomy). We define the subsystem flow, which measures the departure from this error correction regime. Away from autonomy, the infinitesimal flow quantifies relative-entropy loss, while the history algebra and persistent core describe which observables the modular orbit generates and retains. We call an inclusion ergodic when modular orbits generate everything and retain nothing. Analytic continuation of subsystem flow yields Petz recovery maps and a family of positive inclusion parent Hamiltonians. We show that, in general, evolving with these parent Hamiltonians allows $\mN$ to leave $\mM$, and in some cases, it can generate the entire  $\mM\vee\mM'$. Inclusions with a common cyclic and separating vector generate everything, and standard inclusions of factors, such as those of nested subregions in QFT, are ergodic. Moreover, the $L^2$-nuclearity property implies that the spectrum of parent Hamiltonians is discrete. We define a dissipative semigroup by repeating the Petz recovery map, whose fixed-point algebra is the modular persistent core and whose spectrum is gapped under the assumption of $L^2$-nuclearity. We conjecture that the eigenvalues and eigenvectors of inclusion parent Hamiltonians can be used to define local spectral measures of quantum chaos.}

\maketitle
\section{Introduction}\label{sec:intro}

A subsystem of a many-body quantum system or of a quantum field theory need not be a tensor factor of the Hilbert space. The observables accessible in a spacetime region, or along an observer's worldline, form a local von Neumann algebra $\mN$ inside the algebra $\mM$ of observables of the whole spacetime. In continuum QFT the local algebras are type III factors, so neither a density matrix nor a tensor factorization is intrinsic to a region. The inclusion $\mN\subset\mM$ is what remains of the notion of a subsystem.

Every faithful state $\omega$ on $\mM$ comes with a canonical dynamics, its modular flow $\sigma_t^{\omega;\mM}$. For a full-rank density matrix $\rho$, this dynamics is conjugation by $\rho^{it}$. The restricted state on a subsystem $\mN$ defines a second modular flow $\sigma_t^{\omega;\mN}$ intrinsic to the subsystem. There are therefore two ways to move a subsystem observable: with the dynamics of the full system, or with the dynamics inferred from the reduced state alone. When the two agree, the subsystem's information remains confined inside it, and we say $\mN$ evolves autonomously. When they disagree, the subsystem exchanges information with the rest of the system. This paper is about the mismatch, which we call the \emph{subsystem flow}, and about what it says about error correction, information loss, and the emergence of dynamics from an inclusion and a state. 

Modular flow already has a geometric meaning in familiar QFT examples: vacuum wedge flow is a Lorentz boost~\cite{BisognanoWichmann1975}, the modular flow of a ball in vacuum CFT is a conformal transformation~\cite{CasiniHuertaMyers2011}, half-sided modular inclusions reconstruct null translations~\cite{Wiesbrock1993,Borchers1992}. Recent work~\cite{furuya2023information,ouseph2024local,LashkariLeungMoosaOuseph2025} suggests that the emergence of these geometric symmetries is tied to chaotic modular flows. In holography, modular flow connects entanglement-wedge reconstruction to quantum error correction~\cite{AlmheiriDongHarlow2015,JafferisLewkowyczMaldacenaSuh2016}. In this work, we develop the information-theoretic and algebraic structures that connect modular dynamics, QEC and ergodic modular flows.

\paragraph{The subsystem flow:}
Let $\ket{\Omega}$ be the vector representative of state $\omega$ on the GNS Hilbert space $\mH_\mM$. The subsystem has its own GNS Hilbert space $\mH_\mN$, with cyclic vector $\ket{\Omega_\mN}$. The canonical isometry $V:\mH_\mN\to\mH_\mM$, $V\,n\ket{\Omega_\mN}=n\ket{\Omega}$, embeds it as the closed subspace $\overline{\mN\ket{\Omega}}$, and $e_\mN=VV^\dagger$ is the projection onto this subspace.  The subsystem algebra $\mN$ acts on $\mH_\mN$ by its GNS representation and on $\mH_\mM$ by inclusion, and $V$ intertwines the two actions, $nV=Vn$ (Section~\ref{subsec:subsystem-flow-background}). The modular operators $\Delta_\mM$ and $\Delta_\mN$ act on $\mH_\mM$ and $\mH_\mN$, respectively. The subsystem flow is
\begin{equation}
 D_{\mN\subset\mM}(t)
 :=\Delta_\mM^{-it}V\Delta_\mN^{it}V^\dagger .
 \label{eq:intro-compressed-flow}
\end{equation}
It moves a subsystem vector forward with the intrinsic flow of $\mN$ and back with the ambient flow $\mM$. The isometry $V$ is fixed by the inclusion and the reference state. When $\ket{\Omega}$ is also cyclic for $\mN$, the algebras have a common cyclic and separating vector, $V=1$, and $D(t)=\Delta_\mM^{-it}\Delta_\mN^{it}$ is Borchers' characteristic function of the inclusion~\cite{revolutionizing}.

To build intuition, it is helpful to start with finite-dimensional systems. Take $\mM=B(\mH_B\otimes\mH_R)$, $\mN=B(\mH_B)\otimes1$ and a full-rank density matrix $\rho$ with marginal $\rho_B$. The GNS space of $\mM$ is the space of operators on $\mH_B\otimes\mH_R$ with the Hilbert--Schmidt inner product, $\ket{\Omega}=\ket{\rho^{1/2}}$, and the modular superoperator $\Delta_\mM (x)=\rho x\rho^{-1}$. The GNS space $\mH_\mN$ of $\mN$ is the space of operators on $\mH_B$, and $V$ embeds it in $\mH_\mM$ through $b\ket{\rho_B^{1/2}}\mapsto(b\otimes1)\ket{\rho^{1/2}}$. The image is spanned by the operators $(b\otimes1)\ket{\rho^{1/2}}$, has dimension $(\dim\mH_B)^2$, and is a proper subspace whenever $\dim\mH_R>1$, even for a product state. We also have $\Delta_\mN (y)=\rho_By\rho_B^{-1}$. Then,
\begin{equation}
 D(t)\,(b\otimes1)\ket{\Omega}=u_t\,(b\otimes1)\,u_t^{*}\,\ket{\Omega},
 \qquad u_t=\rho^{-it}\bigl(\rho_B^{it}\otimes1\bigr),
 \label{eq:intro-matrix-flow}
\end{equation}
so the subsystem flow conjugates $b$ by the mismatch between the two density-matrix evolutions. Its generator is the commutator with $h_\rho-h_{\rho_B}\otimes1$, where $h=-\log\rho$: the part of the modular Hamiltonian of $\rho$ that couples $B$ to $R$. The flow is trivial, $D(t)=e_\mN$, exactly when $\rho=\rho_B\otimes\chi_R$ is a product state (Corollary~\ref{cor:KL-conormal}). For a general inclusion, the compression by $V$ cannot be dropped\footnote{See Example~\ref{ex:two-qubits} for a two-qubit example with $V\ne1$. We call it a compression, because $V V^\dagger$ is a projection.}.

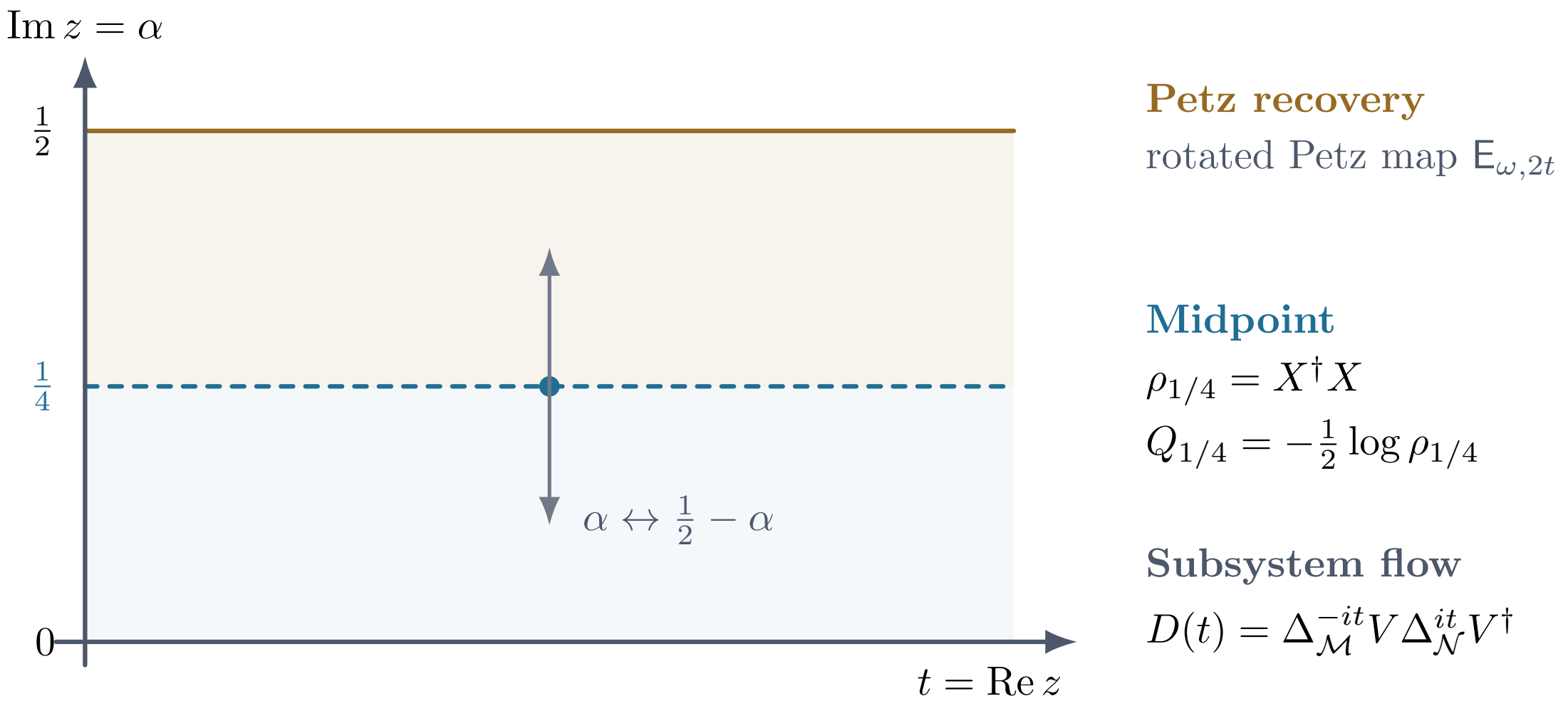
\begin{figure}[!htbp]
\centering
\resizebox{0.98\linewidth}{!}{\begin{tikzpicture}[x=1cm,y=1cm,>=Latex,
 every node/.style={font=\small},line cap=round,line join=round]
 \definecolor{stripblue}{RGB}{31,110,147}
 \definecolor{stripgray}{RGB}{76,87,106}
 \definecolor{stripgold}{RGB}{151,106,35}
 \fill[stripblue!5] (0,0) rectangle (8,4.4);
 \fill[stripgold!8] (0,2.2) rectangle (8,4.4);
 \draw[stripgray,line width=1.0pt] (0,0)--(8,0);
 \draw[stripgold,line width=1.1pt] (0,4.4)--(8,4.4);
 \draw[stripblue,dashed,line width=1.1pt] (0,2.2)--(8,2.2);
 \filldraw[stripblue] (4,2.2) circle (2.3pt);
 \draw[->,stripgray,line width=1.1pt] (0,-0.2)--(0,5.05) node[above,black] {$\operatorname{Im}z=\alpha$};
 \draw[->,stripgray,line width=1.1pt] (-0.25,0)--(8.55,0);
 \node[anchor=north east,black] at (8.55,-0.08) {$t=\operatorname{Re}z$};
 \node[anchor=east] at (-0.12,4.4) {$\tfrac12$};
 \node[anchor=east,text=stripblue] at (-0.12,2.2) {$\tfrac14$};
 \node[anchor=east] at (-0.12,0) {$0$};
 \draw[<->,stripgray!80,line width=0.9pt] (4,1.0)--(4,3.4);
 \node[anchor=west,text=stripgray] at (4.15,1.05) {$\alpha\leftrightarrow\tfrac12-\alpha$};
 \node[anchor=west,font=\bfseries\small,text=stripgold] at (9.0,4.65) {Petz recovery};
 \node[anchor=west,text=stripgray] at (9.0,4.15) {rotated Petz map $\mathsf E_{\omega,2t}$};
 \node[anchor=west,font=\bfseries\small,text=stripblue] at (9.0,2.75) {Midpoint};
 \node[anchor=west] at (9.0,2.25) {$\rho_{1/4}=X^\dagger X$};
 \node[anchor=west] at (9.0,1.70) {$Q_{1/4}=-\tfrac12\log\rho_{1/4}$};
 \node[anchor=west,font=\bfseries\small,text=stripgray] at (9.0,0.65) {Subsystem flow};
 \node[anchor=west] at (9.0,0.10) {$D(t)=\Delta_{\mathcal M}^{-it}V\Delta_{\mathcal N}^{it}V^\dagger$};
\end{tikzpicture}}
\caption{The analytic strip $z=t+i\alpha$ of the subsystem flow $D(z)$ of an inclusion $\mN\subset\mM$ with canonical isometry $V$. The upper edge carries the rotated Petz map. The midpoint operator is $\rho_{1/4}=X^\dagger X$ with $X=D(i/4)V$, and $Q_{1/4}=-\tfrac12\log\rho_{1/4}$ is taken on the support. The reflection $\alpha\to\tfrac12-\alpha$ acts as $J_\mN Q_\alpha J_\mN=Q_{1/2-\alpha}$ under full support.}
\label{fig:subsystem-analytic-strip}
\end{figure}

\paragraph{One function, three readings:}
Modular operators of an inclusion satisfy a monotonicity condition\footnote{Consider an inclusion $\mathcal{N}\subset\mathcal{M}$. If the reference state is common cyclic and separating, then it is well-known that $\Delta_\mathcal{N} \geq \Delta_\mathcal{M}$ \cite{revolutionizing}. When the state is not cyclic for $\mathcal{N}$, the same monotonic relation holds: $V^\dagger\Delta_\mathcal{M}V\le\Delta_\mathcal{N}$ in the sense of quadratic forms on $\mathcal{H}_\mathcal{N}$, where $V: \mathcal{H}_\mathcal{N} \rightarrow\mathcal{H}_\mathcal{M}$ is the isometry.}. In addition, the modular operators are positive \cite{Takesaki1972}. Hence the subsystem flow $D(t)$ continues analytically in $t$ to the strip $0\le\operatorname{Im}z\le\tfrac12$, where it is a contraction; see Figure~\ref{fig:subsystem-analytic-strip}. On the real axis $D(t)$ measures information loss and organizes an ergodic hierarchy of inclusions (Section~\ref{sec:ergodic}). On the upper boundary, it implements Petz recovery of $\mM$ from $\mN$ (Section~\ref{subsec:analytic-subsystem-flow}). In the interior, it defines positive Hamiltonians that kill $\ket{\Omega}$, whose ground spaces, orbits, and spectra we study in Section~\ref{sec:probing-outside-M}.

\paragraph{Main results:}
\begin{enumerate}
\item \emph{State-preserving error correction is modular autonomy} (Theorem~\ref{thm:correctable}). In quantum error correction (QEC), we assume control of the global state and aim to engineer an encoding such that the erased environment learns nothing about the logical input. In QFT or holography, nature provides us with the inclusion $\mN\subset\mM$ and the reference state $\omega$ on $\mM$, and the question becomes whether $\mN$ is the logical algebra of an erasure code whose encoded states contain $\omega$. Such a code is a normal conditional expectation $\mE:\mM\to\mN$ with $\omega\circ\mE=\omega$~\cite{furuya2022real,Faulkner2020}. By Takesaki's theorem~\cite{Takesaki1972} it exists exactly when $\mN$ is preserved by the ambient modular flow, and we show that this is the same as the trivial subsystem flow:
\begin{equation}
 \sigma_t^{\omega;\mM}(\mN)=\mN\quad\text{for all }t
 \quad\Longleftrightarrow\quad
 D_{\mN\subset\mM}(t)=e_\mN\quad\text{for all }t.
 \label{eq:intro-autonomy}
\end{equation}
For a tensor-factor inclusion, this says that the reference is a product state $\rho=\rho_B\otimes\chi_R$. When $\mN$ and its relative commutant $\mR=\mN'\cap\mM$ generate $\mM$, it is the Knill--Laflamme condition~\cite{KnillLaflamme1997} with $\mR$ as the environment. 
Proper inclusions of local algebras of QFT with a common cyclic vector never satisfy it (Corollary~\ref{cor:common-cyclic-noncorrectable}).

\item \emph{The generator measures information loss} (Proposition~\ref{prop:sufficiency}). Away from autonomy, the derivative of the flow at $t=0$ is the difference of the two modular Hamiltonians, $K_\mM -VK_\mN V^\dagger$ with $K=-\log\Delta$, and its expectation in a unit vector $\ket{\Psi}\in\mH_\mN$ is a sum of two nonnegative relative-entropy losses,
\begin{align}
 -i\left.\partial_t\langle\Psi|V^\dagger D(t)V|\Psi\rangle\right|_{t=0}
 &=\bigl[S_\mM(\varphi\|\omega)-S_\mN(\varphi\|\omega)\bigr]
 +\bigl[S_{\mN'}(\varphi\|\omega)-S_{\mM'}(\varphi\|\omega)\bigr],
 \label{eq:intro-entropy}
\end{align}
the loss under restriction to $\mN$ and the loss under the complementary compression of the commutant. Each bracket vanishes exactly when the corresponding pair of states is recoverable by a Petz map~\cite{Petz1986,Petz1988}\footnote{In the operator algebraic literature, this is often called the sufficiency condition.}.

\item \emph{Ground states and orbits of the inclusion Hamiltonians} (Theorem~\ref{thm:Q-kernel-core} and Proposition~\ref{prop:Halpha-full-generation}). For $0<\alpha<\tfrac12$ the interior of the strip defines a positive contraction $\rho_\alpha=V^\dagger D(i\alpha)^\dagger D(i\alpha)V$ on $\mH_\mN$ and a positive {\it parent Hamiltonian}
\begin{equation}
 Q_\alpha:=-\frac{1}{2\sin(2\pi\alpha)}\log\rho_\alpha\geq0,\qquad Q_\alpha\ket{\Omega}=0,
 \label{eq:intro-Halpha}
\end{equation}
which we call an inclusion parent Hamiltonian because the reference vector is its ground state. Its entire ground space is the GNS closure of the persistent core $\mN_{\rm core}=\bigcap_t\sigma_t^{\omega;\mM}(\mN)$, the largest closed autonomous subalgebra inside $\mN$:
\begin{equation}
 \ker(1-\rho_\alpha)=\overline{\mN_{\rm core}\ket{\Omega}}.
 \label{eq:intro-ground-core}
\end{equation}
Hence $\ket{\Omega}$ is the unique ground state if and only if $\mN$ retains no closed autonomous subalgebra, and in that case, for full-support $\rho_\alpha$, the orbit of $\mN$ under $e^{itQ_\alpha}$ generates all of $B(\mH_\mN)$. For a standard inclusion of factors, such as nested regions in QFT, $\mH_\mN=\mH_\mM$ and the orbit generates $B(\mH_\mM)$: $\mN$ leaves the ambient algebra $\mM$. The modular flow keeps $\mN$ inside $\mM$ by construction. The inclusion Hamiltonians probe beyond $\mM$. In special QFT settings, one can show that these inclusion parent Hamiltonians are indeed proportional to the global Hamiltonian \cite{chen2026local}\footnote{We caution the readers that in \cite{chen2026local} we defined $\rho_\alpha= D(i\alpha) D(i\alpha)^\dagger$ instead of the definition $D^\dagger D$ we are using here. The distinction is a matter of convenience with no crucial consequences in our analysis. See Appendix \ref{app:Q-dual-inclusion} for the relation between the two definitions.}. We will explore these inclusion parent Hamiltonians in QFT and holography in more depth in a companion paper \cite{companionQFT}.
\end{enumerate}

\paragraph{Ergodic inclusions:}
The history $\mN_{\rm hist}=\bigvee_t\sigma_t^{\omega;\mM}(\mN)$ is the smallest closed autonomous subalgebra containing $\mN$, and the persistent core $\mN_{\rm core}$, the observables of $\mN$ that remain in $\mN$ at every modular time, is the largest one inside it. We call an inclusion \emph{ergodic} when the modular orbit generates everything and retains nothing, $\mN_{\rm hist}=\mM$ and $\mN_{\rm core}=\mathbb C1$. We will show that inclusions with a common cyclic and separating vector are generating. In addition, when $\ket{\Omega}$ is also cyclic for the relative commutant (a standard inclusion of factors), the inclusion is ergodic. Ergodic half-sided inclusions are the modular $K$-systems. Section~\ref{subsec:QEC-hierarchy} organizes these notions into a hierarchy of correctability, with autonomous inclusions at one end and modular $K$-inclusions at the other.

\paragraph{Further consequences:}
At the upper boundary of the strip, the analytically continued subsystem flow implements the rotated Petz maps\footnote{The  weighted average of rotated Petz maps is the universal recovery map of~\cite{JungeRennerSutterWildeWinter2018,FaulknerHollandsSwingleWang2022}.}. At the midpoint $\alpha=\tfrac14$, $\rho_{1/4}$ is the transfer operator for inclusion followed by Petz recovery, and repeating the recovery step defines a dissipative semigroup whose fixed-point algebra is the persistent core (Section~\ref{subsec:recovery-Poisson}). The assumption of $L^2$-nuclearity (see Section~\ref{sec:nuclearity} for a definition), which holds in many QFT setups, makes the spectrum of $Q_\alpha$ discrete with finite multiplicities and implies a gap in the semigroup. This leads us to conjecture that the level statistics and eigenvectors of inclusion Hamiltonians can be used as a measure of local chaos intrinsic to a local subsystem of a QFT, formulated without access to the global Hamiltonian (Section~\ref{sec:local-chaos}).

Section~\ref{section:basics} sets up the algebraic language and defines the subsystem flow. Readers familiar with modular theory can skip to Definition~\ref{def:subsystem-flow} and use Table~\ref{tab:notation} for notation and terminology. Section~\ref{sec:QEC} proves the key error-correction theorem. Section~\ref{sec:ergodic} studies the subsystem flow at real modular time with information loss, histories and cores, and the ergodic hierarchy. Section~\ref{sec:probing-outside-M} analytically continues subsystem flow to imaginary modular time and studies recovery maps, inclusion parent Hamiltonians, and their spectral consequences. Its assumptions are summarized in a table at the start of that section. Section~\ref{sec:local-chaos} is the chaos proposal. The appendices contain more operator-algebra background, proofs of several results in the paper, QEC examples, and a parallel story in classical dynamical systems and classical ergodic theory where the role of modular flow is played by Hamiltonian flows (see Appendix~\ref{sec:classical}). 

\begin{table}[!htbp]
\centering
\small
\begin{tabular}{@{}>{\raggedright\arraybackslash}p{0.30\linewidth}>{\raggedright\arraybackslash}p{0.66\linewidth}@{}}
\toprule
Symbol or term & Meaning\\
\midrule
$\mN\subset\mM$, $\mR=\mN'\cap\mM$ & subsystem inside the ambient algebra, and its relative commutant (the environment)\\
$\omega$, $\ket{\Omega}$ & reference state and its GNS vector\\
$\mH_\mN$, $V:\mH_\mN\to\mH_\mM$ & GNS space of $\mN$, and the canonical isometry onto $\overline{\mN\ket{\Omega}}=e_\mN\mH_\mM$, with $e_\mN=VV^\dagger$, intertwining the actions of $\mN$: $nV=Vn$\\
$\Delta_\mA$, $K_\mA=-\log\Delta_\mA$ & modular operator and modular Hamiltonian of $\mA=\mN,\mM$\\
$D(z)$ & subsystem flow \eqref{eq:intro-compressed-flow} and its continuation to the strip $0\le\operatorname{Im}z\le\tfrac12$\\
$\rho_\alpha$, $Q_\alpha$ & positive inclusion operator and inclusion parent Hamiltonian \eqref{eq:intro-Halpha}\\
$\mathsf E_\omega$, $\Gamma$ & Petz map $\mM\to\mN$ and its implementer on the GNS Hilbert space\\
$\mN_{\rm hist}$, $\mN_{\rm core}$ & history and persistent core of $\mN$ under the ambient modular flow\\
\midrule
autonomous & $\mN$ is preserved by the ambient modular flow\\
common cyclic & $\ket{\Omega}$ is cyclic for $\mN$ as well as $\mM$, so $V=e_\mN=1$\\
standard & $\ket{\Omega}$ is cyclic and separating for $\mN$, $\mM$ and $\mR$\\
ergodic & $\mN_{\rm hist}=\mM$ and $\mN_{\rm core}=\mathbb C1$\\
\bottomrule
\end{tabular}
\caption{Notation and terminology used throughout the paper.}
\label{tab:notation}
\end{table}

\section{Setup: inclusions, modular theory, and the subsystem flow}\label{section:basics}

The observables of a physical quantum system form a von Neumann algebra. The center of this algebra describes its classical observables (superselection sectors). A quantum system with no nontrivial classical superselection observables is a von Neumann \emph{factor}. For a finite quantum system, such factors are
finite matrix algebras\footnote{The factor associated to a $d$-level quantum system is the algebra of $d\times d$ complex matrices, $M(d,\mathbb{C})$.}. In many-body finite quantum systems, the algebra of observables is the tensor product of single-body algebras. For a composite system $A=BR$ with $\mathcal H_{BR}=\mathcal H_B\otimes\mathcal H_R$ and density matrices $\rho_{BR},\rho_B,\rho_R$, the algebras are $\mN=B(\mathcal H_B)\otimes1$, $\mR=1\otimes B(\mathcal H_R)$ and $\mM=\mN\vee\mR=B(\mathcal H_{BR})$.\footnote{Throughout the paper an inclusion of observable algebras is written $\mN\subset\mM$, $\mR=\mN'\cap\mM$ is its relative commutant, and $\mM'$, $\mN'$ are the commutants on the Hilbert space. Roman capitals are reserved for physical systems.} 
The algebra of $R$ is the \emph{relative commutant} of this inclusion, $\mR=\mN'\cap\mM$.
The partial trace over $R$ in a fixed uncorrelated state $\chi_R$ of the environment, read in the Heisenberg picture, is a unital completely positive projection (conditional expectation)
\begin{align}
    \mE_{\chi_R}:\mM\to\mN,\qquad
    \mE_{\chi_R}(b\otimes r)=\chi_R(r)\,(b\otimes 1)\ ,
    \label{eq:slice-expectation-intro}
\end{align}
and a simple example of a coarse-graining to a subsystem. In infinite-dimensional quantum systems (von Neumann algebras) an inclusion of factors (subfactors) $\mN\subset\mM$ plays the role of a subsystem. However, before generalizing to infinite dimensions, it is helpful to remind ourselves of finite-dimensional (matrix) algebra subalgebras of $B(\mH)$. Following the QEC terminology, we classify them as:
\begin{enumerate}
 \item {\bf Subspace inclusions:} Let $\mathcal H=P\mathcal H\oplus(1-P)\mathcal H$ for some projection $P$, and consider the unital subalgebra
 \[
  \mN_P=P B(\mathcal H)P\oplus\mathbb C(1-P).
 \]
 Its center is $\mathbb CP\oplus\mathbb C(1-P)$. For a fixed state $\chi_{1-P}$
 on the complementary corner, the conditional expectation is
 \[
  \mE_{\chi_{1-P}}(a)=PaP+
  \chi_{1-P}\bigl((1-P)a(1-P)\bigr)(1-P).
 \]
 It preserves precisely the block-diagonal states whose complementary block
 is proportional to the density matrix of $\chi_{1-P}$.
 Block diagonality alone does not suffice for a fixed expectation.
In the Hilbert space of a single quantum particle moving on some ambient space $X$, a net of inclusions of regions in $X$, $I_i\subset I_{i+1}$, defines a net of inclusions of projections $P_i\leq P_{i+1}$ and a net of subspace inclusions $\mN_{P_i}\subseteq\mN_{P_{i+1}}$. 
 
 \item {\bf Subsystem (subfactor) inclusions:} In $\mathcal H=\mathcal H_B\otimes\mathcal H_R$ we consider the subalgebra $B(\mathcal H_B)\otimes1_R$.
 Its expectations are those of \eqref{eq:slice-expectation-intro}, which preserve the product states $\rho_A=\rho_B\otimes\chi_R$. The Hilbert space of a many-body quantum system on a lattice associates nested local observable algebras $\mN_{I_i}\subseteq\mN_{I_j}$ to nested regions $I_i\subseteq I_j$. A compatible family of expectations $\mE_{I_i}:\mM\to\mN_{I_i}$, for example the expectations in a fixed product state, satisfies $\mE_{I_i}\mE_{I_j}=\mE_{I_i}$ for $I_i\subseteq I_j$. The algebraic nesting is independent of this choice of expectations. In continuum QFT the local algebras still form a net even when normal expectations onto them do not exist. For adjacent regions this can reflect the absence of faithful normal product states, as discussed in Section~\ref{subsec:quantum-coarse}.
 
 \item {\bf Mixed inclusions:} A general unital finite-dimensional subalgebra takes the block diagonal form
 \[
  \mathcal H=\bigoplus_a\mathcal H_B^{(a)}\otimes\mathcal H_R^{(a)},\qquad
  \mN=\bigoplus_a B(\mathcal H_B^{(a)})\otimes1_R^{(a)}.
 \]
 The center records the block label $a$. A choice of normalized hidden state $\chi_R^{(a)}$ specifies an
 expectation in each block. The states preserved under the projections by conditional expectation take the Markovian form
 \[
  \rho=\bigoplus_a p_a\,\rho_B^{(a)}\otimes\chi_R^{(a)}.
 \]
 With the label stored in a classical register $Z$, these have $I(B:R|Z)=0$.
\end{enumerate}
Note that the classification above is central to the algebraic formulation of QEC where $\mN$ specifies the logical observable algebra to be protected against the chosen noise \cite{BenyKempfKribs2007}\footnote{The subspace construction resolves alternatives among microstates, while a tensor
subsystem retains simultaneously specified degrees of freedom. The same distinction
organizes the classical constructions in Appendix~\ref{sec:classical}.
}. In QEC jargon, the corresponding codes are called {\it subspace codes}, {\it subsystem codes}, and {\it mixed codes}, respectively.

For the general block decomposition above, the relative commutant is
\begin{equation}
 \mR=\mN'\cap\mM
 =\bigoplus_a1_B^{(a)}\otimes B(\mathcal H_R^{(a)}).
\end{equation}
It contains $Z(\mN)$, whose projections select the blocks. A single hidden Hilbert space $\mathcal H_R$ is intrinsic only in the tensor-product case. Mixed inclusions have a hidden factor in each block. 

Since $\mM=B(\mH)$ and $\mN$ is a von Neumann subalgebra, in all the examples above, its double relative commutant is equal to itself:
\begin{align}
    \mN=(\mN'\cap \mM)'\cap\mM
\end{align}
We say that such an inclusion equals its double relative commutant.\footnote{In subfactor theory this property is called normality of the inclusion. We avoid the word, since normal is reserved for states and maps in this paper.} When $\mM$ has a center, this property can fail, as for inclusions of Abelian algebras. Here the relative commutant $\mN'\cap\mM$ is taken inside $\mM$, so that the double relative commutant does not depend on the Hilbert space on which $\mM$ is represented, whereas the commutant $\mN'$ itself does.

However, when the algebra $\mN$ has a nontrivial center (cases 1 and 3), the algebra and its relative commutant do not additively generate the whole $\mM$. In the second case, by contrast, we have $\mM=\mN\vee \mR$ and we say the inclusion is {\it conormal}.

Outer automorphisms can arise from classical block labels, and an outer action of a discrete group $G$ on a factor $\mN$ gives an inclusion $\mN\subset\mN\rtimes G$ with trivial relative commutant, an \emph{irreducible} inclusion. See Appendix~\ref{app:crossed-products} for some examples.

In this work, we are primarily interested in the infinite-dimensional factors describing
subsystems of many-body quantum systems in the thermodynamic and continuum limit.
In these systems, inclusions can have features that are impossible in
finite-dimensional tensor-product examples: 
\begin{itemize}
\item A proper inclusion of factors can have a trivial relative commutant. Such an inclusion does not equal its double relative commutant. We call such an inclusion irreducible.

\item We can have inclusions of factors with nontrivial relative commutant $\mR$ and conormal $\mM=\mN\vee \mR$, but no normal conditional expectations. This reflects the absence of uncorrelated normal states of $\mN$ and $\mR$.

\item We can have inclusions with common cyclic and separating vectors, or, more strongly, standard inclusions meaning that $\mN$, $\mM$ and $\mR$ all share a common cyclic and separating vector.
\end{itemize}
We introduce the modular notions needed for these examples briefly below.
For completeness, the general definitions of von Neumann algebras, the
types of factors, and the distinction between arbitrary unitary dynamics
and modular dynamics are reviewed in Appendix~\ref{app:extra-operator-algebra}.

\subsection{Modular theory in brief}
For a von Neumann algebra $\mM$ with a faithful normal state $\omega$, the GNS representation has a cyclic and separating vector $\ket{\Omega}$ and the Hilbert space $\mH_\mM=\overline{\mM\ket{\Omega}}$. On the dense subspace $\mM\ket{\Omega}$, a linear map $\Phi:\mM\to\mM$ is represented by an operator $F$ on $\mH_\mM$ through
\begin{align}
    F a\ket{\Omega}=\Phi(a)\ket{\Omega},\qquad a\in\mM
\end{align}
The Tomita super-operator $\mathcal{S}(a)=a^*$ is represented by the Tomita operator $S_\omega a\ket{\Omega}=a^*\ket{\Omega}$ with polar decomposition $S_\omega=J_\omega\Delta_\omega^{1/2}$, where $\Delta_\omega$ is the modular operator and $J_\omega$ the modular conjugation. The modular Hamiltonian $K_\omega=-\log\Delta_\omega$ implements the modular automorphisms through $e^{-itK_\omega}=\Delta_\omega^{it}$:
\begin{align}
\sigma_t^\omega(a)=\Delta_\omega^{it}a\Delta_\omega^{-it},\qquad a\in\mM
\end{align}
Modular conjugation maps the algebra to its commutant:
\begin{align}
&j_\omega(a)=J_\omega a J_\omega\in \mM',\qquad a\in\mM
\end{align}
 Modular flow is the unique state-preserving automorphism of the algebra $\mM$ that satisfies the KMS condition. In a finite matrix algebra, it reduces to conjugation by $\rho_\omega^{it}$, whereas local QFT algebras are typically type III$_1$ factors for which no intrinsic density matrix is available.

The relative entropy of normal states is defined using relative modular operators, and decreases upon restriction to a subalgebra~\cite{Araki1976,Araki1977}. This monotonicity is the information-theoretic input for the sufficiency decomposition that is central to this paper. Readers familiar with these notions can go directly to the inclusion theory in the next subsection. For completeness, we have included detailed definitions, the types of factors, and a comparison of general dynamics with modular dynamics in Appendix~\ref{app:extra-operator-algebra}.

\subsection{General quantum subsystems (subfactors)}
\label{subsec:quantum-coarse}

Since the center of an algebra does not evolve with the modular flow of any state, to discuss ergodicity, we focus on subfactors $\mN\subset \mM$. The modular flow of a faithful state $\omega$ on $\mM$ is a closed, automorphic dynamics of $\mM$. From the point of view of the subalgebra $\mN$, however, it is an open dynamics. The flow can transport $\mN$ outside itself and move it around $\mM$. In tensor-product or conormal examples the relative commutant $\mR$ plays the role of the environment inside $\mM$. However, for a general non-conormal inclusion, such as proper irreducible inclusions (trivial relative commutant), the relative commutant does not contain all discarded information. In this paper, we often use the intuitive language of mixing the degrees of freedom in the subalgebra $\mN$ and the relative commutant $\mR$, which strictly speaking holds only in conormal inclusions. Note that the inclusion of algebras of QFT, associated to the inclusion of regions $A=BR$ where the boundaries of regions $B$ and $R$ touch, is believed to be conormal. In QFT literature, this is often called the strong additivity condition \cite{Wiesbrock1993b,GuidoLongoWiesbrock1998}.

In this subsection, we review the basic notions of subfactors and conditional expectations in general von Neumann algebras. We introduce the notion of subsystem flow, the central object of the paper.

\paragraph{Conditional expectations:} An inclusion $\mN\subseteq\mM$ specifies the retained observables.
When it exists, a \emph{normal conditional expectation}
$\mE_{\mN}:\mM\to\mN$ is a completely positive, unital projection
with the bimodule property
\begin{equation}
 \mE_{\mN}(bac)=b\,\mE_{\mN}(a)\,c,
 \qquad a\in\mM,\quad b,c\in\mN;
 \label{eq:CE-bimodular}
\end{equation}
it is \emph{faithful} if $\mE_{\mN}(a^*a)=0$ implies $a=0$. A conditional expectation specifies how a state on $\mN$ is extended to $\mM$\footnote{Unlike the classical conditional expectation in Appendix~\ref{sec:classical}, which exists for every coarse algebra,
a normal expectation onto an infinite-dimensional subfactor need not exist, as we show below.}. The two results below establish that when conditional expectations exist (for conormal inclusions), there is a tensor product decomposition $\mM=\mN\otimes \mR$ and there are uncorrelated states of the form $\omega=\omega_\mN\otimes\omega_\mR$.

\begin{lemma}\label{lem:expectation-slice}
For subfactors $\mN\subset\mM$, every normal faithful conditional expectation $\mE:\mM\to\mN$ determines a faithful normal state $\chi_{\mR}$ on
the relative commutant $\mR$ such that
\begin{equation}
 \mE(c)=\chi_{\mR}(c)1,
 \qquad \mE(b c)=\chi_{\mR}(c)b,
 \qquad b\in\mN,\quad c\in\mR.
\end{equation} 
If the inclusion is conormal and an expectation exists, this gives a one-to-one correspondence between normal faithful expectations and faithful normal states on $\mR$\footnote{In fact, in the operator algebraic formulation of this problem, we do not need the assumption of conormality here \cite{Longo1989}.}.
\end{lemma}
The proof is given in Appendix~\ref{app:expectation-proofs}.

\begin{lemma}[conormality and tensor product]\label{prop:conormal-factorization}
Under the hypotheses of Lemma~\ref{lem:expectation-slice}, multiplication
extends to a normal $*$-isomorphism
\begin{equation}
 m:\mN\overline\otimes\mR
 \overset{\cong}{\longrightarrow}\mN\vee\mR,
 \qquad m(b\otimes c)=b c.
\end{equation}
If the inclusion is conormal, then
\begin{equation}\label{eq:conormal-factorization}
\mM\cong
 \mN\overline\otimes\mR .
\end{equation}
\end{lemma}
The proof is given in Appendix~\ref{app:expectation-proofs}.

\subsection{Subsystem flow}
\label{subsec:subsystem-flow-background}

Let $\ket{\Omega}$ be the cyclic and separating vector corresponding to the state $\omega$ on $\mM$ in the GNS Hilbert space $\mH_\mM$. Let $\mH_\mN$ be the GNS Hilbert space of $(\mN,\omega_\mN)$, with cyclic vector $\ket{\Omega_\mN}$, on which $\mN$ acts by its GNS representation. The canonical isometry $V$ and its adjoint are
\begin{align}
 &V:\mH_\mN\hookrightarrow\mH_\mM,\qquad
   V(n\ket{\Omega_\mN})=n\ket{\Omega},\qquad
   V^\dagger:\mH_\mM\to\mH_\mN,\nonumber\\
 &V^\dagger V=1_{\mH_\mN},\qquad VV^\dagger=e_\mN.
\end{align}
The range of $V$ is the closed subspace $\overline{\mN\ket{\Omega}}=e_\mN\mH_\mM$, and $e_\mN$ is the projection onto it, often called the Jones projection~\cite{Jones1983}.  The algebra $\mN$ acts on $\mH_\mN$ by its GNS representation and on $\mH_\mM$ by inclusion, and $V$ intertwines the two actions,
\begin{equation}
 nV=Vn,\qquad n\in\mN.
 \label{eq:V-intertwining}
\end{equation}
Hence $V^\dagger nV=n$ and $VnV^\dagger=ne_\mN$, so that $V\mN V^\dagger=\mN e_\mN$, and $e_\mN$ commutes with $\mN$. We write $\ket{\Omega_\mN}$ when distinguishing the vector in the smaller space from $V\ket{\Omega_\mN}=\ket{\Omega}$. The modular operator $\Delta_{\omega;\mM}$ is defined on $\mH_\mM$, and the modular operator $\Delta_{\omega;\mN}$ is defined on $\mH_\mN$.

Unital CP maps preserving $\omega$ define contractions on the GNS Hilbert space $\mH_\mM$. However, not every contraction on $\mH_\mM$ comes from a unital CP map on $\mM$ \cite{furuya2022real}. The compression to the subsystem is carried by the co-isometry $V^\dagger$,
\begin{equation}
    \operatorname{Ad}_{V^\dagger}:\mM\to B(\mH_\mN),\qquad a\longmapsto V^\dagger aV.
\end{equation}
This map is unital and CP, but its image need not lie in $\mN$. Equivalently, the supported compression $a\mapsto e_\mN a e_\mN$ has unit $e_\mN$ and need not take values in $\mN e_\mN$, or in $\mM$. On the commutant, however, compression always defines a unital CP map~\cite{AccardiCecchini1982}:
\begin{equation}
    \mE'_\mN:\mM'\to\mN',\qquad \mE'_{\mN}(a')=V^\dagger a'V,\qquad a'\in\mM',
\end{equation}
where $\mN'$ acts on $\mH_\mN$. Conjugating this compression with the two modular conjugations therefore always defines a unital CP map from $\mM$ to $\mN$, the Petz map of the inclusion~\cite{Petz1986,Petz1988}:
\begin{align}
 &\mathsf E_\omega(a)=j_\mN\circ\mE'_\mN\circ j_\mM(a):\mM\to\mN,\nonumber\\
 &\mathsf E_\omega(a)\ket{\Omega_\mN}
 =J_\mN V^\dagger J_\mM a\ket{\Omega}\in\mH_\mN,
 \qquad a\in\mM.
 \label{eq:background-Petz-dual}
\end{align}
It preserves the reference state, but need not be a conditional expectation. This map is called the Petz dual of the inclusion or the Accardi--Cecchini generalized conditional expectation~\cite{AccardiCecchini1982}. We use the name {\it Petz map} throughout and write $\Gamma=J_\mN V^\dagger J_\mM$ for the operator on the GNS Hilbert space that implements it. In Section~\ref{subsec:analytic-subsystem-flow} we will see that $\Gamma$ is the adjoint of the upper boundary value of the analytically continued subsystem flow.

Write $\Delta_\mM=\Delta_{\omega;\mM}$ and $\Delta_\mN=\Delta_{\omega;\mN}$. The two modular operators act on different Hilbert spaces. The embedding $V$ lets us compare their dynamics without identifying these spaces.

\begin{definition}[Subsystem flow]\label{def:subsystem-flow}
For an inclusion $\mN\subset\mM$ with reference vector $\ket{\Omega}$, define the subsystem flow on the ambient GNS Hilbert space by
\begin{equation}
 D_{\mN\subset\mM}(t)
 :=\Delta_\mM^{-it}V\Delta_\mN^{it}V^\dagger
 \in B(\mH_\mM),\qquad t\in\mathbb R.
 \label{eq:subsystem-flow-general}
\end{equation}
\end{definition}
For real $t$, $D(t)$ is a partial isometry with
\begin{equation}
 D(0)=e_\mN,\qquad
 D(t)^\dagger D(t)=e_\mN,\qquad
 D(t)D(t)^\dagger=\Delta_\mM^{-it}e_\mN\Delta_\mM^{it}.
\end{equation}
It fixes $\ket{\Omega}$ and vanishes on $(1-e_\mN)\mH_\mM$. Its restriction to the code subspace is an isometry. If $\ket{\Omega}$ is also cyclic for $\mN$ on $\mH_\mM$, then $e_\mN=1$, the canonical identification gives $V=1$, and $D(t)$ is unitary.

At the level of observables, this operator compares the intrinsic and ambient modular motions:
\begin{equation}
 \mathcal D_t=\sigma_{-t}^{\omega;\mM}\circ\sigma_t^{\omega;\mN},\qquad
 D(t)\,n\ket{\Omega}=\mathcal D_t(n)\ket{\Omega},
 \quad n\in\mN.
\end{equation}

For a general GNS inclusion, $D(t)$ implements this comparison on the transported code corner $e_\mN\mH_\mM$ rather than on all of $\mH_\mM$,
\[
 D(t)bD(t)^\dagger=\mathcal D_t(b)e_{\mN}(t),\qquad
 e_{\mN}(t)=\Delta_{\mM}^{-it}e_{\mN}\Delta_{\mM}^{it},
\]
where $b\in\mN$ acts on the ambient Hilbert space on both sides and no conditional expectation is assumed. A key identity that subsystem flows satisfy is
\begin{equation}
 D(t+s)=\Delta_\mM^{-it}D(s)V\Delta_\mN^{it}V^\dagger = \Delta_\mM^{-it}D(s)\Delta_\mM^{it}D(t).
 \label{eq:subsystem-flow-cocycle}
\end{equation}
In general, the flow need not be a one-parameter group or belong to $\mM$.

The infinitesimal subsystem flow compares the two modular Hamiltonians. On vectors $\ket{\Psi}\in\operatorname{Dom}K_\mN$ with $V\ket{\Psi}\in\operatorname{Dom}K_\mM$,
\begin{align}
 -i\left.\partial_tD(t)V\ket{\Psi}\right|_{t=0}
 &=(K_\mM V-VK_\mN)\ket{\Psi},\nonumber\\
 -i\left.V^\dagger\partial_tD(t)V\ket{\Psi}\right|_{t=0}
 &=(V^\dagger K_\mM V-K_\mN)\ket{\Psi},
 \quad K_\mathcal A=-\log\Delta_\mathcal A.
 \label{eq:subsystem-flow-generator}
\end{align}
Subsystem flow is a generating function for relative entropies:
\begin{align}
 -i\left.\partial_t\langle\Omega|a^*D(t)a|\Omega\rangle\right|_{t=0}
 =&S_\mM(\ket{\xi}\|\ket{\Omega})-S_\mN(\ket{\xi}\|\ket{\Omega})\nonumber\\
 &+S_{\mN'}(\ket{\xi}\|\ket{\Omega})-S_{\mM'}(\ket{\xi}\|\ket{\Omega})\nonumber\\
 &+\langle\Omega|a^*(1-e_\mN)K_\mM e_\mN a|\Omega\rangle,
\end{align}
where $\ket{\xi} = a\ket{\Omega}- (1-e_\mN)a\ket{\Omega}$. Each of the first two lines is nonnegative by data processing. The last line is a matrix element of $K_\mM$ between the part of $a\ket{\Omega}$ outside the code subspace and the part inside it, so it has no definite sign and is complex in general.
For $\ket{\xi}\ne0$, assume that $\ket{\Psi}=V^\dagger\ket{\xi}/\|\ket{\xi}\|$ and $V\ket{\Psi}=\ket{\xi}/\|\ket{\xi}\|$ satisfy the cyclicity, finiteness and generator-domain hypotheses of Proposition~\ref{prop:sufficiency}. Entropies of the unnormalized vector $\ket{\xi}$ are extended homogeneously from its normalized vector state. The $S_\mN$ and $S_{\mN'}$ terms use $V^\dagger\ket{\xi}$ on $\mH_\mN$. The first difference is the data-processing loss for $\mN\subset\mM$, and the second difference is the data-processing loss for the commutant compression $a'\mapsto V^\dagger a'V$, rather than a literal inclusion when $V\ne1$. 

Monotonicity of the modular operators allows us to analytically continue $D(t)$ to a contraction in the strip $0\leq\operatorname{Im}z\leq\tfrac12$ whose upper boundary $\operatorname{Im}z=\tfrac12$ implements the Petz recovery map and whose midpoint $z=\frac{i}{4}$ defines a positive operator central to our work. The continuation is taken up in Section~\ref{subsec:analytic-subsystem-flow}.

As we stated earlier, infinite-dimensional von Neumann algebras allow more exotic inclusions of factors than those in finite-dimensional matrix algebras. Here, we review a few of these inclusions that enter the ergodic hierarchy of Section~\ref{subsec:QEC-hierarchy}.

\paragraph{Common cyclic and separating vector:} For an inclusion we
call $\ket{\Omega}$ a \emph{common} cyclic and separating vector if $e_{\mN}=1$. These inclusions can occur in algebras of all types I, II and III. However, they are more commonplace in type III algebras\footnote{Every $\sigma$-finite type III factor admits a standard representation~\cite{Blackadar2006}. And under the usual separability hypotheses, the set of vectors that are simultaneously cyclic and separating for finitely many standardly represented algebras is a dense $G_\delta$ set~\cite{dixmiermarechal}.}. In QFT, the vacuum is cyclic and separating for the local algebras by the Reeh--Schlieder theorem~\cite{ReehSchlieder1961}. Since $V=1$, the subsystem flow reduces to Borchers characteristic function \cite{revolutionizing}.

\paragraph{Standard and split inclusions:}
An inclusion $\mN\subset\mM$ is
\emph{standard} with respect to $\ket{\Omega}\in\mathcal H$ if $\ket{\Omega}$ is cyclic
and separating for $\mM$, $\mN$ and the relative commutant $\mR=\mN'\cap \mM$. It is \emph{split} if a type I factor $\mathcal F$ exists with
$\mN\subset\mathcal F\subset\mM$ \cite{DoplicherLongo1984}.

\begin{theorem}[Split inclusions and normal expectations]\label{thm:split-no-expectation}
If $\mN$ is a type III factor and
$\mN\subset\mM$ is split, then no faithful normal conditional expectation
$\mE:\mM\to\mN$ exists.
\end{theorem}
See Appendix~\ref{app:expectation-proofs} for a proof. Since the split property is common in QFT, the theorem above establishes that normal conditional expectations are not always available.

\section{Quantum error correction and modular autonomy}\label{sec:QEC}

In standard quantum error correction (QEC), an encoding is engineered so that the erased environment learns nothing about the logical input. In the setting of this paper, the inclusion $\mN\subset\mM$ and the faithful reference state $\omega$ are given by a region or by an observer. The question is whether restriction to $\mN$ admits an exactly recoverable encoding whose family of encoded states contains $\omega$. We state the answer first: by Takesaki's theorem, such an encoding exists if and only if $\mN$ is preserved by the modular flow of $\mM$ with $\omega$, and this is the same as triviality of the subsystem flow of Section~\ref{subsec:subsystem-flow-background}. Section~\ref{subsec:finite-QEC} then recalls, for a finite-dimensional code, how exact recovery singles out a logical algebra inside the retained system together with a conditional expectation onto it. In Section~\ref{subsec:symmetries-QEC-subtheories}, we give examples of algebras with symmetries as basic examples of inclusions with the QEC property above.

\subsection{Reference-preserving erasure codes and modular autonomy}
\label{subsec:QEC-takesaki}

We start with a comment on the notation. Throughout this section, $\mR=\mN'\cap\mM$ denotes the relative commutant of the retained algebra, as in Section~\ref{subsec:quantum-coarse}. Recall that for the tensor-product inclusion of Section~\ref{section:basics} it is the algebra of the erased system. In an operator algebraic setting of an inclusion $\mN\subset \mM$, the erasure map in the Schrödinger picture is the restriction of normal states
\begin{equation}
 \mathcal T:\mM_*\to\mN_*,
 \qquad\mathcal T(\varphi)=\varphi|_{\mN},
 \label{eq:erasure-channel}
\end{equation}
where $\mM_*$ and $\mN_*$ are the preduals, which, in addition to states, include all normal functionals that are not necessarily positive.
In the Heisenberg picture, the adjoint is the inclusion $\mN\hookrightarrow\mM$. Write $S_{\rm n}(\mM)$ for the normal states, to distinguish them from all normal functionals in $\mM_*$. A normal conditional expectation $\mE:\mM\to\mN$ defines an encoding (or state extension)
\begin{equation}
 \mE_*:\mN_*\to\mM_*,
 \qquad\varphi_{\mN}\mapsto\varphi_{\mN}\circ\mE.
 \label{eq:encoding}
\end{equation}
Since $\mE|_\mN=\operatorname{id}$, its encoded family
\begin{equation}
 \mathsf C_{\mE}
 =\mE_*\bigl(S_{\rm n}(\mN)\bigr)
 =\{\varphi\in S_{\rm n}(\mM):\varphi\circ\mE=\varphi\}
 \label{eq:code}
\end{equation}
is exactly recovered after restriction: $\mathcal T\circ\mE_*=\operatorname{id}$ on $\mN_*$ and $\mE_*\circ\mathcal T=\operatorname{id}$ on $\mathsf C_{\mE}$. The same map $\mE_*$ prepares the specified extension and reconstructs it from the retained state. The additional condition
\begin{equation}
 \omega\circ\mE=\omega
 \quad\Longleftrightarrow\quad\omega\in\mathsf C_{\mE}
 \quad\Longleftrightarrow\quad S(\omega\|\omega\circ\mE)=0
 \label{eq:QEC-entropy}
\end{equation}
requires the prescribed reference to belong to this code. Preservation of logical expectation values under restriction is automatic. Reconstructing their specified extension to all of $\mM$, including its correlations, is the extra important requirement.

For $\mM=\mN\overline\otimes\mR$, the slice expectation gives
\begin{equation}
 \mE_\chi(n\otimes r)=\chi(r)n,\qquad
 \mathsf C_{\mE_\chi}
 =\{\varphi_{\mN}\otimes\chi:\varphi_{\mN}\in S_{\rm n}(\mN)\}.
 \label{eq:reference-product-code}
\end{equation}
A faithful reference is in this code precisely when it has the same fixed hidden state and no logical--hidden correlations. In matrices, for $\mH_A=\mH_B\otimes\mH_R$ as in Section~\ref{section:basics},
\begin{equation}
 S(\rho_A\|\rho_{B}\otimes\chi_{R})
 =I_\rho(B:R)+S(\rho_{R}\|\chi_{R}).
 \label{eq:quantum-coarse-entropy-decomposition}
\end{equation}
This requirement is relative to the specified logical algebra and reference state. Even if $\omega$ is not preserved, exact erasure recovery may still hold with respect to a different reference state or a different logical subalgebra.

\begin{theorem}[Reference-preserving correctable inclusions]
\label{thm:correctable}
Let $\mN\subset\mM$ be an inclusion of von Neumann algebras and let $\ket{\Omega}$ represent a faithful normal state $\omega$ in the GNS Hilbert space $\mH_{\mM}$. Let $\mH_\mN$ be the GNS Hilbert space of $\mN$, $V:\mH_\mN\hookrightarrow\mH_\mM$ the canonical isometry, and $e_\mN=VV^\dagger$ (Section~\ref{subsec:subsystem-flow-background}). The following are equivalent.
\begin{enumerate}
 \item[\rm(i)] \emph{Modular autonomy:} $\sigma_t^{\omega;\mM}(\mN)=\mN$ for every $t\in\mathbb R$.
 \item[\rm(ii)] \emph{Expected inclusion:} there is a normal conditional expectation $\mE:\mM\to\mN$ with $\omega\circ\mE=\omega$.
 \item[\rm(iii)] \emph{Trivial subsystem flow:} $D_{\mN\subset\mM}(t)=e_{\mN}$ for every $t$.
 \item[\rm(iv)] \emph{Knill--Laflamme condition:} $e_{\mN}\mM e_{\mN}\subseteq\mN e_{\mN}$.
\end{enumerate}
When these conditions hold, $\mE$ is unique and defines the exact erasure code \eqref{eq:code} containing $\omega$. It is characterized by
\begin{equation}
 \mE(x)\ket{\Omega}=e_{\mN}x\ket{\Omega},\qquad
 e_{\mN}xe_{\mN}=\mE(x)e_{\mN},\qquad x\in\mM.
 \label{eq:expectation-code-compression}
\end{equation}
\end{theorem}
\begin{proof}
The equivalence (i)$\Leftrightarrow$(ii), uniqueness, and the first identity in \eqref{eq:expectation-code-compression} are Takesaki's theorem~\cite{Takesaki1972}. Under these conditions the intrinsic and ambient modular evolutions agree on $\mN$, and
\begin{align}
 \sigma_t^{\omega;\mM}(n)&=\sigma_t^{\omega;\mN}(n),\qquad n\in\mN,\nonumber\\
 \Delta_{\mM}^{it}V&=V\Delta_{\mN}^{it}.
 \label{eq:takesaki-intertwining}
\end{align}

This gives (iii). Conversely, $D(t)=e_\mN$ implies
$D(t)V=V$, hence $\Delta_\mM^{-it}V=V\Delta_\mN^{-it}$.
Applying this equality to $n\ket{\Omega}$ and using separatingness of $\ket{\Omega}$
for $\mM$ gives (i).

For (ii)$\Rightarrow$(iv), bimodularity and state preservation give
\[
 \langle\Omega|n^*xn'|\Omega\rangle
 =\omega(n^*xn')
 =\omega\bigl(n^*\mE(x)n'\bigr),\qquad n,n'\in\mN.
\]
Density yields the second identity in \eqref{eq:expectation-code-compression}. Conversely, (iv) defines $\mE(x)$ by that identity. The representation $n\mapsto ne_{\mN}$ is normal and faithful because $\ket{\Omega}\in e_{\mN}\mH_{\mM}$ is separating for $\mN$. Its inverse, composed with compression, makes $\mE$ normal, UCP and $\mN$-bimodular. Since $e_{\mN}\ket{\Omega}=\ket{\Omega}$, it preserves $\omega$, proving (ii). The code identities follow from \eqref{eq:encoding}--\eqref{eq:code}.
\end{proof}

The canonical Petz map of the inclusion exists even when these conditions fail. It preserves the reference, but it need not fix every element of $\mN$. It becomes a conditional expectation precisely in the autonomous case, when it coincides with $\mE$~\cite{AccardiCecchini1982,Petz1988}\footnote{See \eqref{eq:QEC-Petz-dual} for the modular-conjugation form of the Petz map.}. Thus, the nontrivial question is whether the reference-compatible recovery fixes the entire prescribed logical algebra, rather than only a smaller recoverable part. We remind the reader that the modular autonomy condition above is the invariance of the subsystem, and crucially different from fixed-point subalgebras, which would require the modular flow to fix each logical observable pointwise.

\paragraph{Knill--Laflamme conditions and the relative commutant:}
\label{subsec:QEC-KL}
Condition (iv) in Theorem~\ref{thm:correctable} has the interpretation of the Knill--Laflamme conditions for correctability. For an inclusion of factors, bimodularity makes the coefficient on the relative commutant scalar: $\mE(r)=\chi(r)1$, where $\chi=\omega|_{\mR}$. For conormal inclusions, this gives the usual environment-side criterion.

\begin{corollary}[Conormal inclusions]
\label{cor:KL-conormal}
Let $\mN\subset\mM$ be an inclusion of factors, with $\mR=\mN'\cap\mM$. If $\mN\vee\mR=\mM$, conditions {\rm(i)--(iv)} are equivalent to
\begin{align}
 e_{\mN}re_{\mN}&=\chi(r)e_{\mN},\qquad r\in\mR,
 \label{eq:reference-relative-KL}\\
 \omega(nr)&=\omega(n)\chi(r),\qquad n\in\mN,\ r\in\mR.
 \label{eq:reference-factorization}
\end{align}
Equivalently, the restriction to $\mR$ is constant on the unit vectors of $\mH_{\mN}$.
\end{corollary}
\begin{proof}
For $n,n'\in\mN$ and $r\in\mR$, the compression identity reads
$\omega(n^*rn')=\chi(r)\omega(n^*n')$. Since $r$ commutes with $\mN$, this is equivalent to \eqref{eq:reference-factorization}. Polarization gives the assertion about unit vectors. An $\omega$-preserving expectation implies factorization by Lemma~\ref{lem:expectation-slice}. Conversely, in a conormal inclusion the faithful normal product state in \eqref{eq:reference-factorization} implements the spatial product $\mM\simeq\mN\overline\otimes\mR$, and $\operatorname{id}\otimes\chi$ is the required normal expectation~\cite{DoplicherLongo1984}.
\end{proof}

Without conormality, the relative commutant need not contain all discarded information, and its scalar compression condition is insufficient. For an irreducible inclusion $\mR=\mathbb C1$, that condition is automatic, while modular autonomy can fail. The full criterion remains (iv). For the relative-entropy comparison in Section~\ref{subsec:QEC-sufficiency}, the relevant complementary inclusion is $\mM'\subset\mN'$. It retains information not accounted for by $\mR$ alone. For a discussion of the connection to the conventional finite-dimensional Knill--Laflamme condition and complementary-channel dictionary, see Appendix~\ref{app:finite-KL}.

\paragraph{Code isometries and the subsystem flow:}
The canonical isometry $V$ in Theorem~\ref{thm:correctable} is fixed by the inclusion and the reference state $(\mN\subset\mM,\omega)$. It is not an encoding isometry chosen for a noise channel, such as $V_{\rm enc}$ of Section~\ref{subsec:finite-QEC} below. Definition~\ref{def:subsystem-flow} of subsystem flow compares the intrinsic and ambient modular evolutions using the canonical isometry $V$. The theorem identifies $D(t)=e_{\mN}$ for every $t$ with exact recovery of the specified reference-preserving code. The projection $e_{\mN}$ alone records only the GNS subspace. For a common cyclic vector, since $e_\mN=1$, the code subspace is the entire Hilbert space $\mH_\mM$, and we will see that the modular autonomy condition is never satisfied in a proper inclusion.

\subsection{The finite-dimensional dictionary}
\label{subsec:finite-QEC}

Consider a finite physical system $A=BR$, with $\mH_A=\mH_B\otimes\mH_R$ and the algebras $\mM=B(\mH_A)$, $\mN=B(\mH_B)\otimes1$ and $\mR=1\otimes B(\mH_R)$ of Section~\ref{section:basics}. We retain $B$ and erase $R$. A logical system $L$, with observable algebra $\cL=B(\mH_L)$, is encoded into $A$ by an isometry,
\begin{equation}
 V_{\rm enc}:\mH_L\longrightarrow\mH_B\otimes\mH_R,
 \qquad
 \mH_{\rm code}=V_{\rm enc}\mH_L,\qquad
 P=V_{\rm enc}V_{\rm enc}^{\dagger},
 \label{eq:physical-code-isometry}
\end{equation}
and the encoded logical algebra is $V_{\rm enc}\cL V_{\rm enc}^{\dagger}\subset P\mM P$. The channel is the encoding followed by the erasure
\begin{equation}
 \mathcal F_*(\rho_L)
 =\Tr_{R}\bigl(V_{\rm enc}\rho_LV_{\rm enc}^{\dagger}\bigr).
 \label{eq:two-cut-noise}
\end{equation}
This is the Stinespring form of a noise channel~\cite{Stinespring1955}: every channel on $\cL$ is an isometry into a larger system followed by the erasure of part of it. The logical subsystem carrying the information need not coincide with the physical subsystem retained after the erasure. When $A$ also admits a logical--auxiliary factorization $\mH_A\simeq\mH_L\otimes\mH_{R_{\rm aux}}$, the encoding has a unitary realization, $V_{\rm enc}\ket{\psi}=U(\ket{\psi}\otimes|0\rangle_{R_{\rm aux}})$. The auxiliary system prepares the encoding, whereas the erased system is specified by the physical decomposition $A=BR$. The two decompositions of $A$ need not coincide. The auxiliary state may also be mixed, in which case its spectral decomposition gives a mixture of encodings into orthogonal subspaces. We use the pure-state encoding for the erasure criterion below.

\begin{figure}[!htbp]
\centering
\resizebox{0.8\linewidth}{!}{\begin{tikzpicture}[x=1cm,y=1cm,line cap=round,line join=round,font=\small]
\definecolor{qblue}{RGB}{31,110,147}
\definecolor{qgreen}{RGB}{51,119,103}
\definecolor{qred}{RGB}{176,60,60}
\definecolor{qgray}{RGB}{74,85,103}
\tikzset{
 wire/.style={draw=qgray,line width=1.05pt},
 Viso/.style={draw=qblue,line width=1.05pt,fill=qblue!12},
 Wiso/.style={draw=qgreen,line width=1.05pt,fill=qgreen!12},
 opbox/.style={draw=qgray,line width=0.95pt,fill=white,minimum width=0.62cm,minimum height=0.5cm,inner sep=1.5pt},
 lab/.style={font=\footnotesize,text=qgray},
 panel/.style={font=\bfseries\small,anchor=north west},
 formula/.style={font=\small,text=qgray,anchor=north,align=center},
 V/.pic={
   \draw[Viso] (0,0) -- (1.0,0.85) -- (-1.0,0.85) -- cycle;
   \node[font=\footnotesize] at (0,0.57) {$V_{\rm enc}$};
 },
 Vdag/.pic={
   \draw[Viso] (-1.0,0) -- (1.0,0) -- (0,0.85) -- cycle;
   \node[font=\footnotesize] at (0,0.28) {$V_{\rm enc}^{\dagger}$};
 },
 W/.pic={
   \draw[Wiso] (0,0) -- (0.8,0.85) -- (-0.8,0.85) -- cycle;
   \node[font=\footnotesize] at (0,0.57) {$W$};
 },
 state/.pic={
   \draw[Wiso] (0,0) -- (0.8,0.85) -- (-0.8,0.85) -- cycle;
   \node[font=\footnotesize] at (0,0.57) {$|\xi\rangle$};
 }
}

\begin{scope}
 \node[panel] at (-0.2,5.6) {(a)};
 \draw[wire] (1.6,0.3) -- (1.6,0.9);  \node[lab,below] at (1.6,0.3) {$L$};
 \pic at (1.6,0.9) {V};
 \draw[wire] (1.2,1.75) -- (1.2,2.35); \node[opbox] at (1.2,2.6) {$y$}; \draw[wire] (1.2,2.85) -- (1.2,3.5);
 \draw[wire] (2.0,1.75) -- (2.0,3.5);
 \node[lab,left] at (1.0,2.0) {$B$}; \node[lab,right] at (2.05,2.6) {$R$};
 \pic at (1.6,3.5) {Vdag};
 \draw[wire] (1.6,4.35) -- (1.6,4.9); \node[lab,above] at (1.6,4.9) {$L$};
 \node[font=\large] at (3.4,2.6) {$=$};
 \draw[wire] (4.6,0.3) -- (4.6,2.35); \node[opbox] at (4.6,2.6) {$\Phi(y)$}; \draw[wire] (4.6,2.85) -- (4.6,4.9);
 \node[lab,below] at (4.6,0.3) {$L$}; \node[lab,above] at (4.6,4.9) {$L$};
 \node[formula] at (3.0,-0.35) {$\Phi(y)=V_{\rm enc}^{\dagger}(y\otimes1_R)V_{\rm enc}$};
\end{scope}

\begin{scope}[shift={(7.4,0)}]
 \node[panel] at (-0.2,5.6) {(b)};
 \draw[wire] (1.6,0.3) -- (1.6,0.9);  \node[lab,below] at (1.6,0.3) {$L$};
 \pic at (1.6,0.9) {V};
 \draw[wire] (1.2,1.75) -- (1.2,3.5);
 \draw[wire] (2.0,1.75) -- (2.0,2.35); \node[opbox] at (2.0,2.6) {$r$}; \draw[wire] (2.0,2.85) -- (2.0,3.5);
 \node[lab,left] at (1.15,2.6) {$B$}; \node[lab,right] at (2.3,2.0) {$R$};
 \pic at (1.6,3.5) {Vdag};
 \draw[wire] (1.6,4.35) -- (1.6,4.9); \node[lab,above] at (1.6,4.9) {$L$};
 \node[font=\large] at (3.4,2.6) {$=$};
 \node[anchor=west] at (3.9,2.6) {$\chi_R(r)$};
 \draw[wire] (5.3,0.3) -- (5.3,4.9);
 \node[lab,below] at (5.3,0.3) {$L$}; \node[lab,above] at (5.3,4.9) {$L$};
 \node[formula] at (3.0,-0.35) {$V_{\rm enc}^{\dagger}(1_B\otimes r)V_{\rm enc}=\chi_R(r)\,1_L$};
\end{scope}

\begin{scope}[shift={(0,-6.6)}]
 \node[panel] at (-0.2,5.6) {(c)};
 \draw[wire] (1.6,0.3) -- (1.6,0.9);  \node[lab,below] at (1.6,0.3) {$L$};
 \pic at (1.6,0.9) {V};
 \draw[wire] (1.2,1.75) -- (1.2,2.5);  \node[lab,left] at (1.1,2.1) {$B$};
 \pic at (1.2,2.5) {W};
 \draw[wire] (0.9,3.35) -- (0.9,4.9);  \node[lab,above] at (0.9,4.9) {$L$};
 \draw[wire] (1.5,3.35) -- (1.5,4.9);  \node[lab,above] at (1.5,4.9) {$S$};
 \draw[wire] (2.0,1.75) -- (2.0,4.9);  \node[lab,above] at (2.0,4.9) {$R$};
 \node[font=\large] at (3.4,2.6) {$=$};
 \draw[wire] (4.3,0.3) -- (4.3,4.9); \node[lab,below] at (4.3,0.3) {$L$}; \node[lab,above] at (4.3,4.9) {$L$};
 \pic at (5.5,2.7) {state};
 \draw[wire] (5.2,3.55) -- (5.2,4.9); \node[lab,above] at (5.2,4.9) {$S$};
 \draw[wire] (5.8,3.55) -- (5.8,4.9); \node[lab,above] at (5.8,4.9) {$R$};
 \node[formula] at (3.0,-0.35) {$(W\otimes1_R)V_{\rm enc}=1_L\otimes|\xi\rangle_{SR}$};
\end{scope}

\begin{scope}[shift={(7.4,-6.6)}]
 \node[panel] at (-0.2,5.6) {(d)};
 \draw[wire] (1.6,0.3) -- (1.6,0.75); \node[opbox] at (1.6,1.0) {$b$}; \draw[wire] (1.6,1.25) -- (1.6,1.7);
 \node[lab,below] at (1.6,0.3) {$L$};
 \pic at (1.6,1.7) {V};
 \draw[wire] (1.2,2.55) -- (1.2,4.9); \node[lab,above] at (1.2,4.9) {$B$};
 \draw[wire] (2.0,2.55) -- (2.0,4.9); \node[lab,above] at (2.0,4.9) {$R$};
 \node[font=\large] at (3.4,2.6) {$=$};
 \draw[wire] (4.9,0.3) -- (4.9,1.7); \node[lab,below] at (4.9,0.3) {$L$};
 \pic at (4.9,1.7) {V};
 \draw[wire] (4.5,2.55) -- (4.5,3.25); \node[opbox] at (4.5,3.5) {$\mathcal D(b)$}; \draw[wire] (4.5,3.75) -- (4.5,4.9);
 \node[lab,above] at (4.5,4.9) {$B$};
 \draw[wire] (5.3,2.55) -- (5.3,4.9); \node[lab,above] at (5.3,4.9) {$R$};
 \node[formula] at (3.0,-0.35) {$V_{\rm enc}\,b=(\mathcal D(b)\otimes1_R)\,V_{\rm enc}$};
\end{scope}
\end{tikzpicture}}
\caption{Tensor-network dictionary for exact erasure correction. (a) Erasing $R$ pulls $y$ back to $\Phi(y)$. (b) An operator $r$ on the erased leg reduces to the number $\chi_R(r)$. (c) A unitary $W$ splits $B$ into a logical leg $L$ and a syndrome leg $S$, and $S$ and $R$ end in a fixed state $|\xi\rangle_{SR}$. (d) A logical operator $b$ equals $\mathcal D(b)=W^{\dagger}(b\otimes1_S)W$, the recovery of \eqref{eq:logical-reconstruction}.}
\label{fig:qec-tensor}
\end{figure}
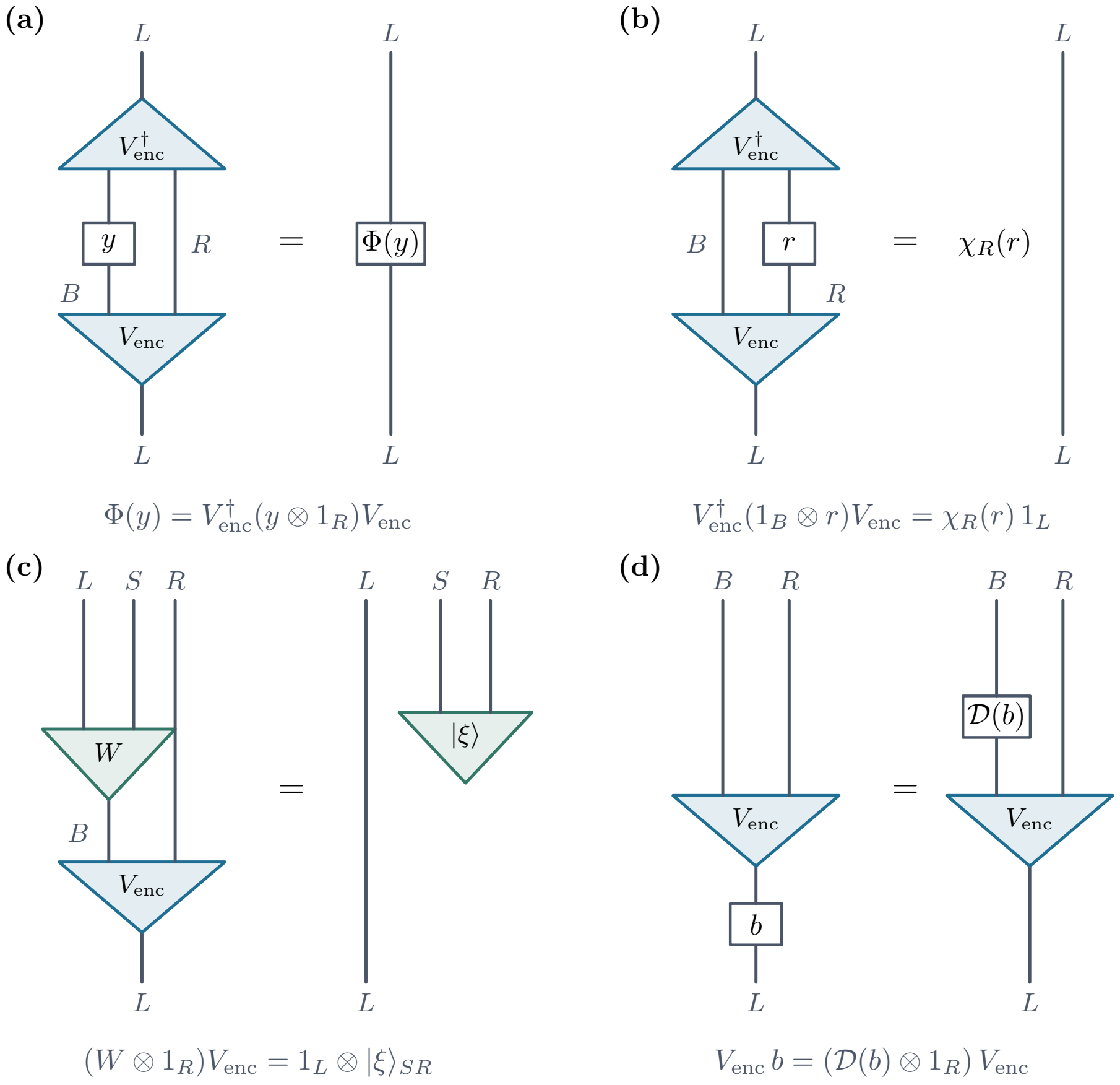

\paragraph{Exact reconstruction:}
The code corrects the erasure if there exists a recovery channel $\mathcal D_*$ acting only on $B$ that recovers every logical state:
\begin{equation}
 \mathcal D_*\circ\mathcal F_*=\operatorname{id}.
 \label{eq:Schrodinger-exact-QEC}
\end{equation}
In the Heisenberg picture, put $\Phi=\mathcal F_*^\dagger$ and $\mathcal D=\mathcal D_*^\dagger$. The map $\Phi$ pulls retained observables back to the logical system, and exact recovery becomes
\begin{gather}
 \Phi(y)=V_{\rm enc}^{\dagger}(y\otimes1_{R})V_{\rm enc},
       \qquad y\in B(\mH_B),\nonumber\\
 \Phi\circ\mathcal D(b)=b,\qquad b\in\cL.
 \label{eq:logical-reconstruction}
\end{gather}
The map $\mathcal D$ reconstructs each logical observable as a retained observable with the same action on the code. From this point onward we work in the Heisenberg picture, except when identifying a reference density matrix or the encoded family of states. The environment-side statement is the erasure Knill--Laflamme condition,
\begin{equation}
 V_{\rm enc}^{\dagger}(1_{B}\otimes r)V_{\rm enc}
 =\chi_{R}(r)\,1_{L},\qquad r\in B(\mH_R),
 \label{eq:erasure-code-KL}
\end{equation}
for a state $\chi_{R}$ independent of the logical input~\cite{KnillLaflamme1997}. Equivalently, the complementary channel sends every logical state to the same state on $R$: the erased system has learned nothing about the information stored in $L$. For a logical algebra on a code subspace, the right-hand side scalar is replaced by the identity element in the commutant of that algebra~\cite{BenyKempfKribs2007Observables}. The full code embedding is essential in this condition. The inclusion of the retained observables alone would not specify which input information must be protected.
Figure~\ref{fig:qec-tensor} draws the four identities of this subsection as tensor networks.

\paragraph{The occupied retained algebra:}
Choose a faithful logical reference density matrix $\sigma_L$ and let $Q$ be the projection onto the support of $\tau_B$ (the output of the noise map)
\begin{equation}
 \tau_{B}=\mathcal F_*(\sigma_L),\qquad
 Q=\operatorname{supp}\tau_{B},\qquad
 \mM_{\rm ret}=Q\,B(\mH_B)\,Q.
 \label{eq:retained-reference-support}
\end{equation}
The projection $Q$ records the occupied output support and is the smallest projection with $P\leq Q\otimes1$. Since the range of $V_{\rm enc}$ lies in $Q\mH_{B}\otimes\mH_{R}$, compressing a recovery to $Q$ preserves \eqref{eq:logical-reconstruction}. The Petz map of this finite-dimensional code, which realizes such a recovery, is written out in Appendix~\ref{app:finite-KL}.

\paragraph{Reconstructed logical algebras:}
For an exactly correctable code, a choice of coordinates on the occupied retained space reveals the surviving logical subsystem and a syndrome system $S$,
\begin{equation}
 Q\mH_{B}\simeq\mH_{L}\otimes\mathcal H_S,\qquad
 \mathcal D(b)=b\otimes1_S,\qquad
 \Phi(b\otimes s)=\chi_S(s)\,b .
 \label{eq:decoded-subsystem-form}
\end{equation}
In these coordinates the encoded erasure channel is $\mathcal F_*(\rho_L)=\rho_L\otimes\chi_S$: all input dependence remains in the logical factor, while the syndrome state $\chi_S$ is fixed by the noise. The output reference is $\sigma_L\otimes\chi_S$, whose modular evolution preserves the logical algebra,
\begin{equation}
 (\sigma_L\otimes\chi_S)^{it}(b\otimes1_S)
 (\sigma_L\otimes\chi_S)^{-it}
 =\sigma_L^{it}b\,\sigma_L^{-it}\otimes1_S .
 \label{eq:decoded-modular-autonomy}
\end{equation}
The following proposition establishes this structure without choosing coordinates. Consider $\Phi$ as a map $\mM_{\rm ret}\to\cL$. It is faithful on positive operators: for $y\geq0$, $\Phi(y)=0$ implies $\Tr(\tau_{B}y)=\Tr(\sigma_L\Phi(y))=0$, hence $y=0$ on $Q$. Faithfulness here concerns positive operators. It does not require linear injectivity of the noise map.

\begin{proposition}[Exact reconstruction gives an expected inclusion]
\label{prop:recovery-expectation}
Let $\Phi:\mM_{\rm ret}\to\cL$ be a faithful normal UCP map and let $\mathcal D:\cL\to\mM_{\rm ret}$ be a normal UCP map with $\Phi\circ\mathcal D=\operatorname{id}$. Then $\mathcal D$ is an injective $*$-homomorphism, and
\begin{equation}
 \widetilde{\cL}=\mathcal D(\cL)\subseteq\mM_{\rm ret},
 \qquad E_{\rm log}=\mathcal D\circ\Phi
 \label{eq:reconstructed-expectation}
\end{equation}
is a normal conditional expectation onto the reconstructed logical algebra. For every faithful normal state $\eta$ on $\cL$, it preserves the faithful normal state $\omega_{\rm ret}=\eta\circ\Phi$.
\end{proposition}

\begin{proof}
Two applications of Kadison--Schwarz inequalities~\cite{Choi1974,Paulsen2002} for the two maps give
\[
 b^*b = \Phi\mathcal{D}(b^*)\Phi\mathcal{D}(b)\leq\Phi\bigl(\mathcal D(b)^*\mathcal D(b)\bigr)
 \leq\Phi\mathcal D(b^*b)=b^*b.
\]
The positive difference $\mathcal D(b^*b)-\mathcal D(b)^*\mathcal D(b)$ is therefore annihilated by the faithful map $\Phi$, so it vanishes. Applying the same argument to $bb^*$ shows that $\mathcal D$ is multiplicative, and $\Phi\mathcal D=\operatorname{id}$ gives injectivity. The map $E_{\rm log}$ has image $\mathcal D(\cL)$ and fixes this algebra pointwise, so it is a conditional expectation. Finally,
$\omega_{\rm ret}E_{\rm log}=\eta\Phi\mathcal D\Phi=\eta\Phi=\omega_{\rm ret}$.
\end{proof}

The recovery in the proposition is the Petz map for these compatible reference states~\cite{furuya2022real}. The composite $E_{\rm log}=\mathcal D\circ\Phi$ is idempotent, fixes the reconstructed logical algebra pointwise, and includes the recovery step. There are two inclusions in this construction. The spatial inclusion $\mN\subset\mM$ specifies what is available after erasure. The reconstructed inclusion $\widetilde\cL\subset\mM_{\rm ret}$ specifies the protected logical information within it. Exact correction of $L$ does not require the whole retained algebra to be modularly autonomous in $\mM$ for an independently prescribed reference, and $V_{\rm enc}\sigma_LV_{\rm enc}^{\dagger}$ is supported on $P$ and is not faithful on $\mM$ when the code is proper. The faithful reference used in the proposition lives on the occupied retained algebra. Theorem~\ref{thm:correctable} concerns the first inclusion, with the reference prescribed in advance: the proposition shows that exact recovery always produces an expected inclusion, and Takesaki's theorem says which expected inclusions a given reference admits.

\subsection{Symmetries and QEC}\label{subsec:symmetries-QEC-subtheories}

Internal symmetries provide concrete examples of modularly autonomous subsystems. Averaging over a compact symmetry group selects its invariant observable algebra. When the reference state shares the symmetry, this averaging defines a reference-preserving erasure code.

\paragraph{Invariant algebras and symmetry averaging:}
Let a compact group $G$ act normally and continuously on $\mM$ by
$*$-automorphisms $\alpha_g$, and let the faithful normal state $\omega$
be $G$-invariant, $\omega\circ\alpha_g=\omega$. The neutral, or
fixed-point, algebra and its Haar expectation are
\begin{align}
 \mN=\mM^G
 &=\{x\in\mM:\alpha_g(x)=x,\;\forall g\in G\},\nonumber\\
 \mE_G(x)&=\int_G\!\dd g\,\alpha_g(x),
 \qquad \omega\circ\mE_G=\omega .
 \label{eq:symmetry-Haar-expectation}
\end{align}
The modular automorphism group of a faithful state is unique. Therefore
\begin{align}
 \alpha_g\circ\sigma_t^{\omega;\mM}
 &=\sigma_t^{\omega;\mM}\circ\alpha_g,
 \qquad
 \sigma_t^{\omega;\mM}(\mN)=\mN,
 \qquad
 \mE_G\circ\sigma_t^{\omega;\mM}
 =\sigma_t^{\omega;\mM}\circ\mE_G .
 \label{eq:symmetry-modular-intertwining}
\end{align}
The fixed-point observables need not be modular zero modes: they can evolve
nontrivially into other elements of $\mN$. By Theorem~\ref{thm:correctable},
the expectation $\mE_G$ defines an exactly recoverable family
$\mathsf C_{\mE_G}=\{\varphi:\varphi=\varphi\circ\mE_G\}$ under
restriction to $\mN$. In other words, information encoded in these
symmetry-invariant states is retained by the neutral algebra. Haar averaging removes the non-invariant components of a state and fixes precisely the family $\mathsf C_{\mE_G}$.

The state-preserving assumption is essential. Compactness of the group ensures that we have a
normal conditional expectation. In a matrix
algebra $\mM=B(\cH)$, with $\alpha_g=\operatorname{Ad}U(g)$ and faithful
density matrix $\rho$, the condition reads $[\rho,U(g)]=0$ for every $g$.
Consequently $[-\log\rho,U(g)]=0$. The representation of $G$ on $\cH$ decomposes into isotypic components as
\begin{align}
 \cH=\bigoplus_{\lambda} V_\lambda\otimes\mathbb C^{m_\lambda},
 \qquad
 \mM^G\simeq\bigoplus_{\lambda}
       1_{V_\lambda}\otimes B(\mathbb C^{m_\lambda}),
 \label{eq:symmetry-isotypic}
\end{align}
where $V_\lambda$ are inequivalent irreducible representations; Schur's
lemma shows that an invariant faithful density matrix has the form
$\rho=\bigoplus_\lambda1_{V_\lambda}\otimes r_\lambda$,
with $r_\lambda>0$ and $\sum_\lambda\dim(V_\lambda)\Tr r_\lambda=1$.
Modular evolution acts within the multiplicity factors\footnote{Note that the neutral algebra in general has a nontrivial center. The general von Neumann algebra statement of Theorem~\ref{thm:correctable} includes this general case.}.

Stabilizer codes, topological codes, and noiseless subsystems give concrete finite-dimensional illustrations of the above setup~\cite{Gottesman1997,KribsLaflammePoulin2005,BenyKempfKribs2007}.

\paragraph{From symmetry sectors to local autonomous subalgebras:}
For a vacuum QFT with a compact unbroken internal zero-form symmetry,
let $\mathfrak F(O)$ be a field algebra and $U(g)\ket{\Omega}=\ket{\Omega}$. Internal
symmetry preserves every region, so the neutral subalgebra and its Haar expectation
\begin{align}
 \mN(O):=\mathfrak F(O)^G,\qquad
 \mE_O(x)=\int_G\dd g\,\operatorname{Ad}U(g)(x)
 \label{eq:internal-neutral-net}
\end{align}
define vacuum-preserving, modular-covariant inclusions. Moreover,
Reeh--Schlieder for the field net implies
\begin{align}
 \overline{\mN(O)\ket{\Omega}}=P_G\cH,
 \qquad P_G=\int_G\dd g\,U(g),
 \label{eq:internal-coherent-projection}
\end{align}
independent of $O$. The family is therefore coherent in the sense of
Borchers' construction of local sub-theories~\cite{revolutionizing}:
its restriction to $P_G\cH$ is a Poincar\'e-covariant neutral theory\footnote{Note that the proper neutral algebra is not cyclic on the full field Hilbert space, consistent with Corollary~\ref{cor:common-cyclic-noncorrectable}.}. Geometric and higher-form symmetry examples require additional care about the region and the reference state. They are discussed in a companion QFT paper~\cite{companionQFT}.

\section{Information loss and ergodic inclusions}\label{sec:ergodic}

In Section~\ref{sec:QEC}, we identified the condition for exact reference-preserving QEC with modular autonomy. In this section, we study the subsystem flow at real modular time away from autonomy. Its generator at the origin $t=0$ measures the relative-entropy loss of individual state pairs. The history and the persistent core of the modular orbit describe which observables the orbit generates and retains. The latter define the ergodic inclusions, which in the ergodic hierarchy sit between the autonomous inclusions of Section~\ref{sec:QEC} and the half-sided inclusions that are the modular $K$-systems. The analytic continuation to imaginary modular time, which gives the recovery maps and the inclusion parent Hamiltonians, is the subject of Section~\ref{sec:probing-outside-M}.

\subsection{Relative entropy loss and the flow generator}\label{subsec:QEC-sufficiency}

We say the restriction to $\mN$ is sufficient for a pair of faithful normal states $(\varphi,\omega)$ on $\mM$ if one normal recovery channel reconstructs both states from their restrictions. In the Heisenberg picture, this means a normal unital CP (UCP) map $\beta:\mM\to\mN$ with $\varphi|_\mN\circ\beta=\varphi$ and $\omega|_\mN\circ\beta=\omega$~\cite{Petz1986,Petz1988,OhyaPetz1993}.

\begin{theorem}[Petz]\label{thm:Petz}
For faithful normal states $\varphi,\omega$ on $\mM$ with $S(\varphi\|\omega)<\infty$, the restriction to $\mN$ is sufficient for $(\varphi,\omega)$ if and only if
\[
 S_{\mM}(\varphi\|\omega)=S_{\mN}(\varphi|_{\mN}\|\omega|_{\mN}) .
\]
In that case, a recovery channel is given by the Petz map of the inclusion with respect to $\omega$. Equivalently, the Connes cocycle $[D\varphi:D\omega]_t=\Delta_{\varphi|\omega}^{it}\Delta_{\omega}^{-it}$ belongs to $\mN$ for every $t\in\mathbb R$~\cite{Petz1988,JencovaPetz2006}.
\end{theorem}

The two modular Hamiltonians $K_\mM=-\log\Delta_\mM$ and $K_\mN=-\log\Delta_\mN$ act on $\mH_\mM$ and $\mH_\mN$, respectively. The canonical isometry $V$ compares their expectations. In this subsection, $\mM'$ acts on $\mH_\mM$, while $\mN'$ denotes the commutant of $\mN$ in $B(\mH_\mN)$. The normal UCP map
\begin{equation}
 \mE'_\mN:\mM'\longrightarrow\mN',\qquad
 \mE'_\mN(a')=V^\dagger a'V
 \label{eq:sufficiency-complementary-compression}
\end{equation}
is the complementary comparison. It becomes the inclusion $\mM'\hookrightarrow\mN'$ when $V=1$.

We use the channel form of Petz's sufficiency theorem~\cite{Petz1988}: for a normal UCP map $\alpha:\mathcal B\to\mathcal A$ and faithful normal states $\varphi,\omega$ with finite $S_{\mathcal A}(\varphi\|\omega)$ and faithful pullbacks, equality in data processing is equivalent to the existence of a normal UCP map $\beta:\mathcal A\to\mathcal B$ satisfying $\varphi\circ\alpha\circ\beta=\varphi$ and $\omega\circ\alpha\circ\beta=\omega$. The cocycle-membership criterion in Theorem~\ref{thm:Petz} is the specialization to an inclusion.

\begin{proposition}[Sufficiency decomposition of the subsystem-flow generator]\label{prop:sufficiency}
Let $\ket{\Omega}$ be cyclic and separating for $\mM$, and let $V:\mH_\mN\hookrightarrow\mH_\mM$ be the canonical isometry of Section~\ref{subsec:subsystem-flow-background}, with $V\ket{\Omega_\mN}=\ket{\Omega}$. Let $\ket{\Psi}\in\mH_\mN$ be a unit vector cyclic and separating for $\mN$, and assume that $V\ket{\Psi}$ is cyclic and separating for $\mM$. The induced states are
\begin{align}
 \varphi_\mM(m)&=\langle\Psi|V^\dagger mV|\Psi\rangle,
 &\varphi_\mN&=\varphi_\mM|_\mN,\nonumber\\
 \varphi_{\mN'}(b')&=\langle\Psi|b'|\Psi\rangle,
 &\varphi_{\mM'}&=\varphi_{\mN'}\circ\mE'_\mN.
\end{align}
Define the reference states $\omega_{\mathcal A}$ analogously from $\ket{\Omega}$ and $\ket{\Omega_\mN}$. Assume that all four relative entropies $S_{\mathcal A}(\varphi_{\mathcal A}\|\omega_{\mathcal A})$, for $\mathcal A=\mM,\mN,\mN',\mM'$, are finite and that
\begin{equation}
 \ket{\Psi}\in\Dom K_\mN,\qquad V\ket{\Psi}\in\Dom K_\mM.
\end{equation}
Abbreviating the relative entropies to $S_{\mathcal A}(\varphi\|\omega)$, one has
\begin{align}
 -i\left.\partial_t\langle\Psi|V^\dagger D(t)V|\Psi\rangle\right|_{t=0}
 &=\langle\Psi|(V^\dagger K_\mM V-K_\mN)|\Psi\rangle\nonumber\\
 &=\underbrace{S_\mM(\varphi\|\omega)-S_\mN(\varphi\|\omega)}_{\text{loss under restriction to }\mN}\nonumber\\
 &\quad+\underbrace{S_{\mN'}(\varphi\|\omega)-S_{\mM'}(\varphi\|\omega)}_{\text{loss under the channel }\mE'_\mN}.
 \label{eq:sufficiency-decomposition}
\end{align}
Both losses are nonnegative. Their sum vanishes if and only if restriction to $\mN$ is sufficient for $(\varphi_\mM,\omega_\mM)$ and the channel $\mE'_\mN$ is sufficient for $(\varphi_{\mN'},\omega_{\mN'})$. For common cyclicity, $V=1$ and the second channel is the commutant inclusion $\mM'\subset\mN'$.
\end{proposition}

\begin{proof}
The first equality follows from \eqref{eq:subsystem-flow-generator} on the stated domains. Apply Lemma~\ref{lem:modular-energy} to $(\mM,\ket{\Omega},V\ket{\Psi})$ on $\mH_\mM$ and to $(\mN,\ket{\Omega_\mN},\ket{\Psi})$ on $\mH_\mN$. It gives
\begin{align}
 \langle\Psi|V^\dagger K_\mM V|\Psi\rangle
 &=S_\mM(\varphi\|\omega)-S_{\mM'}(\varphi\|\omega),\nonumber\\
 \langle\Psi|K_\mN|\Psi\rangle
 &=S_\mN(\varphi\|\omega)-S_{\mN'}(\varphi\|\omega).
\end{align}
Subtracting proves the decomposition. The retained states are restrictions of the ambient ones, while $\varphi_{\mM'}=\varphi_{\mN'}\circ\mE'_\mN$ and $\omega_{\mM'}=\omega_{\mN'}\circ\mE'_\mN$. Data processing for the inclusion and for $\mE'_\mN$ makes both losses nonnegative. Petz's sufficiency theorem, in its inclusion and channel forms, characterizes their separate vanishing. Their nonnegative sum vanishes exactly when both do.
\end{proof}

The modular-energy identity used in the proof compares the information carried by an algebra with that carried by its commutant:

\begin{lemma}[Modular energy as an entropy imbalance]\label{lem:modular-energy}
Let $\mathcal A$ be a von Neumann algebra on $\mathcal H$, let $\ket{\Omega}$ and $\ket{\Psi}$ be unit vectors that are cyclic and separating for $\mathcal A$, with induced states $\omega$ and $\varphi$, and suppose that $S_{\mathcal A}(\varphi\|\omega)$ and $S_{\mathcal A'}(\varphi\|\omega)$ are finite. Then $t\mapsto\langle\Psi|\Delta_{\omega;\mathcal A}^{it}|\Psi\rangle$ is differentiable at $t=0$ and
\begin{equation}
 -i\,\partial_t\langle\Psi|\Delta_{\omega;\mathcal A}^{it}|\Psi\rangle\big|_{t=0}
 =S_{\mathcal A'}(\varphi\|\omega)-S_{\mathcal A}(\varphi\|\omega).
 \label{eq:modular-energy}
\end{equation}
If moreover $\ket{\Psi}$ lies in the form domain of $\log\Delta_{\omega;\mathcal A}$, the left side is the quadratic form $\langle\Psi|\log\Delta_{\omega;\mathcal A}|\Psi\rangle$.
\end{lemma}

The proof is given in Appendix~\ref{app:modular-energy-proof}.

\begin{corollary}\label{cor:positive-generator}
Assume the hypotheses of Proposition~\ref{prop:sufficiency} and common cyclicity, so that $V=1$. If $\mN\subset\mM$ is a proper inclusion, then the quadratic form $\ket{\Psi}\mapsto\langle\Psi|K_{\mM}-K_{\mN}|\Psi\rangle$ is nonnegative and does not vanish identically on any dense subspace of $\operatorname{Dom}K_{\mM}\cap\operatorname{Dom}K_{\mN}$ that is a core for $K_{\mM}$: there exist states $\varphi$ for which the erasure \eqref{eq:erasure-channel} is not sufficient, and the subsystem flow registers them.
\end{corollary}
\begin{proof}
Nonnegativity is \eqref{eq:sufficiency-decomposition}. If the form vanished on a dense subspace $D\subseteq\operatorname{Dom}K_{\mM}\cap\operatorname{Dom}K_{\mN}$, polarization would give $K_{\mM}\ket{\Psi}=K_{\mN}\ket{\Psi}$ for $\ket{\Psi}\in D$. If $D$ is a core for $K_{\mM}$, the closure of $K_{\mM}|_D$ is $K_{\mM}$ and is contained in the self-adjoint operator $K_{\mN}$. Hence $K_{\mM}=K_{\mN}$ and $\Delta_{\omega;\mM}=\Delta_{\omega;\mN}$. Then $\sigma^{\omega;\mM}_t(\mN)=\mN$ for all $t$, which contradicts Corollary~\ref{cor:common-cyclic-noncorrectable}.
\end{proof}

\paragraph{Logical unitary perturbations and the two gaps:}
For $\ket{\Psi}=u\ket{\Omega_\mN}$ with $u\in\mN$ unitary, one has $V\ket{\Psi}=u\ket{\Omega}$. The induced states agree with the reference states on $\mN'$ and $\mM'$, so the complementary gap vanishes. The Connes cocycle on $\mM$ is $u\,\sigma_t^{\omega;\mM}(u)^*$. Under the finiteness assumptions above, Theorem~\ref{thm:Petz} therefore makes the retained gap vanish precisely when $\sigma_t^{\omega;\mM}(u)\in\mN$ for every $t$. Requiring this modular condition for every logical unitary is modular autonomy.

Conversely, in the common-cyclic case $V=1$, a perturbation $\ket{\Psi}=v'\ket{\Omega}$ with $v'\in\mM'$ unitary leaves the state on $\mM$ unchanged. Assuming finite relative entropies, the retained gap therefore vanishes, while the complementary gap need not vanish. Thus the two gaps concern different restrictions. They are state-pair quantities, whereas the information--disturbance trade-off compares recovery and complementary channels uniformly over a code~\cite{KretschmannSchlingemannWerner2008}. Example~\ref{ex:two-qubits} at the end of the next subsection evaluates both gaps in closed form for two qubits.

\subsection{History, persistent cores and generation}\label{subsec:QEC-history-memory}

The modular orbit of $\mN$ records which observables become accessible as the subsystem moves. Its history collects the entire orbit, while its persistent core consists of the observables retained at every modular time. Write $\alpha_t=\sigma_t^{\omega;\mM}$.

\begin{definition}[History and persistent core]\label{def:history-core-two-sided}
For a subalgebra $\mN\subset\mM$, put
\begin{equation}
 \mN_{\rm hist}:=\bigvee_{t\in\mathbb R}\alpha_t(\mN),
 \qquad
 \mN_{\rm core}:=\bigcap_{t\in\mathbb R}\alpha_t(\mN).
\end{equation}
The inclusion is generating if $\mN_{\rm hist}=\mM$, and has trivial persistent core if $\mN_{\rm core}=\mathbb C1$. It is \emph{ergodic} for $\omega$ if it is generating and has trivial persistent core.\footnote{The notions of history and persistent core are well-studied concepts in quantum stochastic analysis~\cite{Parthasarathy1992}. From this perspective, the modular evolution of the subsystem produces a stationary quantum observation process in the sense of Accardi, Frigerio and Lewis~\cite{QS-AFL1982}. The algebra $\mN$ supplies the observables being followed, their embeddings $\alpha_t|_\mN$ at different modular times specify the process, and the state $\omega$ assigns the joint correlation functions. Because $\omega$ is modular invariant, the process is stationary.}
\end{definition}

Both the history and persistent core algebras are modularly invariant and therefore admit $\omega$-preserving normal expectations by Takesaki's theorem. The core is the largest autonomous subalgebra contained in $\mN$, while the history is the smallest autonomous algebra containing it. Equivalently, they are the largest reference-preserving correctable algebra inside $\mN$ and the smallest such algebra containing $\mN$, in the sense of Theorem~\ref{thm:correctable}.

\begin{lemma}[History--core duality]\label{lem:history-memory}
Let $\alpha_t=\operatorname{Ad}U_t$ be a one-parameter group of automorphisms of $B(\mathcal H)$ with $U_t\mM U_t^*=\mM$, hence also $U_t\mM'U_t^*=\mM'$. For every subalgebra $\mN\subset\mM$:
\begin{enumerate}
 \item[\rm(a)] On $B(\mathcal H)$ the duality is exact,
 \begin{equation}
  (\mN_{\rm hist})'=(\mN')_{\rm core},\qquad (\mN_{\rm core})'=(\mN')_{\rm hist}.
  \label{eq:history-memory-BH}
 \end{equation}
 \item[\rm(b)] Inside $\mM$, with $\mR=\mN'\cap\mM$,
 \begin{equation}
  (\mN_{\rm hist})'\cap\mM=\mR_{\rm core},\qquad
  \mR_{\rm hist}\subseteq(\mN_{\rm core})'\cap\mM ,
  \label{eq:history-memory}
 \end{equation}
 with equality in the second relation when $\mN$ and $\mR_{\rm hist}$ each coincide with their double relative commutant in $\mM$. This holds in particular when $\mM=\mN\overline\otimes\mR$ and $\mR$ is autonomous.
\end{enumerate}
\end{lemma}
The proof is given in Appendix~\ref{app:history-core-proof}.

\begin{corollary}[Generation forces a trivial environment persistent core]\label{cor:generating-forgetful}
If $\mM$ is a factor and $\mN\subset\mM$ is generating, then $\mR_{\rm core}=\mathbb C1$: the relative commutant has trivial persistent core under the modular flow. In particular this holds whenever $\ket{\Omega}$ is a common cyclic and separating vector for $\mN$ and $\mM$ {\rm(Lemma~\ref{lem:common-cyclicity-generating})}.
\end{corollary}
\begin{proof}
$\mR_{\rm core}=(\mN_{\rm hist})'\cap\mM=\mM'\cap\mM=\mathbb C1$.
\end{proof}

Inside $\mM$, generation leaves no nontrivial modularly invariant algebra in $\mR$. On the full GNS space, Lemma~\ref{lem:history-memory}(a) instead compares $\mN$ with its full commutant $\mN'$: generation is equivalent to $(\mN')_{\rm core}=\mM'$. Here $\mN'$ is the complementary algebra in this representation and contains $\mM'$.

A classical illustration of the concepts of generation and persistent core is given for the baker map in Figure~\ref{fig:ink-history-core}\footnote{For a more detailed connection to classical ergodic theory, see Appendix~\ref{subsec:classical-ergodic-hierarchy}.}. Note that this is an analogy with ordinary classical dynamics, and completely unrelated to modular flows of quantum systems.
\begin{figure}[!htbp]
\centering
\resizebox{0.7\linewidth}{!}{\input{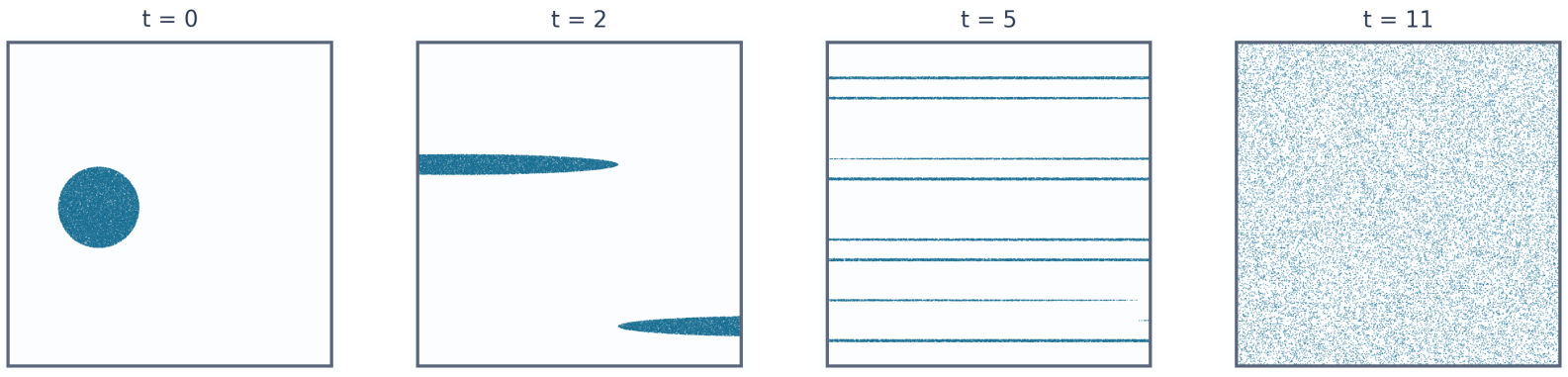}}
\caption{A drop of ink transported by the invertible, area-preserving baker map $T(x,y)=(2x\bmod 1,(y+\lfloor 2x\rfloor)/2)$ looks mixed although no information is lost. For $\mathcal N=\operatorname{vN}(1_{\{x<1/2\}})\subset\mathcal M=L^\infty([0,1]^2)$ and $\alpha_t(f)=f\circ T^t$, one finds $\mathcal N_{\rm hist}=\mathcal M$ and $\mathcal N_{\rm core}=\mathbb C1$.}
\label{fig:ink-history-core}
\end{figure}

The two-qubit example below shows the definitions at work in the simplest setup.
\begin{example}[Two qubits]\label{ex:two-qubits}
Let $\mM=M_2\otimes M_2$, $\mN=M_2\otimes1$ and $\rho=(1-\epsilon)\ket{\Phi^+}\bra{\Phi^+}+\epsilon\,\mathbbm1/4$ with $0<\epsilon\leq1$ and $\ket{\Phi^+}$ maximally entangled. The subsystem space $\overline{\mN\ket{\Omega}}$ has dimension $4$ inside the $16$-dimensional GNS space, so $V\neq1$ for every $\epsilon$. Both marginals are $\mathbbm1/2$, so $\rho$ is a product only at $\epsilon=1$, and by Corollary~\ref{cor:KL-conormal} the inclusion is autonomous only there: for $\epsilon<1$ the modular Hamiltonian $-\log\rho$ contains the interaction $X\otimes X-Y\otimes Y+Z\otimes Z$, which moves $\mN$. For the logical rotation $\ket{\Psi}=(Z\otimes1)\ket{\Omega_\mN}$, the complementary gap of Proposition~\ref{prop:sufficiency} vanishes, and the retained gap is $S\bigl((Z\otimes1)\rho(Z\otimes1)\,\|\,\rho\bigr)=(1-\epsilon)\log\frac{4-3\epsilon}{\epsilon}$, which diverges for the pure Bell state and vanishes at the product point. Finally, if $1\otimes r\in\mR_{\rm core}$ then $[-\log\rho,1\otimes r]=\sum_kc_k\,\sigma_k\otimes[\sigma_k,r]$, with $c_k\neq0$, must lie in $1\otimes M_2$, which forces $r$ to be a scalar. Hence $\mR_{\rm core}=\mathbb C1$, and since every intermediate algebra has the form $M_2\otimes\mathcal P$, Lemma~\ref{lem:history-memory}(b) gives $\mN_{\rm hist}=\mM$.
\end{example}

\subsection{Common cyclicity and standard inclusions}
\label{subsec:common-cyclic}
\begin{lemma}[Common cyclicity implies generation]\label{lem:common-cyclicity-generating}
If the inclusion $\mN\subset\mM$ comes with a common cyclic and separating vector $\ket{\Omega}$, then
$\bigvee_{t\in\mathbb R}\sigma^{\omega;\mM}_t(\mN)=\mM$.
\end{lemma}
\begin{proof}
Let $\mP=\bigvee_t\sigma^{\omega;\mM}_t(\mN)$. It is modular invariant, so Takesaki's theorem
gives an $\ket{\Omega}$-preserving expectation $\mE_{\mP}:\mM\to\mP$.
Since $\mN\subset\mP$ and $\ket{\Omega}$ is cyclic for $\mN$, the projection onto
$\overline{\mP\ket{\Omega}}$ is $1$. Hence $\mE_{\mP}(a)\ket{\Omega}=a\ket{\Omega}$ for every $a\in\mM$.
Since $\ket{\Omega}$ is separating for $\mM$, this gives $\mE_{\mP}(a)=a$.
\end{proof}
\begin{corollary}\label{cor:common-cyclic-noncorrectable}
If, in addition to the hypotheses of Theorem~\ref{thm:correctable}, $\ket{\Omega}$ is cyclic for $\mN$ and the inclusion is proper, then none of {\rm(i)--(iv)} of that theorem holds: there is no $\omega$-preserving conditional expectation, and hence no code of the form \eqref{eq:code} representing all normal logical states and containing $\omega$.
\end{corollary}

\begin{proof}
With $e_{\mN}=1$, (ii) of Theorem~\ref{thm:correctable} would give $\mE(m)\ket{\Omega}=m\ket{\Omega}$ and, by separation, $\mE(m)=m$ for all $m$, forcing $\mN=\mM$ (Lemma~\ref{lem:common-cyclicity-generating}).
\end{proof}

\begin{lemma}[Standard inclusions of factors are ergodic]
\label{lem:standard-forgetful}
Let $\mN \subset \mM$ be a standard inclusion of factors. Then the inclusion is generating and has trivial persistent core, $\mN_{\rm core} = \mathbb C1$: it is ergodic in the sense of Definition~\ref{def:history-core-two-sided}.
\end{lemma}
\begin{proof}
Common cyclicity and Lemma~\ref{lem:common-cyclicity-generating} give $\mN_{\rm hist}=\mM$ and $\mR_{\rm hist}=\mM$. By Lemma~\ref{lem:history-memory}(b), $\mR_{\rm hist}\subseteq\mN_{\rm core}'\cap\mM$, hence $\mN_{\rm core}\subseteq\mM'\cap\mM=\mathbb C1$.
\end{proof}

Common cyclicity and standardness thus place an inclusion at definite points on a scale. A common cyclic vector forces generation, standardness forces ergodicity, and Corollary~\ref{cor:common-cyclic-noncorrectable} excludes autonomy in both cases. The next subsection arranges this scale, from autonomous inclusions to their opposite, into a hierarchy.

\subsection{Modular ergodic hierarchy}\label{subsec:QEC-hierarchy}

The results so far single out several ways in which a subsystem can sit inside the ambient algebra under the modular flow. It can be preserved, it can generate everything, it can retain nothing, or both, and its orbit can be nested in one direction of modular time. Table~\ref{tab:hierarchy} collects these properties, ordered from the most to the least retentive, together with their meaning for the subsystem and the sufficient conditions established above. We refer to this as the quantum ergodic hierarchy. It mirrors the classical ergodic hierarchy reviewed in Appendix~\ref{subsec:classical-ergodic-hierarchy}, with the closed autonomous subalgebras at one end and the analogues of $K$-systems at the other.

\begin{table}[!htbp]
\centering
\small
\begin{tabular}{@{}>{\raggedright\arraybackslash}p{0.15\linewidth}>{\raggedright\arraybackslash}p{0.23\linewidth}>{\raggedright\arraybackslash}p{0.27\linewidth}>{\raggedright\arraybackslash}p{0.26\linewidth}@{}}
\toprule
Property & Modular formulation & Meaning for the subsystem & Sufficient condition\\
\midrule
Autonomous & $\sigma_t^{\omega;\mM}(\mN)=\mN$ for all $t$ & $\mN_{\rm hist}=\mN_{\rm core}=\mN$: a closed autonomous subalgebra, exactly correctable (Theorem~\ref{thm:correctable}) & product reference on a tensor-product inclusion (Corollary~\ref{cor:KL-conormal})\\
Generating & $\mN_{\rm hist}=\mM$ & the orbit reaches every observable of $\mM$ & common cyclic vector (Lemma~\ref{lem:common-cyclicity-generating})\\
Trivial persistent core & $\mN_{\rm core}=\mathbb C1$ & no closed autonomous subalgebra survives inside $\mN$ & standard inclusion of factors (Lemma~\ref{lem:standard-forgetful})\\
Ergodic & $\mN_{\rm hist}=\mM$ and $\mN_{\rm core}=\mathbb C1$ & generates everything and retains nothing & standard inclusion of factors (Lemma~\ref{lem:standard-forgetful})\\
Half-sided & $\sigma_t^{\omega;\mM}(\mN)\subseteq\mN$ for one sign of $t$ & the orbit is nested in one direction of modular time & half-sided modular inclusions~\cite{Wiesbrock1993}\\
Modular $K$-inclusion & half-sided and ergodic (Definition~\ref{def:K-inclusion}) & everything is generated in one direction of modular time and lost in the other & standard half-sided inclusion (Corollary~\ref{cor:standard-K}) and interval algebras of a conformal net (Example~\ref{ex:local-algebras})\\
\bottomrule
\end{tabular}
\caption{The modular ergodic hierarchy of an inclusion $\mN\subset\mM$ in the state $\omega$.}
\label{tab:hierarchy}
\end{table}

Two remarks are in order. First, ergodicity and mixing of the modular flow of the state $\omega$, namely triviality of the centralizer $\mM^\omega$ and decay of the connected modular correlation function 
for every pair of observables, are properties of $\omega$ on $\mM$ rather than of an inclusion. They neither imply nor follow from the rows of the table (Remark~\ref{rem:no-mixing}). Second, the rows of the table above are not linearly ordered. A proper inclusion cannot be both autonomous and generating, so the first row excludes the others, but generation and a trivial persistent core are independent conditions, and half-sidedness is an orientation that can accompany any of them. At the autonomous end, a product reference on $\mN\overline\otimes\mR$ keeps the subsystem flow trivial and makes the history and the core both equal to $\mN$. At the other end, the one-sided notions below add the orientation.

\begin{definition}[Future and past algebras]\label{def:history-memory}
The one-sided versions of Definition~\ref{def:history-core-two-sided} are
\begin{align}
 \mN_{\rm hist}^{\pm}&:=\bigvee_{\pm t>0}\alpha_t(\mN),&
 \mN_{\rm core}^{\pm}&:=\bigcap_{\pm t>0}\alpha_t(\mN).
\end{align}
We call the inclusion future or past generating when the corresponding history is $\mM$, and say that its future or past persistent core is trivial when it is $\mathbb C1$. Each one-sided generation condition implies two-sided generation, and each trivial one-sided core implies a trivial two-sided core. The reverse implications require additional hypotheses.
\end{definition}

The commutant identities of Lemma~\ref{lem:history-memory}(a) also hold separately for the future and past algebras. Under one-sided nesting, the one-sided core is the tail algebra, the observables that remain accessible arbitrarily far into the remote past or future\footnote{The classical $K$-property combines half-sided inclusion, one-sided generation and a trivial tail~\cite{CornfeldFominSinai1982,ouseph2024local}. Its modular counterpart uses Wiesbrock's half-sided modular inclusions~\cite{Wiesbrock1993}. The positive-energy translations and their QFT interpretation are developed in the companion paper~\cite{companionQFT}.}.

\begin{definition}[Half-sided and modular $K$-inclusions]\label{def:K-inclusion}
The inclusion is \emph{half-sided} for $\omega$ if $\alpha_t(\mN)\subseteq\mN$ for all $t\leq0$, or for all $t\geq0$. A \emph{modular $K$-inclusion} is a half-sided inclusion that is ergodic.
\end{definition}

\begin{lemma}[One-sided and two-sided algebras of a half-sided inclusion]\label{lem:half-sided-one-sided}
If $\alpha_t(\mN)\subseteq\mN$ for all $t\leq0$, then $t\mapsto\alpha_t(\mN)$ is increasing, and
\begin{equation}
 \mN_{\rm hist}=\mN_{\rm hist}^{+},\qquad
 \mN_{\rm core}=\mN_{\rm core}^{-} .
 \label{eq:half-sided-one-sided}
\end{equation}
Hence a half-sided inclusion is a modular $K$-inclusion if and only if it is future generating with trivial past core, which are the one-sided conditions of the classical $K$-property \eqref{eq:classical-K-algebra}. For the opposite orientation, the signs are exchanged.
\end{lemma}
\begin{proof}
For $s\leq t$, $\alpha_s(\mN)=\alpha_t(\alpha_{s-t}(\mN))\subseteq\alpha_t(\mN)$. Thus $\alpha_t(\mN)\subseteq\mN\subseteq\alpha_s(\mN)$ for $t\leq0\leq s$, so the join over all $t$ equals the join over $t>0$, and the intersection over all $t$ equals the intersection over $t<0$.
\end{proof}

\begin{corollary}[Standard half-sided inclusions]\label{cor:standard-K}
A standard half-sided inclusion of factors is a modular $K$-inclusion. In particular, a half-sided modular inclusion of factors in the sense of Wiesbrock~\cite{Wiesbrock1993} is a modular $K$-inclusion whenever the reference vector is also cyclic for the relative commutant.
\end{corollary}
\begin{proof}
Lemma~\ref{lem:standard-forgetful} gives ergodicity, and half-sidedness is assumed.
\end{proof}

\begin{remark}[Emergent symmetry of a modular $K$-inclusion]\label{rem:K-groups}
A half-sided modular inclusion carries more structure than the hierarchy records. For a half-sided modular inclusion $\mN\subset\mM$ with common cyclic and separating vector, Wiesbrock's theorem~\cite{Wiesbrock1993,Borchers1992} provides a positive-energy translation group $U(a)=e^{iaP}$, $P\geq0$, fixing $\ket{\Omega}$, with
\begin{equation}
 \Delta_\mM^{it}U(a)\Delta_\mM^{-it}=U(e^{-2\pi t}a),\qquad J_\mM U(a)J_\mM=U(-a),\qquad \mN=U(1)\mM U(-1).
 \label{eq:hsm-relations}
\end{equation}
The modular orbit of the subsystem is therefore the family of translates $\alpha_t(\mN)=U(e^{-2\pi t})\mM U(-e^{-2\pi t})$. Modular time is the logarithm of a translation parameter. The modular flow and the translations together form a unitary representation of the affine group $t\mapsto at+b$ of the line, with dilations generated by the modular Hamiltonian and translations by $P$. In this language, the two halves of the $K$-property are explicit. The history is the join of the translates as the translation parameter decreases to zero, which is $\mM$ by strong continuity, and the persistent core is their intersection as the parameter grows without bound, which is trivial for a standard inclusion (Corollary~\ref{cor:standard-K}). The one-sided loss of access in Table~\ref{tab:hierarchy} is loss under a positive-energy translation, and the inclusion parent Hamiltonian of Section~\ref{sec:probing-outside-M} is this generator, $Q_\alpha=P$ (Proposition~\ref{prop:half-sided-normalization}). A single half-sided inclusion yields only the affine group. However, with a standard inclusion one obtains a representation of $PSL(2,\mathbb R)$~\cite{Wiesbrock1997,LashkariLeungMoosaOuseph2025}. In Example~\ref{ex:local-algebras} below, the modular groups of $\mathcal A(I_1)$ and $\mathcal A(I)$ generate the affine subgroup of the M\"obius group fixing the shared endpoint, and adjoining the modular group of a third interval generates all of $PSL(2,\mathbb R)$~\cite{Wiesbrock1993b,GuidoLongoWiesbrock1998}.
\end{remark}

\begin{remark}[Half-sided nesting does not imply a trivial core]\label{rem:half-sided-spectator}
If $\mN_1\subset\mM_1$ is half-sided, then $\mN_1\overline\otimes\mathcal P\subset\mM_1\overline\otimes\mathcal P$ is also half-sided for a faithful product reference, while its persistent core contains $1\otimes\mathcal P$. For a proper half-sided modular inclusion with a common cyclic and separating vector, the modular orbit generates $\mM$, and no expectation preserves the reference (Lemma~\ref{lem:common-cyclicity-generating}, Corollary~\ref{cor:common-cyclic-noncorrectable}), but a trivial core needs an additional assumption. Standardness suffices by Lemma~\ref{lem:standard-forgetful}, and then the inclusion is a modular $K$-inclusion.
\end{remark}

The hierarchy is realized in full by the local algebras of a quantum field theory.

\begin{example}[Local algebras of a conformal net]\label{ex:local-algebras}
In a local conformal net $\{\mathcal A(I)\}$ on the circle with vacuum vector $\ket{\Omega}$, let $I_1\subset I$ be open intervals sharing one endpoint, and let $I_2$ be the interior of $I\setminus I_1$. Haag duality and strong additivity give $\mathcal A(I_1)'\cap\mathcal A(I)=\mathcal A(I_2)$ and $\mathcal A(I_1)\vee\mathcal A(I_2)=\mathcal A(I)$, so $\mN=\mathcal A(I_1)\subset\mM=\mathcal A(I)$ is conormal with $\mR=\mathcal A(I_2)$. By the Reeh--Schlieder theorem, the vacuum is cyclic and separating for all three algebras. The inclusion is therefore standard and, by Lemma~\ref{lem:standard-forgetful}, ergodic: the modular orbit of the smaller interval generates the algebra of the larger one and retains no closed autonomous subalgebra. Since the intervals share an endpoint, the inclusion is a half-sided modular inclusion in the sense of Wiesbrock~\cite{Wiesbrock1993}, and Corollary~\ref{cor:standard-K} makes it a modular $K$-inclusion, the last row of Table~\ref{tab:hierarchy}. At the same time, no faithful normal conditional expectation $\mathcal A(I)\to\mathcal A(I_1)$ exists, for any faithful normal state and not only for the vacuum. The adjacent pair admits no spatial tensor-product factorization in a nontrivial conformal net~\cite{Wiesbrock1993}, and Lemma~\ref{prop:conormal-factorization} would produce one. The obstruction concerns normal product structure. It does not exclude error correction for other channels or chosen families of states.
\end{example}

\begin{definition}[Asymptotic forgetfulness]\label{def:forgetful-dynamics}
For a specified normal UCP family $\Phi_t:\mN\to\mN$ with faithful invariant state $\omega_\mN$, future forgetfulness means
\begin{equation}
 \lim_{t\to+\infty}\|\varphi\circ\Phi_t-\omega_\mN\|_{\mN_*}=0
 \quad\text{for every normal state }\varphi\in\mN_*.
 \label{eq:forgetful-dynamics}
\end{equation}
Past forgetfulness is defined analogously.
\end{definition}

\begin{remark}[What generation and a trivial core do not imply]\label{rem:no-mixing}
Note that generation and a trivial persistent core are statements about which observables an orbit reaches or retains, and concern the time-averaged behavior of modular flow. None of them implies mixing of the modular flow, asymptotic forgetfulness in the sense of \eqref{eq:forgetful-dynamics}, a decay rate for connected correlations, state factorization, or an arrow of time.
\end{remark} 

To get intuition about asymptotic forgetfulness, it is helpful to first consider quantum channels acting on a finite-dimensional Hilbert space. For the iterates $\Phi^n$ of a channel on a finite-dimensional algebra with a faithful stationary state, an analog of asymptotic forgetfulness is equivalent to primitivity~\cite{Wolf2012}. In spectral terms, the only eigenvalue of $\Phi$ on the unit circle is $1$, and it is simple\footnote{The spectrum of a quantum channel lies in the closed unit disk.}. Primitivity combines irreducibility, the absence of a nontrivial invariant hereditary subalgebra, with aperiodicity, the absence of other peripheral eigenvalues. It is equivalent to the existence of an $n$ for which $\Phi^n$ maps every state to a full-rank state~\cite{Wolf2012}. When it holds, the convergence is exponential at a rate set by the second largest eigenvalue modulus. 
To generalize this spectral analysis to infinite-dimensional algebras, in Section~\ref{subsec:recovery-Poisson} we introduce the recovery semigroup. A reduced family such as $\Phi_t=\mE\circ\sigma_t^{\omega;\mM}|_\mN$ must have its stationary state checked separately: if $\mE$ preserves $\omega$, Theorem~\ref{thm:correctable} makes $\mN$ autonomous and the family is simply its intrinsic modular evolution.

\section{\texorpdfstring{Probing outside of $\mM$ using inclusion parent Hamiltonians}{Probing outside of M using inclusion parent Hamiltonians}}
\label{sec:probing-outside-M}

The modular flow of $\mM$ is an automorphism group of $\mM$. It moves the subsystem $\mN$ around inside $\mM$, but never out of it, so the real-time analysis of the previous sections describes $\mN$ only from the point of view of the ambient algebra. To ask how the subsystem sits relative to what lies outside $\mM$, we need a dynamics that does not preserve $\mM$. Our key observation in this section is that subsystem flow at imaginary modular time provides such dynamics. The analytically continued subsystem flow is a contraction rather than an automorphism, and its squared modulus is a positive operator that fixes the reference vector. We use this analytic continuation to define {\it inclusion parent Hamiltonians}. They generate unitary dynamics whose orbits of $\mN$ leave $\mM$ (Remark~\ref{rem:orbit-leaves}). In some cases, the orbits generate the entire $B(\mH_\mM)$\footnote{For example, for half-sided modular inclusions, this orbit is the family of null translations, which carries the algebra of the smaller region beyond the larger one for one sign of the translation~\cite{companionQFT}.}. In this section, we study two evolutions associated with the inclusion: the reversible unitary orbit generated by the inclusion parent Hamiltonian and the dissipative recovery dynamics.

The section has five parts. Section~\ref{subsec:analytic-subsystem-flow} analytically continues the subsystem flow to imaginary time and identifies the Petz maps on the upper boundary of the strip. Section~\ref{subsec:Q-construction} defines the inclusion parent Hamiltonians and fixes their normalization. Section~\ref{subsec:Q-ground-space} shows that their ground space is the persistent core and that their orbits leave the ambient algebra. Section~\ref{subsec:recovery-Poisson} identifies the operator at the midpoint with a single step of recovery and builds the recovery semigroup. Section~\ref{sec:nuclearity} studies the spectrum, gap, entropy, and entanglement. Table~\ref{tab:section5-summary} summarizes the key results of this section and the assumptions under which each holds. Technical details and proofs are in Appendix~\ref{app:inclusion-Hamiltonian-details}. For completeness, Appendix~\ref{subsec:canonical-tower-common-cyclic} describes the canonical tower of a common-cyclic inclusion, and Appendix~\ref{app:Q-dual-inclusion} describes the commutant-dual inclusion and the two orderings of its characteristic function.

The analytic continuation, the Petz map, universal recovery, and the implication from $L^2$-nuclearity to the split property are known results in the operator-algebra and QEC literature~\cite{revolutionizing,JungeRennerSutterWildeWinter2018,BDL2007}. The inclusion parent Hamiltonian with its ground-space and generation theorems, the recovery semigroup, and the spectral statements built on them are our key new results. 

\begin{table}[!htbp]
\centering
\small
\renewcommand{\arraystretch}{1.25}
\begin{tabular}{@{}>{\raggedright\arraybackslash}p{0.34\linewidth}>{\raggedright\arraybackslash}p{0.38\linewidth}>{\raggedright\arraybackslash}p{0.20\linewidth}@{}}
\toprule
Result & Assumptions & Reference\\
\midrule
The ground space of $Q_\alpha$ is the persistent core & none  & Theorem~\ref{thm:Q-kernel-core}\\
The reference vector is the unique ground state & trivial core & Theorem~\ref{thm:Q-kernel-core}\\
The orbit of $\mN$ under $e^{itQ_\alpha}$ generates $B(\mH_\mN)$ & trivial core & Proposition~\ref{prop:Halpha-full-generation}\\
The future orbit and the past orbit each generate $B(\mH_\mN)$ & trivial core, and $Q_\alpha$ has pure point spectrum on $\mH_\mN$ & Proposition~\ref{prop:Halpha-pure-point-one-sided-generation}\\
The Petz map is the canonical endomorphism, and the recovery semigroup relaxes to the core & common cyclic vector, with a spectral gap for the rate if $X_{1/4}$ is trace class & Theorem~\ref{thm:main-recovery-relaxation}\\
A positive attenuation gap $-\log\|X_\alpha|_{\ket{\Omega}^\perp}\|$ above the ground state, largest at $\alpha=\tfrac14$ & trivial core and compact $X_\alpha$ & Section~\ref{sec:nuclearity}\\
Discrete spectrum of $Q_\alpha$ with finite multiplicities, and split property & common cyclic vector and $X_{1/4}$ trace class ($L^2$-nuclearity, \eqref{eq:L2-nuclearity-alpha}) & Section~\ref{sec:nuclearity}\\
\leavevmode Entanglement bound $E_R\leq\log\Tr e^{-Q_{1/4}}$ and inclusion-entropy bound $S_{\rm inc}^{(1/4)}\leq2\log\Tr e^{-Q_{1/4}}$ & \leavevmode standard inclusion of factors and $X_{1/4}$ trace class & \leavevmode Section~\ref{sec:nuclearity}
\\
\bottomrule
\end{tabular}
\caption{Summary of the main results of Section~\ref{sec:probing-outside-M} and the assumptions behind each. Results that use the unitary group $e^{itQ_\alpha}$ assume full support of $\rho_\alpha$ (Section~\ref{subsec:Q-construction}).}
\label{tab:section5-summary}
\end{table}

\subsection{Imaginary modular time and the Petz map}
\label{subsec:analytic-subsystem-flow}

The subsystem flow of Definition~\ref{def:subsystem-flow} admits an analytic continuation to complex modular time in the strip $0\leq\operatorname{Im}z\leq\tfrac12$:
\begin{equation}
    D_{\mN\subset\mM}(z)
    :=
    {
    \Delta_{\mM}^{-iz}V\Delta_{\mN}^{iz}}\,V^\dagger
    \in B(\mH_\mM) .
    \label{eq:Wz-definition}
\end{equation}
To simplify our notation, we write $D(z)=D_{\mN\subset\mM}(z)$ and $D_\alpha=D(i\alpha)$. The analytic continuation rests on the monotonicity of the modular operators, with the two spaces related by $V$,
\begin{equation}
 \|\Delta_\mM^\alpha V\ket{\psi}\|
 \leq\|\Delta_\mN^\alpha\ket{\psi}\|,\qquad
 0\leq\alpha\leq1/2,\quad
 \ket{\psi}\in\operatorname{Dom}\Delta_\mN^\alpha.
 \label{eq:modular-monotonicity-main}
\end{equation}
 For $z=t+i\alpha$ with $0\leq\alpha\leq1/2$,
\begin{align}
 D(t+i\alpha)&=\Delta_\mM^{-it}D_\alpha
                       V\Delta_\mN^{it}V^\dagger,\qquad 
 D_\alpha=D(i\alpha)
 ={\Delta_\mM^\alpha V\Delta_\mN^{-\alpha}}\,V^\dagger.
 \label{eq:W-alpha-strip}
\end{align}

The operator $D(z)$ is a contraction on the strip, meaning
\begin{equation}
    \|D(z)\|\leq1,
    \qquad
    0\leq\operatorname{Im}z\leq\frac12 .
    \label{eq:W-strip-bound}
\end{equation}
Both boundaries are isometries on the code subspace and vanish on its orthogonal complement, with $D_0=e_\mN$ and $D_{1/2}=J_\mM VJ_\mN V^\dagger$\footnote{See Appendix~\ref{subsec:beyond-Borchers} for more details on this analytic continuation.}.
More generally, the upper boundary is the lower one conjugated by the two modular conjugations,
\begin{equation}
 D(t+i/2)=J_\mM D(t)VJ_\mN V^\dagger .
 \label{eq:subsystem-flow-contraction}
\end{equation}
When $\ket{\Omega}$ is common cyclic and separating for the inclusion, $V=1$ and $D(t)$ is Borchers\textquotesingle{} characteristic function~\cite{revolutionizing}\footnote{See Appendix~\ref{app:characteristic-function} for a characterization of inclusions by common cyclic and separating vectors in terms of the properties of this analytic function.}.
Figure~\ref{fig:subsystem-analytic-strip} in the introduction displays the analytic strip and the distinguished role of its two boundaries and midpoint $\alpha=1/4$.

\paragraph{The Petz map from the upper boundary:}
The adjoint of the subsystem flow at the upper boundary $(D(i/2) V)^\dagger $ implements the Petz map $\mathsf E_\omega$ of \eqref{eq:background-Petz-dual}:
\begin{equation}
 \begin{gathered}
 \mathsf E_\omega(m)=\Gamma m\Gamma^\dagger,\\
 \Gamma=J_{\mN}V^{\dagger}J_{\mM}
 =V^\dagger D_{\mN\subset\mM}(i/2)^\dagger=(D_{\mN\subset\mM}(i/2)V)^\dagger.
 \end{gathered}
 \label{eq:QEC-Petz-dual}
\end{equation}
At general real time, $D(t+i/2)$ implements the rotated Petz maps\footnote{Recall that as we discussed earlier, the co-isometry $V^\dagger$ alone need not implement a CP map into $\mN$. The modular conjugations in $\Gamma$ are essential.}:
\begin{equation}
 D(t+i/2)^\dagger a\ket{\Omega}
 =\left(
 \sigma_{-t}^{\omega;\mN}\circ\mathsf E_\omega
                  \circ\sigma_t^{\omega;\mM}(a)\right)\ket{\Omega} ,\qquad a\in\mM .
 \label{eq:flow-Petz-boundary}
\end{equation}

\paragraph{Universal recovery:}
The Petz map is the $t=0$ member of the family of rotated Petz maps
\begin{equation}
 \mathsf E_{\omega,t}(m):=\sigma_{-t/2}^{\omega;\mN}\circ\mathsf E_\omega\circ\sigma_{t/2}^{\omega;\mM}(m)
 =V^\dagger D\!\left(\tfrac{t+i}{2}\right)^\dagger m\,D\!\left(\tfrac{t+i}{2}\right)V ,
 \label{eq:rotated-Petz-D}
\end{equation}
each normal, unital and completely positive. Averaging them against $\beta_0(t)=\pi/(2[1+\cosh\pi t])$ gives the universal recovery map $\mathsf E_\omega^{\rm univ}$ of~\cite{JungeRennerSutterWildeWinter2018}, a weighted integral of the completely positive maps induced by the upper boundary and not a linear integral of $D(t)$. It depends only on the inclusion and $\omega$, recovers $\omega$ exactly, and bounds the information loss, $S_\mM(\varphi\|\omega)-S_\mN(\varphi\|\omega)\geq-2\log F_\mM(\varphi,\varphi|_\mN\circ\mathsf E_\omega^{\rm univ})$ with $F_\mM$ the root fidelity~\cite{JungeRennerSutterWildeWinter2018,FaulknerHollandsSwingleWang2022}.

\subsection{Inclusion parent Hamiltonians}
\label{subsec:Q-construction}

For $0<\alpha<\tfrac12$, let $X_\alpha:=D_\alpha V:\mH_\mN\to\mH_\mM$ be the strip operator restricted to the subsystem space (for a common cyclic vector $X_\alpha=D_\alpha$). The positive inclusion operator is
\begin{equation}
 \rho_{\alpha,\mN\subset\mM}:=X_\alpha^\dagger X_\alpha=V^\dagger D_\alpha^\dagger D_\alpha V ,
 \label{eq:rho-alpha-general}
\end{equation}
an operator on $\mH_\mN$. By the strip bound \eqref{eq:W-strip-bound}, and because every modular operator fixes the reference vector,
\begin{equation}
 0\leq\rho_{\alpha,\mN\subset\mM}\leq1_{\mH_\mN},\qquad
 \rho_{\alpha,\mN\subset\mM}\ket{\Omega}=\ket{\Omega}.
 \label{eq:rho-alpha-contraction}
\end{equation}
We define the \emph{inclusion parent Hamiltonian}, or inclusion energy, as the rescaled logarithm of the attenuation $\rho_\alpha$,
\begin{equation}
 Q_{\alpha,\mN\subset\mM}:=-\frac{1}{2\sin(2\pi\alpha)}\log\rho_{\alpha,\mN\subset\mM},\qquad 0<\alpha<\tfrac12 .
 \label{eq:H-alpha-general}
\end{equation}
It measures the cost of the mismatch between the intrinsic and ambient modular motions. For a fixed inclusion, we use the simplified notation $\rho_\alpha$ and $Q_\alpha$\footnote{Note that this differs from the definition in \cite{chen2026local} which was proportional to $\log (X_\alpha X_\alpha^\dagger)$. We comment on this in \cite{companionQFT}. See Appendix \ref{app:inclusion-Hamiltonian-details} for a discussion of the relations between $X^\dagger X$, $XX^\dagger$, and the Petz maps $J_\mN J_\mM$.}. 
Two important properties of $Q_\alpha$ are\footnote{See Appendices~\ref{app:Q-reflection} and~\ref{app:Q-restriction} for a proof. The limit $\alpha\to 0$ is in operator norm for bounded modular Hamiltonians and a formal expansion otherwise. It is the compression to $\mH_\mN$ of the generator $K_\mM-VK_\mN V^\dagger$ of the subsystem flow (Proposition~\ref{prop:sufficiency}), divided by $2\pi$.}:
\begin{align}
    J_\mN Q_\alpha J_\mN=Q_{1/2-\alpha},\qquad \lim_{\alpha\to 0} Q_\alpha=(V^\dagger K_\mM V-K_\mN)/2\pi\ .
    \label{eq:Q-reflection-limit}
\end{align}

When taking the logarithm in \eqref{eq:H-alpha-general}, one must pay attention to the following two points. First, in this range of $\alpha$ there is no branch-cut ambiguity. The operator $\rho_\alpha$ is positive with spectrum in $[0,1]$, so $\log\rho_\alpha$ is the real logarithm of the spectral calculus and is nonpositive. This is why $Q_\alpha$ is built from the squared modulus $X_\alpha^\dagger X_\alpha$ of the strip operator and not from its boundary values at $\alpha=0$ or $\alpha=1/2$. The logarithm of an isometry (unitary in the case $V=1$) requires a choice of branch, whereas the logarithm of a positive contraction does not. Consequently
\begin{equation}
 Q_{\alpha,\mN\subset\mM}\geq0,\qquad Q_{\alpha,\mN\subset\mM}\ket{\Omega}=0,
 \label{eq:H-alpha-positive}
\end{equation}
which is why we call it a parent Hamiltonian.

Second, the logarithm is defined only on the support of $\rho_\alpha$. Let $s(\rho_\alpha)$ be the support projection, the orthogonal projection onto $(\ker\rho_\alpha)^\perp$. On $s(\rho_\alpha)\mH_\mN$ the operator $Q_\alpha$ is self-adjoint and nonnegative. On $\ker\rho_\alpha$ the inclusion energy is formally $+\infty$, and the unitary group $e^{itQ_\alpha}$ is not defined there. We say that $\rho_\alpha$ has \emph{full support} when $s(\rho_\alpha)=1$. Then $Q_\alpha$ is a densely defined self-adjoint operator on all of $\mH_\mN$ and $e^{itQ_\alpha}$ is a unitary group on $\mH_\mN$. Whenever a statement uses $Q_\alpha$ as an operator on all of $\mH_\mN$, we assume full support and call this the Full Support (FS) assumption\footnote{Faithfulness of $\omega$ makes $X_\alpha$ well defined but not injective. $X_\alpha$ is the continuous extension of $\Delta_\mM^\alpha V\Delta_\mN^{-\alpha}$, which is injective on $\operatorname{Ran}\Delta_\mN^\alpha$. A kernel vector would be a normalizable subsystem state orthogonal to $\ket{\Omega}$ and outside $\operatorname{Ran}\Delta_\mN^\alpha$, with infinite intrinsic modular energy and infinite inclusion energy. An equivalent condition is that $X(z):=D(z)V$ has no zeros on the line $\operatorname{Im}z=\alpha$. $X(t)$ and $X(t+i/2)$ are isometries and carry no zeros. (FS) holds in finite dimensions, whenever $K_\mN$ is bounded, and for half-sided modular inclusions, where $D(z)=e^{i(e^{2\pi z}-1)P}$ has no zeros. It is not known to follow from nuclearity or compactness of $X_\alpha$. We know of no counterexample among standard inclusions of type III factors, and of no proof that it always holds.}. This assumption covers the unitary group $e^{itQ_\alpha}$ of Section~\ref{subsec:Q-ground-space}, the identity $L=1-e^{-2Q_{1/4}}$ of Section~\ref{subsec:recovery-Poisson}, and the gaps of Section~\ref{sec:nuclearity}. The ground-space theorem and the recovery semigroup do not need it. 

We consider a half-sided modular inclusion (reviewed below) as a running example for the rest of Section~\ref{sec:probing-outside-M}: its orbit is studied in Section~\ref{subsec:Q-ground-space}, the shift $\operatorname{Ad}\Gamma=\operatorname{Ad}U(2)$ in Section~\ref{subsec:recovery-Poisson}, and its continuous spectrum, in contrast with the discrete spectra of nuclear inclusions, in Section~\ref{sec:nuclearity}. 

\begin{proposition}[Half-sided modular inclusions]\label{prop:half-sided-normalization}
Let $\mN\subset\mM$ be a half-sided modular inclusion with common cyclic and separating vector $\ket{\Omega}$, in the conventions of~\cite{Wiesbrock1993}, and let $U(a)=e^{iaP}$, $P\geq0$, be the Borchers translation group with $\mN=U(1)\mM U(1)^*$. Then
\begin{enumerate}
 \item[\rm(i)] $\Gamma=J_\mN J_\mM=U(2)$, and $K_\mM-K_\mN=2\pi P$ on the common domain,
 \item[\rm(ii)] $Q_{\alpha,\mN\subset\mM}=P$ for every $0<\alpha<\tfrac12$, so that $\rho_\alpha=e^{-2\sin(2\pi\alpha)P}$ and $e^{-2Q_{1/4}}=e^{-2P}$.
\end{enumerate}
Statement {\rm(i)} follows from $J_\mN=U(2)J_\mM$ and $\Delta_\mN^{it}=U(1)\Delta_\mM^{it}U(-1)$, which are Wiesbrock's relations~\cite{Wiesbrock1993}. Statement {\rm(ii)} is the explicit computation of the companion paper~\cite{companionQFT}.
\end{proposition}
For a half-sided inclusion, the parent Hamiltonian is the translation generator at every imaginary modular time, and the three objects $e^{-2Q_{1/4}}$, $e^{2itQ_{1/4}}$, and the discrete shift $\operatorname{Ad}\Gamma$ are all translations. The sine factor in the definition of $Q_\alpha$ is what makes it independent of $\alpha$ in this case, and this fixes the normalization.

\subsection{Inclusion parent Hamiltonian flows}
\label{subsec:Q-ground-space}\label{subsec:Q-generation}

We identify $\mN$ with its faithful representation on $\mH_\mN$. Under the FS assumption, $Q_\alpha$ generates a unitary group and an automorphism group of $B(\mH_\mN)$,
\begin{equation}
 U_t^{(\alpha)}:=e^{itQ_\alpha},\qquad
 \gamma_t^\alpha(a):=U_t^{(\alpha)}aU_t^{(\alpha)*},
 \quad a\in B(\mH_\mN),
 \label{eq:N-alpha-t-general}
\end{equation}
the inclusion parent Hamiltonian flow. Unlike the ambient modular flow, it acts on all of $B(\mH_\mN)$ and need not preserve the ambient observable algebra. We study it through its stationary states. A vector $\ket{\psi}$ with $Q_\alpha\ket{\psi}=0$ is frozen under the flow, $\langle\psi|\gamma_t^\alpha(b)|\psi\rangle=\langle\psi|b|\psi\rangle$ for every $b\in B(\mH_\mN)$ and every $t$, so the ground vectors play the role of conserved charges, and the reference vector is always one of them. The next theorem identifies all of them. We then show that the absence of such charges makes the orbit of $\mN$ under $\gamma_t^\alpha$ generate everything, in analogy with ergodic theory, where a flow without nonconstant invariants explores the whole space.

The ambient persistent core $\mN_{\rm core}=\bigcap_t\sigma_t^{\omega;\mM}(\mN)$ is the largest autonomous subalgebra contained in $\mN$. Its observables have no mismatch between intrinsic and ambient modular motion. The theorem shows that these are the only charges, including those not of the form $a\ket{\Omega}$ with $a$ bounded.

\begin{theorem}[Ground-state space and modular persistent core]
\label{thm:Q-kernel-core}
Let $\mN\subset\mM$ be a unital inclusion, with $\ket{\Omega}$
cyclic and separating for $\mM$, and let $\Delta_\mN$ act on
the GNS space $\mH_\mN$ of $\mN$, where $V\mH_\mN=\overline{\mN\ket{\Omega}}$. For $0<\alpha<1/2$,
\begin{equation}
    \ker(1-\rho_{\alpha,\mN\subset\mM})
    =\overline{\mN_{\rm core}\ket{\Omega}}.
    \label{eq:Q-kernel-core-theorem}
\end{equation}
The kernel of the inclusion parent Hamiltonian (on the support of
$\rho_\alpha$) is the same space. In particular, it is independent
of $\alpha$ in the open modular strip. Hence
\begin{equation}
    \ker Q_{\alpha,\mN\subset\mM}=\mathbb C\ket{\Omega}
    \quad\Longleftrightarrow\quad
    \mN_{\rm core}=\mathbb C1,
    \label{eq:Q-core-unique-equivalence}
\end{equation}
where $\ker Q_\alpha$ is the kernel on the support of $\rho_\alpha$, which is $\mH_\mN$ under the FS assumption. In particular, the parent Hamiltonian of an ergodic inclusion (Definition~\ref{def:history-core-two-sided}), and hence of every standard inclusion of factors (Lemma~\ref{lem:standard-forgetful}), has the reference vector as its unique ground state.
\end{theorem}
The proof is given in Appendix~\ref{app:Q-ground-proof}. Equality in the contraction bound $1-\rho_\alpha\geq 0$ at one interior point forces the intrinsic and ambient modular motions to agree on that vector for every real time. At the midpoint $\alpha=1/4$, the Petz map supplies a state-preserving CP map whose fixed algebra is precisely the core. Its $L^2$ implementation then identifies the full closed ground space. Note that this theorem needs no full-support assumption, since it describes the kernel of $Q_\alpha$ on the support of $\rho_\alpha$. The conserved charges of the flow are therefore the vectors $a\ket{\Omega}$ with $a$ in the persistent core and their closure. The reference vector is the only charge exactly when the core is trivial.

\paragraph{The orbit and full generation:}
The complete history of the transported subsystem is
\begin{equation}
 \mathcal A_\alpha:=\bigvee_{t\in\mathbb R}\gamma_t^\alpha(\mN).
 \label{eq:A-alpha-general}
\end{equation}
The next result says that if the flow has no charge beyond the reference vector, then $\mathcal A_\alpha$ is as large as possible.

\begin{proposition}[{Trivial persistent core implies full generation}]
\label{prop:Halpha-full-generation}
Assume that the persistent core $\mN_{\rm core}=\mathbb C1$ is trivial and that the FS assumption holds, and let $0<\alpha<1/2$. Then
\begin{equation}
    \bigvee_{t\in\mathbb R}
    \gamma_t^\alpha(\mN)
    =
    B(\mH_{\mN}).
    \label{eq:Halpha-full-BHN}
\end{equation}
\end{proposition}

\begin{proof}
By Theorem~\ref{thm:Q-kernel-core}, the trivial core implies $\ker Q_\alpha=\mathbb C\ket{\Omega}$ (using the full-support assumption for the unitary group). Let $\mathcal A_\alpha$ be the algebra generated by the $U_t^{(\alpha)}$-orbit of $\mN$. It is invariant under $U_t^{(\alpha)}$, and its commutant
$\mathcal A_\alpha'$ is invariant as well. Since $\mN\subset\mathcal A_\alpha$ and $\mN\ket{\Omega}$ is dense
in $\mH_\mN$, $\ket{\Omega}$ is separating for $\mathcal A_\alpha'$. On $\mathcal K=\overline{\mathcal A_\alpha'\ket{\Omega}}$, the reference vector is cyclic and separating. The restricted unitary group commutes with its Tomita conjugation $J_{\mathcal K}$, so antiunitarity gives $J_{\mathcal K}Q_\alpha|_{\mathcal K}J_{\mathcal K}=-Q_\alpha|_{\mathcal K}$. Positivity forces this restriction to vanish (Appendix~\ref{app:Q-positive-energy}). Thus $Q_\alpha x\ket{\Omega}=0$ for every $x\in\mathcal A_\alpha'$.
Uniqueness of the ground state forces $x\ket{\Omega}=c_x\ket{\Omega}$.
Separatingness for $\mathcal A_\alpha'$ then gives
$x=c_x1$. Hence $\mathcal A_\alpha'=\mathbb C1$ and
$\mathcal A_\alpha=B(\mH_\mN)$ by the bicommutant theorem.
\end{proof}

\begin{corollary}[Standard inclusions generate $B(\mH_{\mM})$]
\label{cor:standard-Halpha-full-generation}
Let $(\mN\subset\mM,\ket{\Omega})$ be a standard inclusion of factors and let $0<\alpha<1/2$. Under the standing FS assumption,
\begin{equation}
    \bigvee_{t\in\mathbb R}
    \gamma_t^\alpha(\mN)
    =B(\mH_{\mM}).
    \label{eq:standard-Halpha-full-BHM}
\end{equation}
\end{corollary}

\begin{proof}
Standardness implies that $\ket{\Omega}$ is cyclic for $\mN$, hence
$\mH_{\mN}=\mH_{\mM}$ and $e_{\mN}=1$. It also implies a trivial ambient modular persistent core by Lemma~\ref{lem:standard-forgetful}. The result is therefore
Proposition~\ref{prop:Halpha-full-generation} with
$B(\mH_{\mN})=B(\mH_{\mM})$.
\end{proof}

\begin{remark}[The orbit leaves the ambient algebra]\label{rem:orbit-leaves}
The ambient modular flow keeps $\mN$ inside $\mM$ by construction. The flow generated by the inclusion Hamiltonian does not. In the common-cyclic case, full generation implies that the orbit leaves $\mM$ whenever $\mM\neq B(\mH_\mM)$, as for every type III algebra. Hence $Q_\alpha$ probes the complement of the ambient algebra.\footnote{For an inclusion with $e_\mN<1$, the comparison uses the corner $VB(\mH_\mN)V^\dagger=e_\mN B(\mH_\mM)e_\mN$. Moreover, $e_\mN\notin\mM$ in that case, since otherwise $(1-e_\mN)\ket{\Omega}=0$ and separatingness would force $e_\mN=1$. The generation statement is then understood on the subsystem GNS space $\mH_\mN$.} 
Many inclusions of type III algebras realized in QFT have an extra property called the split property \cite{DoplicherLongo1984}\footnote{In fact, the $L^2$-nuclearity assumption we discuss in this paper implies the split property.}. A standard split inclusion has another candidate dynamics, the inner flow of its canonical type-I factor $\mathcal F_{\rm split}$, with $\mN\subset\mathcal F_{\rm split}\subset\mM$. Unlike the inclusion Hamiltonian flows, the flow generated by unitaries in $\mathcal F_{\rm split}$ never leaves $\mM$.
\end{remark}

We define the future and past histories as $\mathcal A_\alpha^\pm:=\bigvee_{\pm t>0}\gamma_t^\alpha(\mN)$. An additional recurrence assumption makes either half of the orbit sufficient:

\begin{proposition}[Pure point spectrum implies future and past generation]
\label{prop:Halpha-pure-point-one-sided-generation}
Assume the hypotheses of Proposition~\ref{prop:Halpha-full-generation}.
Suppose, in addition, that $\mH_{\mN}$ is separable and
$Q_\alpha:=Q_{\alpha,\mN\subset\mM}$ has pure point spectrum, meaning
that its eigenvectors span $\mH_{\mN}$. Then
\begin{equation}
    \mathcal A_\alpha^+
    =\mathcal A_\alpha^-
    =B(\mH_{\mN}).
    \label{eq:Halpha-pure-point-one-sided-generation}
\end{equation}
In particular, this holds if
$\rho_{\alpha,\mN\subset\mM}$ is trace class with full support.
\end{proposition}

For pure point spectrum, there are Poincaré recurrences: arbitrarily large times at which $U_t^{(\alpha)}$ returns strongly to the identity. Every real-time translate can therefore be approximated by translates in either time direction. Appendix~\ref{app:Q-recurrence} gives the proof.

The half-sided inclusion shows that recurrence cannot be dropped. Let $\mN\subset\mM$ be a standard half-sided modular inclusion in the conventions of Proposition~\ref{prop:half-sided-normalization}. The core is trivial and $\rho_\alpha=e^{-2\sin(2\pi\alpha)P}$ has full support, so the full orbit generates $B(\mH_\mM)$ by Corollary~\ref{cor:standard-Halpha-full-generation}. But $Q_\alpha=P$ generates the translations, and $U(t)\mN U(-t)\subseteq\mN$ for $t\geq0$. By strong continuity at $t=0$, the future history is $\mathcal A_\alpha^+=\mN$. The spectrum of $P$ is continuous and there are no recurrences: the future orbit never leaves $\mN$, and the whole of $B(\mH_\mM)$ is reached through past times.

\subsection{Recovery as Euclidean evolution}
\label{subsec:emergent-transfer-time}\label{subsec:recovery-Poisson}

At the midpoint $\alpha=1/4$, the inclusion Hamiltonian has a second meaning: a single step of Petz recovery is a Euclidean evolution generated by $Q_{1/4}$, at inverse temperature $2$. In this subsection and the next one, we write $X=X_{1/4}$ and $\rho=\rho_{1/4}=X^\dagger X=e^{-2Q_{1/4}}$.

In the symmetric $L^2$ GNS Hilbert space picture, an observable $a\in\mA$, with $\mA=\mN,\mM$, is represented by the vector $\Delta_\mA^{1/4}a\ket{\Omega}$. For matrices, this is $\rho_\mA^{1/4}a\rho_\mA^{1/4}$ in the Hilbert--Schmidt representation. Let $\Phi=\mathsf E_\omega|_\mN$ be the Petz map restricted to $\mN$, so that $\Phi(a)=\Gamma a\Gamma^\dagger$ with $\Gamma$ from \eqref{eq:QEC-Petz-dual}. We have the identity
\begin{equation}
 \rho\,\Delta_\mN^{1/4}a\ket{\Omega}=\Delta_\mN^{1/4}\Phi(a)\ket{\Omega}
 =\Delta_\mN^{1/4}\Gamma a\Gamma^\dagger\ket{\Omega},
 \qquad a\in\mN.
 \label{eq:Q-AC-L2}
\end{equation}
Thus, $e^{-2Q_{1/4}}$ implements the Petz recovery in this symmetric representation. We call $\rho=e^{-2Q_{1/4}}$ the \emph{transfer operator} of the inclusion, in analogy with the transfer matrix of statistical mechanics. It advances the system by one step, its powers $\rho^{\,n}$ describe $n$ steps, and its logarithm is a Hamiltonian. It is not a density matrix unless its trace is finite and it is normalized. If $\ket{\Omega}$ is common cyclic and separating, then $V=1$. The adjoint of the upper boundary value $D(i/2)$ is the unitary
\begin{equation}
 \Gamma=D_{\mN\subset\mM}(i/2)^\dagger
 =J_{\mN}J_{\mM},\qquad \Gamma\ket{\Omega}=\ket{\Omega},
 \label{eq:interpolation-Gamma}
\end{equation}
and the Petz map becomes the canonical endomorphism
\begin{equation}
 \gamma_{\rm can}(m):=\Gamma m\Gamma^\dagger,\qquad
 \gamma_{\rm can}(\mM)=J_{\mN}\mM'J_{\mN}\subseteq\mN .
 \label{eq:interpolation-canonical-endomorphism}
\end{equation}
Iteration of \eqref{eq:Q-AC-L2} gives $\rho^{\,n}\Delta_\mN^{1/4}a\ket{\Omega}=\Delta_\mN^{1/4}\Gamma^n a\Gamma^{\dagger n}\ket{\Omega}$. The map $\operatorname{Ad}\Gamma$ shifts the canonical tower by two levels, but this identity does not identify its discrete unitary implementation with the real-time group of $Q_{1/4}$: in general, $\Gamma\neq e^{2iQ_{1/4}}$\footnote{See Appendix \ref{subsec:canonical-tower-common-cyclic} for a discussion of the canonical tower of an inclusion. The exception is the half-sided modular inclusion, where both sides are the translation $U(2)$ (Proposition~\ref{prop:half-sided-normalization}).}.

\paragraph{Recovery semigroup:}
Assume the inclusion has a common cyclic and separating vector, so that $\Phi=\operatorname{Ad}\Gamma|_\mN$ is an $\omega$-preserving endomorphism of $\mN$ whose symmetric $L^2$ implementation is the positive contraction $\rho$. Averaging over a Poisson-distributed number of recovery steps defines a superoperator
\begin{equation}
 P_u:=e^{u(\Phi-\mathrm{id})}
 =e^{-u}\sum_{n=0}^{\infty}\frac{u^n}{n!}\Phi^n,
 \qquad u\geq0 .
 \label{eq:main-recovery-Poisson}
\end{equation}

\begin{proposition}[Markov and symmetry properties]
\label{prop:QS-Markov}
The family $(P_u)_{u\geq0}$ is a uniformly continuous, normal, unital,
completely positive, $\omega_{\mN}$-preserving semigroup. Its symmetric
$L^2$ implementation is 
\begin{equation}
 T_u=e^{-uL},\qquad
 L=1-\rho=1-e^{-2Q_{1/4}}.
 \label{eq:QS-generator}
\end{equation}
The formula involving $Q$ assumes full support. In all cases,
$\operatorname{Fix}(P_u)=\mN_{\rm core}$. 
\end{proposition}
\begin{proof}
Every $\Phi^n$ is a normal state-preserving UCP map, and the coefficients in \eqref{eq:main-recovery-Poisson} form a probability distribution. The exponential
identity gives the semigroup law. Applying \eqref{eq:Q-AC-L2} term by term gives $\Delta_\mN^{1/4}P_u(a)\ket{\Omega}=e^{u(\rho-1)}\Delta_\mN^{1/4}a\ket{\Omega}$.
The implementation is a self-adjoint contraction semigroup, whose fixed
space is $\ker(1-\rho)$. Theorem~\ref{thm:Q-kernel-core} 
identifies it.
\end{proof}
The associated bounded quadratic form is
\begin{equation}
 \mathcal E(\xi)=\langle\xi,L\xi\rangle
 =\|\xi\|^2-\|X\xi\|^2.
\end{equation}
Dissipation is therefore the norm defect of the midpoint inclusion operator.
The construction belongs to the theory of KMS-symmetric quantum Markov
semigroups and their differential structures~\cite{QS-VW2023}.
It guarantees complete positivity, whereas arbitrary fractional powers of $\rho$ need not.

\paragraph{Equilibrium and normal-state convergence:}
The core is globally invariant under the modular flow. For readability, we use $\mathcal C := \mN_{\rm core}$ for this subsection. Let
$E_{\mathcal C}:\mM\to\mathcal C$ be the state-preserving normal expectation,
also denoting its restriction to $\mN$ by $E_{\mathcal C}$. Then, we have the following relaxation result:
\begin{theorem}[Relaxation to persistent information]
\label{thm:QS-relaxation}
Let $e_{\mathcal C}$ be the symmetric implementation of $E_{\mathcal C}$ on $\mH_\mN$, that is, $e_{\mathcal C}\Delta_\mN^{1/4}a\ket{\Omega}=\Delta_\mN^{1/4}E_{\mathcal C}(a)\ket{\Omega}$ for $a\in\mN$. The proof shows that it is the projection onto the fixed space $\ker(1-\rho)=\overline{\mathcal C\ket{\Omega}}$ of the transfer operator.
Then $T_u\to e_{\mathcal C}$ strongly,
$P_u(a)\to E_{\mathcal C}(a)$ ultraweakly, and
\begin{equation}
 \|\varphi\circ P_u-\varphi\circ E_{\mathcal C}\|_{\mN_*}
 \longrightarrow0\qquad(\varphi\in\mN_*).
 \label{eq:QS-predual-convergence}
\end{equation}
\end{theorem}
\begin{proof}
The spectral theorem gives
$e^{-u(1-\rho)}\to1_{\{1\}}(\rho)=e_{\mathcal C}$ strongly. Here $1_{\{1\}}(\rho)$ is the projection onto $\overline{\mathcal C\ket{\Omega}}=\ker(1-\rho)$ (Theorem~\ref{thm:Q-kernel-core}). Since $\mathcal C$ is invariant under both modular groups with equal restrictions, $\overline{\mathcal C\ket{\Omega}}$ reduces $\Delta_\mN$, so $1_{\{1\}}(\rho)\Delta_\mN^{1/4}a\ket{\Omega}=\Delta_\mN^{1/4}E_{\mathcal C}(a)\ket{\Omega}$, which identifies $1_{\{1\}}(\rho)$ with the symmetric implementation $e_{\mathcal C}$ of $E_{\mathcal C}$.
For $\ket{\xi}\in\mH_\mN$, let $f_\xi(a)=\langle\xi|\Delta_\mN^{1/4}a|\Omega\rangle$.
The map $a\mapsto\Delta_\mN^{1/4}a\ket{\Omega}$ is bounded, injective, and weak*-to-weak continuous. The normal functionals $f_\xi$ have norm-dense span in $\mN_*$, since their
annihilator is zero. Self-adjointness gives
\begin{equation}
 \|f_\xi\circ P_u-f_\xi\circ E_{\mathcal C}\|
 \leq\|(T_u-e_{\mathcal C})\ket{\xi}\|.
\end{equation}
Contractivity extends convergence to all normal functionals. The ultraweak
assertion follows as well.
\end{proof}

\paragraph{Common cyclicity and strong convergence:}
To study strong operator convergence, we need to move away from the symmetric implementation. Recall that:
\begin{equation}
 \Gamma=J_{\mN}J_{\mM},\qquad
 \Phi(a)=\Gamma a\Gamma^\dagger,\qquad \Gamma\ket{\Omega}=\ket{\Omega}.
\end{equation}
Thus the ordinary GNS implementation is
\begin{equation}
 P_u(a)\ket{\Omega}=e^{u(\Gamma-1)}a\ket{\Omega}.
 \label{eq:QS-GNS}
\end{equation}
It differs from the symmetric implementation $T_u$.
The unitary spectral theorem gives
$e^{u(\Gamma-1)}\to1_{\{1\}}(\Gamma)$ strongly. Comparing with the
ultraweak limit in Theorem~\ref{thm:QS-relaxation} shows
$1_{\{1\}}(\Gamma)a\ket{\Omega}=E_{\mathcal C}(a)\ket{\Omega}$.
Applying the argument also to $a^*$ and using boundedness and cyclicity for
$\mN'$ yields
\begin{equation}
 P_u(a)\longrightarrow E_{\mathcal C}(a)\quad\text{strongly*}.
 \label{eq:QS-strong-star}
\end{equation}
Collecting the discussion so far, we have the following theorem:
\begin{theorem}[Recovery semigroup and persistent information]
\label{thm:main-recovery-relaxation}
Let $\mN\subset\mM$ have a common cyclic and separating vector
$\ket{\Omega}$. Let
$E_{\mathcal C}:\mM\to\mathcal C = \mN_{\rm core}$ be the $\omega$-preserving
conditional expectation. Then, for every $u>0$,
\begin{equation}
 \operatorname{Fix}(P_u)
 =\operatorname{Fix}(\Phi)=\mathcal C
 \label{eq:main-recovery-fixed-MD}
\end{equation}
Moreover, with $E_{\mathcal C}$ restricted to $\mN$,
\begin{equation}
 P_u(a)\xrightarrow[u\to\infty]{\mathrm{strong}^*}
 E_{\mathcal C}(a),\qquad
 \|\varphi\circ P_u-\varphi\circ E_{\mathcal C}\|_{\mN_*}
 \xrightarrow[u\to\infty]{}0,
 \label{eq:main-recovery-convergence}
\end{equation}
for every $a\in\mN$ and every normal functional
$\varphi\in\mN_*$. Neither nuclearity nor a spectral gap is required.
\end{theorem}
\subsection{Inclusion spectrum: gap, entropy, and entanglement}
\label{sec:nuclearity}\label{subsec:modular-nuclearity-entanglement}

To gain insights into the nature of the dynamics generated by $Q_{1/4}$ we would like to analyze its spectrum. Insight into the spectrum of this operator comes from the assumption of $L^2$-nuclearity, which states that $X_{1/4}$ is trace class (see \eqref{eq:L2-nuclearity-alpha} below). $L^2$-nuclearity implies that this spectrum is discrete, with the gap above the ground state (smallest positive eigenvalue) giving the relaxation rate of the recovery semigroup. We will use the inclusion spectrum to define an {\it inclusion entropy} in \eqref{eq:rho-inclusion-normalized}, and establish that the partition function of $Q_{1/4}$ bounds from above both this entropy and other key measures of entanglement between the subsystem and the exterior.

\paragraph{$L^2$-nuclearity and discrete spectrum:}

With $X_\alpha=D_\alpha V$ as in \eqref{eq:rho-alpha-general}, the standard $L^2$-nuclearity condition is
\begin{equation}
    \|X_{1/4}\|_1<\infty .
    \label{eq:L2-nuclearity-alpha}
\end{equation}
It implies modular nuclearity, and hence the split property~\cite{BDL2007}\footnote{However, the converse implication from modular to $L^2$-nuclearity fails in general.}. More generally, if $X_\alpha$ is trace class at some $0<\alpha<1/2$, its nonzero singular values $\{s_k\}$ determine the {\it inclusion energies}. On the support of $\rho_\alpha$,

\begin{equation}
 \rho_\alpha=X_\alpha^\dagger X_\alpha=e^{-2\sin(2\pi\alpha)Q_\alpha},
 \qquad E_k^{(\alpha)}=-\frac{\log s_k}{\sin(2\pi\alpha)},
\end{equation}
where $E_k^{(\alpha)}$ is the eigenvalue of $Q_\alpha$, the inclusion energy. The lowest nonzero eigenvalue of $Q_\alpha$ corresponds to the largest eigenvalue of $\rho_\alpha$ below one, that is, to the norm of $\rho_\alpha$ on the orthogonal complement of the ground space. The finite energies are discrete, have finite multiplicities, and can accumulate only at $+\infty$. At the midpoint, let $Z(\beta):=\Tr e^{-\beta Q_{1/4}}$ be the partition function, with the trace taken on the support. Then $\|X_{1/4}\|_1=Z(1)$ and $\Tr\rho=Z(2)$, and $Z(\beta)\geq1$ because the reference vector has zero energy. Nuclearity thus controls the spectrum through $Z$.

\paragraph{Midpoint $\alpha=1/4$ maximizes the gap:}
The gap above the reference vector is largest at the midpoint, so $Q_{1/4}$ is the parent Hamiltonian with the largest gap. Under $L^2$-nuclearity and a trivial persistent core this gap is strictly positive. To compare different $\alpha$, we rescale $Q_\alpha$ to $-\tfrac12\log\rho_\alpha=\sin(2\pi\alpha)Q_\alpha$, which equals $Q_\alpha$ at $\alpha=\frac14$. Under the FS assumption its gap above the reference vector, which has zero energy, is
\begin{equation}
 \lambda_{\rm gap}(\alpha)
 :=\inf\sigma\!\left(
 \sin(2\pi\alpha)\,Q_\alpha\big|_{\ket{\Omega}^\perp}\right)
 =-\log\left\|
 X_\alpha\big|_{\ket{\Omega}^\perp}\right\|.
 \label{eq:Q-midpoint-gap}
\end{equation}
It is concave in $\alpha$, symmetric about $\alpha=\frac14$, and zero at the edges of the strip, where nothing is attenuated. Hence
\begin{equation}
 \lambda_{\rm gap}(\tfrac14)
 =\max_{0<\alpha<1/2}\lambda_{\rm gap}(\alpha).
 \label{eq:Q-gap-optimal-midpoint}
\end{equation}
The proof applies Hadamard's three-lines theorem to the second-largest singular value of $D(z)$, which is the norm of its antisymmetrized square (Appendix~\ref{subsec:Q-three-lines}). The largest singular value is always one because of the reference vector. If the persistent core is trivial and $X_{1/4}$ is compact, then $\ket{\Omega}$ is the only singular vector of $X_{1/4}$ with singular value one, so $\lambda_{\rm gap}(1/4)>0$.

Note that the maximality statement above applies to $\sin(2\pi \alpha)Q_\alpha$. The gap of $Q_\alpha$ itself is not in general maximized at the midpoint, because the normalization of Proposition~\ref{prop:half-sided-normalization} makes the half-sided translation generator independent of $\alpha$. The two criteria meet at $\alpha=1/4$.

\paragraph{Recovery relaxation gap:}
The spectral gap also controls the rate of the recovery semigroup of Section~\ref{subsec:recovery-Poisson}. Under the trace-class condition, $\rho$ is compact. Let $\mathcal C=\mN_{\rm core}$ and let $e_{\mathcal C}$ be the projection onto the fixed space of $\rho$, $\ker(1-\rho)=\overline{\mathcal C\ket{\Omega}}$, which is the ground space of $Q_{1/4}$. It is the symmetric implementation of the core expectation $E_{\mathcal C}$ of Theorem~\ref{thm:QS-relaxation}. Assume that its orthogonal complement is nonzero. Then
\begin{equation}
 q:=\bigl\|\rho|_{(1-e_{\mathcal C})\mH_\mN}\bigr\|<1,
 \qquad
 \|T_u-e_{\mathcal C}\|=e^{-u(1-q)}.
 \label{eq:main-recovery-gap}
\end{equation}
When $\rho$ has full support, write $\lambda=-\frac12\log q$ for the gap of $Q_{1/4}$ above its ground sector. The relaxation gap is then $1-e^{-2\lambda}$. For a trivial core, $\lambda=\lambda_{\rm gap}(1/4)$. This is an operator-norm estimate in the symmetric $L^2$ representation. Without additional assumptions, neither a uniform normal-state mixing rate nor exponential relative-entropy decay follows. The half-sided modular inclusion lies at the other extreme. Its midpoint operator is $\rho=e^{-2P}$, and by dilation covariance the spectrum of $P$ on the complement of the reference vector is $[0,\infty)$. Hence $X$ is not compact, $\lambda_{\rm gap}(1/4)=0$, and there is no operator-norm relaxation gap, although the recovery semigroup still relaxes to the core by Theorem~\ref{thm:main-recovery-relaxation}. The concentric double cones inclusion below is an example with a nonzero gap.

We now turn to entropy and entanglement. Below, we assume a standard inclusion of factors ($V=1$).

\paragraph{Inclusion entropy:}
Whenever $X_\alpha$ is trace class, $\rho_\alpha$ is trace class and it defines a density matrix $\widehat\rho_\alpha=\rho_\alpha/\Tr\rho_\alpha$ with entropy
\begin{equation}
 S_{\rm inc}^{(\alpha)}=-\Tr\widehat\rho_\alpha\log\widehat\rho_\alpha .
 \label{eq:rho-inclusion-normalized}
\end{equation}
Since $\rho_\alpha=e^{-2\sin(2\pi\alpha)Q_\alpha}$, this is the thermal entropy of $Q_\alpha$ at inverse temperature $2\sin(2\pi\alpha)$, which equals $2$ at the midpoint, and also the entropy of the normalized squared singular values of $X_\alpha$. The order-$\tfrac12$ R\'enyi entropy bounds it, and $\Tr\rho_\alpha\geq1$ because $\rho_\alpha\ket{\Omega}=\ket{\Omega}$, so
\begin{equation}
 S_{\rm inc}^{(\alpha)}\leq2\log\|X_\alpha\|_1-\log\Tr(X_\alpha^\dagger X_\alpha)\leq2\log\|X_\alpha\|_1 .
 \label{eq:Sinclusion-nuclearity-bound}
\end{equation}
Note that our inclusion entropy differs from the canonical split entropy $S_{\rm split}=-\Tr\sigma_{\rm split}\log\sigma_{\rm split}$ of the reduced state $\sigma_{\rm split}$ of $\ket{\Omega}$ on the canonical type-I factor $\mathcal F_{\rm split}$ of a standard split inclusion \cite{DoplicherLongo1984}.

\paragraph{Entanglement bound:}
Modular nuclearity is an assumption about the map $\Xi:\mN\to\mH_\mM$, $\Xi(a)=\Delta_\mM^{1/4}a\ket{\Omega}$, with $\mN$ in the operator norm (see \cite{HollandsSanders2018} for a review). The map $\Xi_\mN(a)=\Delta_\mN^{1/4}a\ket{\Omega}$ is a contraction, since $\|\Delta_\mN^{1/4}a\ket{\Omega}\|\leq\|a^*\ket{\Omega}\|^{1/2}\|a\ket{\Omega}\|^{1/2}\leq\|a\|$, and $\Xi=X\,\Xi_\mN$, so $\|\Xi\|_{\rm nuc}\leq\|X\|_1=Z(1)$, which is why $L^2$-nuclearity implies modular nuclearity. Hollands and Sanders bound the relative entropy of entanglement $E_R$ between commuting algebras by the nuclearity of such a map~\cite{HollandsSanders2018}. For the commuting algebras $\mN$ and $\mM'$, with environment $\mN'\cap\mM$, the bound for the reference state is
\begin{equation}
 E_R(\omega;\mN:\mM')\leq\log\|\Xi\|_{\rm nuc}\leq\log\|X\|_1=\log Z(1).
 \label{eq:HS-entanglement-inclusion-bound}
\end{equation}

Combining \eqref{eq:Sinclusion-nuclearity-bound} and \eqref{eq:HS-entanglement-inclusion-bound},
\begin{equation}
    E_R\leq\log Z(1),\qquad
    \tfrac12S_{\rm inc}^{(1/4)}\leq\log Z(1)-\tfrac12\log Z(2)\leq\log Z(1).
    \label{eq:Sinc-entanglement-common-bound}
\end{equation}
The partition function of $Q_{1/4}$ at $\beta=1$ therefore bounds the entanglement between $\mN$ and the exterior of $\mM$ and, through the Gibbs state at $\beta=2$, the inclusion entropy. If $Z(1)$ is close to $1$, because the first excited inclusion energy is large, both are small. Conversely, a large inclusion entropy forces $Z(1)\geq e^{S_{\rm inc}^{(1/4)}/2}$, so the entanglement bound is informative only for a sparse inclusion spectrum. 

We consider a concrete example with a gap. 
\begin{example}[Concentric double cones]\label{ex:double-cones}
For vacuum double cones of radii $r<R$ in a CFT, Hollands and Sanders compute, in the proof of Theorem~13 of~\cite{HollandsSanders2018},
\begin{equation}
 X_{1/4}=q^{L_{\rm conf}}q^{iD},\qquad q=r/R=e^{-\tau},\qquad L_{\rm conf}=\tfrac12(P^0+K^0),
 \label{eq:HS-conformal-inclusion-operator}
\end{equation}
with $D$ the dilation generator. Hence, we have $Q_{1/4}\simeq\tau L_{\rm conf}$ and, under the operator-state completeness assumptions of their Theorem~14, $Z(\beta)=\sum_kq^{\beta\Delta_k}$. The sum is over the dimensions of primary and descendant fields. Then, we have $E_R\leq\log\sum_kq^{\Delta_k}$, and $S_{\rm inc}^{(1/4)}$ is the thermal entropy corresponding to the spectrum of the dimensions of the fields at inverse temperature $2\tau$. As $r\to R$, we have $q\to1$ and both bounds grow with the operator content, as expected for entanglement across a sharp boundary. The relaxation gap above is $1-e^{-2\tau\Delta_{\min}}$ for a unique vacuum and the smallest positive scaling dimension $\Delta_{\min}$.
\end{example}

\section{Inclusion spectra as a proposal for local chaos}
\label{sec:local-chaos}\label{subsec:local-spectral-chaos}

In a finite quantum system, a class of late-time measures of quantum chaos is formulated in terms of the eigenvalues and eigenvectors of the global Hamiltonian. A chaotic Hamiltonian is believed to have level statistics that follow random-matrix theory~\cite{BGS1984,GuhrMullerGroelingWeidenmuller1998}, and highly excited eigenvectors that satisfy the eigenstate thermalization hypothesis (ETH)~\cite{Deutsch1991,Srednicki1994,RigolDunjkoOlshanii2008,DAlessio2016}.\footnote{Integrable systems instead show Poisson statistics~\cite{BerryTabor1977}.} 

In continuum theories such as QFT and holography, these measures of chaos are studied by putting the theory in finite volume (spatial cylinder). ETH in conformal field theories in finite volume is studied in~\cite{LDL2018ETHCFT,LDL2018Univ,DLL2018subETH}; see also~\cite{BasuDasDattaPal2017,HikidaKusukiTakayanagi2018,FaulknerWang2018} for tests of ETH and its corrections in two dimensions. In holography, black-hole spectra show random-matrix correlations~\cite{Cotler2017}, and the ramp of the spectral form factor is reproduced by a gravitational saddle and by JT gravity as a random-matrix ensemble~\cite{SaadShenkerStanford2018,StanfordWitten2020}. These diagnostics rely on a global Hamiltonian which is not accessible to a local observer. In many physical contexts, one would like to define measures of quantum chaos, purely in terms of local data accessible to an observer. In relativistic QFT, the local algebras are type III$_1$, which means that the modular Hamiltonian has no point spectrum. The constructions of Section~\ref{sec:probing-outside-M} suggest that, when the inclusion of local algebras is nuclear, the inclusion Hamiltonians have a discrete spectrum that is purely defined in terms of local data. We conjecture that the spectrum of local inclusion parent Hamiltonians can be used to define a diagnostic of quantum chaos intrinsic to a local subsystem. The support for this conjecture comes from calculations showing that, for the inclusion of wedges in a general QFT and of ball-shaped regions in a general conformal field theory, $Q_\alpha$ is indeed proportional to the global Hamiltonian of the theory \cite{chen2026local}.

For a common-cyclic inclusion with $X_\alpha$ trace class, the inclusion parent Hamiltonian
\begin{equation}
    Q_{\alpha,\mN\subset\mM}
    =-\frac{1}{2\sin(2\pi\alpha)}\log\rho_{\alpha,\mN\subset\mM},
    \qquad \rho_{\alpha,\mN\subset\mM}=X_\alpha^\dagger X_\alpha,
    \label{eq:local-chaos-Halpha}
\end{equation}
has a discrete finite-energy spectrum with finite multiplicities, accumulating only at $+\infty$\footnote{Note that a positive rescaling of $Q_\alpha$ changes neither unfolded level statistics nor adjacent-gap ratios, so the normalization in \eqref{eq:H-alpha-general} does not change the physics.}. The equilibrium ensemble is that of the parent Hamiltonian itself,
\begin{equation}
 \omega_{\beta,\alpha}(a)=\frac{\Tr(e^{-\beta Q_\alpha}a)}{\Tr e^{-\beta Q_\alpha}},
\end{equation}
which is finite for $\beta\geq\sin(2\pi\alpha)$ when $X_\alpha$ is trace class. 

For nested regions $O_1\Subset O_2$ and the inclusion $\mN(O_1)\subset\mM(O_2)$ in the state $\ket{\Omega}$, let $\{E_n\}$ be the ordered eigenvalues of $Q_\alpha$ within an irreducible symmetry sector (i.e., fix global charges), after accounting for symmetry-enforced degeneracies. Let $N(E)=\#\{n:E_n\leq E\}$ be the staircase that counts the inclusion levels up to energy $E$. It jumps by one at every level, so on the scale of the level spacing it is irregular. The smoothed counting function $\overline N(E)$ is its average over an energy window that contains many levels but is short compared with the scale on which the spectrum changes. It is the number of inclusion modes below $E$, in the same sense as the integrated density of states of a solid or Weyl's law for a particle in a box. Its derivative $\partial_E\overline N$ is the mean density of levels, that is, the number of levels per unit energy, and its inverse is the mean level spacing. Up to a term that depends only on the width of the window, $\log\partial_E\overline N$ is the microcanonical entropy $S_{\rm loc}(E)$ of the inclusion, so $\partial_E\overline N$ encodes the thermodynamics of $Q_\alpha$. It therefore carries non-universal information about the theory, the state and the geometry of $O_2\setminus O_1$, the region between the two boundaries, whose width sets the spectral gap in the examples of the companion paper~\cite{companionQFT}. In the concentric CFT example the inclusion spectrum is the cylinder spectrum, so its mean growth is fixed by the operator content. A universal diagnostic can only reside in the fluctuations of the levels around this smooth density. The simplest probe is the statistics of the spacings between neighboring levels. Levels of a chaotic system repel each other, like the eigenvalues of a random matrix, so very small spacings are rare. Levels of an integrable system are uncorrelated, like random points on a line, so they cluster and small spacings are common. The distribution of spacings is the standard fingerprint that tells the two apart. To compare different systems one first unfolds the spectrum~\cite{GuhrMullerGroelingWeidenmuller1998}, that is, rescales the levels by the mean density $\partial_E\overline N$ so that the mean spacing is one everywhere. There are also probes such as the ratio of two adjacent spacings, in which the local density cancels, that need no unfolding~\cite{OganesyanHuse2007,Atas2013}.

\paragraph{The conjecture:}
We conjecture that, after resolving conserved charges and antiunitary symmetries, sufficiently chaotic local QFT states have Wigner--Dyson statistics~\cite{BGS1984} for these inclusion spectra, while local integrability or an extensive family of autonomous subtheories gives Poisson-like statistics~\cite{BerryTabor1977}. The random-matrix class is fixed by the symmetries of the parent Hamiltonian itself~\cite{Dyson1962,AltlandZirnbauer1997}. At the midpoint, the antiunitary symmetry $J_\mN Q_{1/4}J_\mN=Q_{1/4}$ can select a real class when it acts within the sector examined. The statistics should persist over parametrically large spectral windows, be stable under small deformations of the regions and, by the reflection $J_\mN Q_\alpha J_\mN=Q_{1/2-\alpha}$, agree at $\alpha$ and $\tfrac12-\alpha$. Whether the unfolded high-energy statistics are independent of $\alpha$ throughout the strip tests whether they are a property of the inclusion rather than of one imaginary modular time.

\paragraph{An inclusion eigenstate thermalization hypothesis:}
Spectral statistics probe eigenvalues only. For eigenvectors, $Q_\alpha|n\rangle=E_n|n\rangle$, and an algebra $\mathcal Q\subset\mN$ of observables localized well inside the inner region, away from the collar, a local inclusion analogue of the eigenstate thermalization hypothesis~\cite{Deutsch1991,Srednicki1994} is the ansatz
\begin{equation}
    \langle m|a|n\rangle
    =a_{\rm loc}(\overline E)\,\delta_{mn}
    +e^{-\frac12 S_{\rm loc}(\overline E)}
    f_a(\overline E,\nu)\,R^{(a)}_{mn},
    \qquad a\in\mathcal Q,
    \label{eq:local-inclusion-ETH}
\end{equation}
with $\overline E=\tfrac12(E_m+E_n)$ and $\nu=E_m-E_n$, $S_{\rm loc}(E)$ the logarithm of the number of inclusion levels in a microcanonical window around $E$, smooth functions $a_{\rm loc}$ and $f_a$, and order-one fluctuations $R^{(a)}_{mn}$ whose statistics depend only on the symmetry class.    Its diagonal part says that highly excited inclusion eigenstates in a narrow window have the same expectation values for observables deep inside $\mN$. Its off-diagonal part says that deviations from this local equilibrium are exponentially suppressed by the spectral entropy of the inclusion. Note that any formulation of ETH specifies a class of simple operators, and in our case, we conjecture it is the operators well-localized inside the subalgebra $\mN$.

\section{Discussion}\label{sec:discussion}

\paragraph{Summary:}
A subsystem and a state come with two intrinsic evolutions. The full algebra moves an observable of $\mN$ with its modular flow, and the reduced state moves it with another. The subsystem flow is the mismatch between the two, and the results of this paper are different readings of that one function.

On the real axis, the flow measures what the subsystem loses to its surroundings. It is trivial exactly when the ambient dynamics never takes an observable of $\mN$ out of $\mN$. By Takesaki's theorem this is the existence of a reference-preserving conditional expectation, that is, exact erasure correction of the reference state (Theorem~\ref{thm:correctable}). In the language of error correction it is a Knill--Laflamme condition with the Jones projection as the code projection, and for conormal inclusions it says that the complementary channel into the relative commutant is forgetful on the code (Section~\ref{subsec:QEC-KL}).

Away from autonomy, the generator of the flow is a sum of two Petz sufficiency gaps, so it counts the information that leaks out (Proposition~\ref{prop:sufficiency}). Following the orbit of $\mN$ under the ambient flow produces two algebras. The history is the smallest correctable algebra containing $\mN$, and the persistent core is the largest correctable algebra inside it. They are dual to each other (Lemma~\ref{lem:history-memory}), and a generating inclusion forces the environment to have a trivial persistent core, which is the algebraic form of decoupling. The ergodic hierarchy of Section~\ref{subsec:QEC-hierarchy}, from autonomous inclusions to modular $K$-inclusions, therefore ranks inclusions by how much of $\mN$ stays correctable.

Imaginary modular time gives the same function a second life. On the upper boundary of the strip it becomes a Petz recovery map, and in the interior it defines a family of positive operators $Q_\alpha$ with the reference vector as a zero-energy ground state. These inclusion parent Hamiltonians see exactly what the real-time picture retains. The ground space of every $Q_\alpha$ is the GNS closure of the persistent core (Theorem~\ref{thm:Q-kernel-core}), so the reference is the unique ground state when the core is trivial. The orbit of $\mN$ under $e^{itQ_\alpha}$ then generates $B(\mH_\mN)$ (Proposition~\ref{prop:Halpha-full-generation}). For a standard inclusion, not only does this flow leave $\mM$, but it also generates $B(\mH_\mM)$ (Corollary~\ref{cor:standard-Halpha-full-generation}). 

At the midpoint $\alpha=1/4$, the Petz map and the parent Hamiltonian are two aspects of one transfer operator, $\rho_{1/4}=e^{-2Q_{1/4}}$. Repeated recovery can be viewed as Euclidean evolution, and it relaxes to the persistent core (Theorem~\ref{thm:main-recovery-relaxation}). With the assumption of $L^2$-nuclearity the inclusion spectrum is discrete, and we conjecture that the level statistics and eigenvector typicality of this discrete spectrum can be used as local diagnostics of quantum chaos (Section~\ref{sec:local-chaos}). 

\paragraph{Ergodicity versus mixing:}
The ergodic hierarchy is easy to over-read. Generation and a trivial persistent core record which observables the modular orbit reaches and retains, and they concern time averages. However, this is a much weaker statement than mixing, which involves late-time limits instead of time-averaging. Ergodicity without mixing is familiar classically. An irrational rotation of the circle has dense orbits, yet nearby points never separate and correlations never decay. We saw that the recovery semigroup relaxes, but the parameter of this semigroup is neither modular nor physical time.

A second set of distinctions concerns factorization. The split property says that $\mN$ and $\mM'$ are statistically independent, and it excludes faithful normal expectations onto a type III subalgebra, but it is a statement about the inclusion and not about dynamics. Conormality concerns $\mN$ and its relative commutant $\mR$, and together with a faithful normal expectation it forces a spatial tensor factorization, which is what fails for adjacent local algebras (Example~\ref{ex:local-algebras}). That failure rules out a normal product structure. It does not rule out expectations preserving other faithful states, or error correction for other channels or families of states, and the general criterion remains Takesaki's. 

\paragraph{Emergence of time:}
The central open question, from a physics standpoint, is whether modular data of an inclusion can be used to define local notions of emergent time. In the case of a half-sided modular inclusion, the inclusion parent Hamiltonian leads to a natural notion of emergent time {\cite{Borchers1992,Wiesbrock1993,companionQFT}}. For a general inclusion, $Q_\alpha$ is a family of positive Hamiltonians, one for each imaginary modular time, and nothing in the construction selects a physical clock among them. 
A companion paper~\cite{companionQFT} is motivated by inclusions in QFT and holography that come with a nested interpolating net. There we argue that $Q_{1/4}$, at least for an infinitesimal time step, is a natural notion of bulk Hamiltonian conjugate for an observer localized in the vicinity of the HRT surfaces corresponding to the inclusion. However, a deeper understanding of this requires moving from ergodicity statements to mixing.

\paragraph{Local quantum chaos:} The Introduction recalled recent work suggesting that geometric and chaotic modular flows are tied. The inclusion level statistics and the inclusion ETH of Section~\ref{sec:local-chaos} turn this into a question about the spectrum of $Q_\alpha$, but they remain conjectures awaiting computation in concrete QFTs. The companion paper computes the subsystem flows of wedges and balls in vacuum QFT and derives the modular chaos bound from their analytic structure. How the translation-like spectrum of a half-sided inclusion relates to the Wigner--Dyson statistics conjectured for generic interacting states is not understood.

Finally, besides the conceptual open problems above, we would like to mention two more technical questions that are natural extensions of this work. First, given that the upper boundary of $D(z)$ gives the rotated Petz maps, it is natural to look for a more direct connection between subsystem flow and the recovery bounds of Fawzi--Renner and of Junge, Renner, Sutter, Wilde and Winter~\cite{JungeRennerSutterWildeWinter2018}. Second, the 
assumption of full support, $s(\rho_\alpha)=1$, was essential in the ergodic interpretation of the results in this paper. It would be interesting to understand the physical interpretation of this assumption better. 

\section*{Acknowledgments}
 We thank Shadi Ali Ahmad, Sulaiman Alvi, Marc Klinger, Hong Liu and Allic Sivaramakrishnan for many insightful conversations. NL and KLL's work is supported by the DOE grant DE-SC0025547, and the Templeton Foundation grant 63670 on “Emergence of time from
chaos in operator algebras”. We have used AI in finding the proof of Proposition \ref{prop:Halpha-full-generation} (initially), in the proof of equation (\ref{eq:Q-gap-optimal-midpoint}), and in editing the draft and making the figures.

\appendix
\section{Additional operator-algebra background}\label{app:extra-operator-algebra}

This appendix sets the operator-algebra notation, collects the background used in the main text, and contains several of its proofs and the compression identities for the subsystem flow. 

\subsection{Review of general quantum systems (von Neumann algebras)}\label{app:vN}

We start by reviewing and fixing our notation for general quantum systems (von Neumann algebras). Readers familiar with standard definitions can skip this subsection.

\paragraph{Algebra of quantum observables:}
Quantum observables are described by a \emph{von Neumann algebra} $\mM$, a unital $*$-subalgebra of the bounded operators $B(\mathcal{H})$ on some separable Hilbert space $\mathcal{H}$ closed in the weak operator topology. A von Neumann algebra $\mM$ is a \emph{factor} if its center is trivial $Z(\mM) = \mM\cap\mM'=\mathbb{C} 1$, where $\mM' = \{b\in B(\mathcal{H}) : [b,a]=0\text{ for all }a\in\mM\}$ is the commutant in $B(\mH)$.  Physically, a factor algebra describes a system with no classical observables or superselection sectors\footnote{Appendix~\ref{sec:classical} develops a classical analogy using the Poisson centralizer.}.

\paragraph{Quantum states:}

A \emph{normal state} on $\mM$ is a positive, normalized, $\sigma$-weakly continuous
linear functional $\omega:\mM\to\mathbb{C}$, with $\omega(1)=1$\footnote{A normal state is the infinite-dimensional generalization of a standard quantum density matrix on the GNS Hilbert space. Infinite-dimensional von Neumann algebras can also have singular states that are not represented by trace-class density operators. For example, fixing a free ultrafilter $\mathcal U$ on $\mathbb N$, the ultralimit $\omega_{\mathcal U}(a)=\lim_{n\to\mathcal U}\braket{n|a|n}$ defines a singular state on $B(\mathcal H)$ that vanishes on compact operators. Requiring a state to be {\it normal} rules out such states.}.
For a finite matrix algebra it has the form $\omega(a)=\Tr(\rho_\omega a)$,
where $\rho_\omega$ is positive with unit trace\footnote{Type III factors are infinite-dimensional and admit no nonzero normal semifinite trace. A normal state is therefore defined intrinsically as a functional, without a density matrix relative to a trace on the algebra.}. A state $\omega$ is \emph{faithful} if $\omega(a^*a)=0$ implies $a=0$.
In finite dimensions, this means that its density matrix has no zero eigenvalues (is full rank)\footnote{Faithfulness is the counterpart of a strictly positive classical probability density
and gives a separating vector in the GNS construction.}. A state $\omega$ is \emph{tracial} if $\omega(ab)=\omega(ba)$ for all $a,b\in\mM$.
\footnote{On a finite algebra, a faithful normal tracial state plays the role of the reference
Liouville measure of classical algebras. Its modular flow is trivial, despite faithfulness. In the type III
algebras relevant to QFT, we have faithful normal states but no trace.}

\paragraph{Dynamics, inner and outer:}
A dynamics is a flow, a pointwise $\sigma$-weakly continuous one-parameter group $t\mapsto\alpha_t\in\operatorname{Aut}(\mM)$. The automorphism $\alpha_t$ is \emph{inner} if it is $\operatorname{Ad}u$ for a unitary $u\in\mM$, and the flow is inner if every $\alpha_t$ is. On a factor with separable predual, an inner flow is implemented by $e^{itH}$ with $H$ self-adjoint and affiliated with $\mM$, meaning that its spectral projections lie in $\mM$, and $H$ need not be bounded. An outer automorphism has no implementing unitary in $\mM$~\cite{Kallman1971}.

\paragraph{GNS Hilbert space and modular operators:}
Given a von Neumann algebra $\mM$ and a faithful normal state $\omega$, the
Gelfand--Naimark--Segal (GNS) construction produces a Hilbert space $\mathcal{H}_\omega$,
a cyclic unit vector $\ket{\Omega}\in\mathcal{H}_\omega$ with
$\omega(a)=\langle\Omega|a|\Omega\rangle$, and a $*$-representation
$\pi:\mM\to B(\mathcal{H}_\omega)$\footnote{The classical analog is $\mathcal{H}=L^2(X,\mu)$
with the constant function $|1\rangle$ as the reference vector.}.

Tomita--Takesaki theory assigns to the pair $(\mM,\ket{\Omega})$ a closable antilinear operator, initially defined on the dense domain
$\mM\ket{\Omega}$ by
\begin{equation}
  S_\omega\, a\ket{\Omega} = a^*\ket{\Omega}, \qquad a\in\mM.
  \label{eq:Tomita-S}
\end{equation}
We denote its closure by $S_\omega$. Its polar decomposition $S_\omega=J_\omega\Delta_\omega^{1/2}$ defines the
\emph{modular conjugation} $J_\omega$, an antiunitary involution, and the
\emph{modular operator} $\Delta_\omega$, a positive self-adjoint operator. 

\paragraph{Modular dynamics:}
The Tomita--Takesaki theorem~\cite{Takesaki1970} guarantees that the one-parameter group
of unitaries $\Delta_\omega^{it}$, $t\in\mathbb{R}$, implements an automorphism group of
$\mM$:
\begin{equation}
  \sigma_t^\omega(a) = \Delta_\omega^{it}\, a\, \Delta_\omega^{-it},
  \qquad a\in\mM,\quad t\in\mathbb{R}.
  \label{eq:modular-flow}
\end{equation}
This is the \emph{modular flow} (or \emph{modular automorphism group}) of $\omega$.
For matrix algebras, this is $\sigma^{\omega}_t(a)=\rho_\omega^{it}a\rho_\omega^{-it}$.
The modular operator $\Delta_\omega$ and $J_\omega$ encode the non-commutativity of
$\mM$. Modular flow is trivial in Abelian algebras. In Appendix~\ref{subsec:classical-state-flow} we argue that the Poisson flow associated with classical states is the correct replacement of modular flow to formulate subsystem flow and ergodic theory\footnote{This construction uses
the Poisson bracket and differs from the trivial modular automorphism group of
an abelian von Neumann algebra. In both constructions the reference state is stationary:
$\omega\circ\sigma_t^\omega=\omega$.}.

With the convention \eqref{eq:modular-flow}, the \emph{KMS condition} reads
\begin{equation}
 F_{a,b}(t)=\omega\bigl(a\,\sigma_t^\omega(b)\bigr),
 \qquad F_{a,b}(t-i)=\omega\bigl(\sigma_t^\omega(b)a\bigr).
 \label{eq:KMS-function}
\end{equation}
The function is bounded and analytic on $-1<\operatorname{Im}z<0$, with continuous
boundary values. Equivalently, $\beta_t^\omega:=\sigma_{-t}^\omega
=\operatorname{Ad}\Delta_\omega^{-it}$ gives the upper-strip condition at inverse
temperature one. Its generator is $K_\omega=-\log\Delta_\omega$, and on its domain
$\partial_t\sigma_t^\omega(b)|_{t=0}=i[\log\Delta_\omega,b]$.

\paragraph{Relative modular flow and relative entropy:}
For faithful normal states $\omega,\psi$ on a von Neumann algebra $\mathcal A$, represented by vectors $\ket{\Omega},\ket{\Psi}$ in standard form, the relative Tomita operator $S_{\psi|\omega;\mathcal A}$ is defined by
\begin{align}
 S_{\psi|\omega;\mathcal A} a\ket{\Omega}=a^*\ket{\Psi},\qquad a\in\mathcal A
\end{align}
After taking the closure of this relative Tomita operator, the relative modular operator is
\begin{align}
 \Delta_{\psi|\omega;\mathcal A}=S_{\psi|\omega;\mathcal A}^\dagger S_{\psi|\omega;\mathcal A},
 \qquad
 \langle\Delta_{\psi|\omega;\mathcal A}^{1/2}a\Omega|\Delta_{\psi|\omega;\mathcal A}^{1/2}b\Omega\rangle=\psi(ba^*),\qquad a,b\in\mathcal A,
 \label{eq:relative-modular}
\end{align}
so that $\Delta_{\omega|\omega;\mathcal A}=\Delta_{\omega;\mathcal A}$. Araki's relative entropy is
\begin{align}
 S_{\mathcal A}(\omega\|\psi)
 =-\langle\Omega|\log\Delta_{\psi|\omega;\mathcal A}|\Omega\rangle
 =i\left.\frac{d}{dt}\langle\Omega|\Delta_{\psi|\omega;\mathcal A}^{it}|\Omega\rangle\right|_{t=0},
 \label{eq:relative-entropy}
\end{align}
with the derivative understood when the logarithmic spectral moment is integrable~\cite{Araki1976,Araki1977}. In matrix algebras, the relative modular superoperator acts as  $\Delta_{\psi|\omega}(b)=\rho_\psi b \rho_\omega^{-1}$ and $S_{\mathcal A}(\omega\|\psi)$ reduces to the usual quantum relative entropy. Relative entropy is nonnegative and monotonic under restriction to a von Neumann subalgebra. These properties underlie the sufficiency gaps used in the main text \cite{Araki1976}.

\subsection{Algebra types and ergodicity of modular flow}\label{subsec:types-modular}

The inner/outer character and the ergodic properties of modular flows distinguish the types of infinite-dimensional factors. To start, we need a few definitions.
The set of operators that are invariant under modular flow (conserved modular charges) forms an algebra of modular charges (fixed-point subalgebra)\footnote{The subalgebra of modular charges is called {the centralizer} in mathematical jargon.}. Ergodicity of modular dynamics requires the absence of conserved modular charges (trivial centralizer) \cite{longo1979notes,ouseph2024local}. If there are classical observables (the algebra has a center), they are inside the intersection of modular charges of all states. Since our focus is modular ergodicity, we focus on von Neumann factors so that there are no operators that are conserved under the modular flow of all states. von Neumann factors can be classified as
\begin{enumerate}
    \item {\bf Type I:} Every faithful normal state has a trace-class density matrix with pure-point spectrum. In other words, the modular unitary group is almost periodic and cannot be ergodic. There is a trace, with $\Tr(1)=n<\infty$ for type I$_n$ and $\Tr(1)=\infty$ for type I$_\infty$.

    \item {\bf Type II:} Every faithful normal state, and more generally every faithful normal semifinite weight, has a density relative to a faithful normal semifinite trace $\tau$: for $x\geq0$, $\omega(x)=\tau(\rho_\omega^{1/2}x\rho_\omega^{1/2})$, and $\sigma_t^\omega=\operatorname{Ad}\rho_\omega^{it}$. Here $\rho_\omega$ is a positive injective operator affiliated with $\mM$. The spectral projections of $\rho_\omega$ lie in the centralizer, so a type-II factor has no ergodic modular state. A finite trace exists in type II$_1$ and has trivial modular flow. Type II$_\infty$ has a trace, but it is not finite: $\Tr(1)=\infty$.
    
    \item {\bf Type III:} Type III algebras admit no trace and no intrinsic density matrices. Within modular ergodic theory, type III$_1$ is the only factor type that can support a faithful normal state with no modular charges (trivial centralizer). Under the standard assumptions of relativistic AQFT, local algebras are typically type III$_1$ factors. The relation between modular mixing, information loss, and emergent type III$_1$ structure was emphasized in~\cite{furuya2023information}.

\end{enumerate}

\subsection{Modular flows versus arbitrary unitary dynamics}\label{sec:modular-vs-dynamics}
Stone's theorem gives a self-adjoint generator for every strongly continuous unitary group, $U_t=e^{itH}$. With the modular convention \eqref{eq:modular-flow}, a Gibbs density matrix $\rho=Z^{-1}e^{-H}$ instead gives
\begin{equation}
 \sigma_t^\rho(a)=e^{-itH}ae^{itH},\qquad
 \operatorname{Ad}U_t=\sigma_{-t}^\rho.
 \label{eq:appendix-Gibbs-time-orientation}
\end{equation}
Thus the two flows agree after reversing the parameter. This sign matters when comparing a physical Hamiltonian with the modular Hamiltonian $-\log\Delta$.

A faithful normal semifinite \emph{weight} generalizes a state by allowing the value $+\infty$ on positive observables. It need not be normalizable. For a semifinite factor with faithful normal semifinite trace $\tau$, a positive injective operator $h$ affiliated with the algebra defines the weight
\begin{equation}
 \varphi_h(x)=\tau(h^{1/2}xh^{1/2}),\qquad x\geq0,
 \qquad \sigma_t^{\varphi_h}(a)=h^{it}ah^{-it}.
\end{equation}
Consequently, if $H$ is affiliated with the algebra, $\operatorname{Ad}e^{itH}$ is the modular flow of the weight with $h=e^H$. It is the modular flow of a normal state precisely when this weight can be normalized, $\tau(e^H)<\infty$. A physical Gibbs weight with $h=e^{-H}$ gives the opposite time orientation.

\paragraph{Type I factors:}
Every automorphism of $B(\mathcal H)$ is inner, and every strongly continuous automorphism group is implemented by $e^{itH}$ for a self-adjoint $H$. It is therefore modular for a weight. In infinite dimensions, the weight can have infinite total mass, $\Tr e^{\pm H}=\infty$. The exponential itself remains a well-defined positive operator. The statement about inner automorphisms concerns factors: a type I algebra with a nontrivial center can also have automorphisms that move its central sectors.

\paragraph{Type II factors:}
The modular flow of every faithful normal semifinite weight is inner. Conversely, for a factor with separable predual, an inner flow has a self-adjoint generator affiliated with the algebra~\cite{Kallman1971} and is modular for a weight as above. Outer dynamics therefore cannot be modular. An example on the tracial infinite spin chain is the global rotation $\alpha_t=\bigotimes_n\operatorname{Ad}e^{it\sigma_z}$: it is outer for $t\notin\pi\mathbb Z$, since the identical site rotations fail the infinite-product innerness criterion. The total magnetization is not an affiliated generator in that representation.

A type $\mathrm{II}_\infty$ factor can admit trace-scaling automorphisms,
\begin{equation}
 \tau\circ\theta_\lambda=\lambda\tau.
\end{equation}
The allowed values of $\lambda$ depend on the factor. For $\mM=\mathcal P\overline\otimes B(\ell^2)$ with $\mathcal P$ a type $\mathrm{II}_1$ factor, they form its fundamental group $\mathcal F(\mathcal P)=\{\lambda>0:\mathcal P^\lambda\cong\mathcal P\}$. Here $\mathcal P^\lambda$ is the amplification, obtained by cutting $\mathcal P\overline\otimes B(\ell^2)$ by a projection of trace $\lambda$, with $\tau_\mathcal P(1)=1$. All positive values occur for the hyperfinite factor, but this is not a property of every type $\mathrm{II}_\infty$ factor. Any nontrivial scaling is outer because inner automorphisms preserve $\tau$.

\paragraph{Type III factors:}
Modular dynamics determines a canonical outer class. For faithful normal semifinite weights $\varphi,\psi$,
\begin{equation}
 \sigma_t^\psi=\operatorname{Ad}[D\psi:D\varphi]_t\circ\sigma_t^\varphi,
 \qquad
 \delta_\mM(t):=[\sigma_t^\varphi]\in\operatorname{Out}(\mM).
\end{equation}
The class $\delta_\mM(t)$ is independent of the weight, and its kernel is Connes' $T$-invariant,
\begin{equation}
 T(\mM)=\{t:\sigma_t^\varphi\in\operatorname{Inn}(\mM)\}.
\end{equation}
For $\mathrm{III}_\lambda$, $0<\lambda<1$, one has $T(\mM)=(2\pi/|\log\lambda|)\mathbb Z$ and periodic faithful normal semifinite weights can be chosen. For $\mathrm{III}_0$, the $T$-invariant is not determined by the type. For $\mathrm{III}_1$, $T(\mM)=\{0\}$, so every nonzero modular time is outer. Its continuous core is type $\mathrm{II}_\infty$ and carries the dual trace-scaling action~\cite{Connes1973,ConnesTakesaki1977}.

Modular flow preserves the local algebra; however, there are potentially many symmetries (spacetime or internal) that could preserve it. To be a modular flow of a faithful normal state, it must represent the canonical outer class and satisfy the corresponding KMS condition. Causal or geometric intuition alone does not establish these requirements.

\paragraph{Ergodic and strongly mixing type III$_1$ modular flows:}
Elements of the center $Z\subset\mM$ commute with every modular flow, $\Delta_\omega^{it}z\Delta_\omega^{-it}=z$ for all $z\in Z$, $t$ and $\omega$. They are conserved charges and obstruct ergodicity, which is why we restrict to factors. For a fixed faithful normal state, modular ergodicity means that the centralizer is trivial,
\begin{align}
    \label{eq:modular-ergodicity}
    \mM^\omega\equiv \{a\in \mM:\Delta_{\omega;\mM}^{it}a\Delta_{\omega;\mM}^{-it}=a\qquad \forall t\}=\mathbb{C} 1.
\end{align}
This is a property of the pair $(\mM,\omega)$, not of the type alone. For a nontrivial factor, a trivial centralizer forces type III$_1$~\cite{longo1979notes}, but a III$_1$ factor can still carry a state with a nontrivial centralizer\footnote{Take an infinite tensor product of identical three-level systems with faithful site density matrix $\rho=\sum_{i=1}^3p_i\ket{i}\bra{i}$. If $\log(p_1/p_3)$ and $\log(p_2/p_3)$ are rationally independent, the Araki--Woods factor is type III$_1$~\cite{ArakiWoods1968}. Yet the single-site spectral projections $\ket{i}\bra{i}_n$ lie in the centralizer of the product state.}. Conversely, every type III$_1$ factor with separable predual has a faithful normal state with trivial centralizer~\cite{MarrakchiVaes2024}. The intuition of information loss is formalized by strong mixing, meaning clustering in modular time. It is strictly stronger than ergodicity~\cite{furuya2023information}, and in a QFT example it must be established for the specified algebra and state, since it does not follow from the type III$_1$ classification.

\subsection{Outer actions and irreducible inclusions}
\label{app:crossed-products}

The following examples, referred to in Section~\ref{section:basics}, show how outer automorphisms and irreducible inclusions arise.

A simple mixed inclusion illustrates how outer automorphisms can arise from classical block labels. Let
\begin{equation}
 \mathcal H=\mathcal H_B\otimes\mathbb C^d,\qquad
 \mN=B(\mathcal H_B)\otimes D_{d\times d} = \bigoplus_{a=1}^{d}B(\mathcal H_B)\otimes\ket{a}\bra{a}
 \subset\mM=B(\mathcal H),
\end{equation}
where $D_{d\times d}$ is the diagonal algebra. The unitaries
\begin{equation}
 U_\sigma=1_B\otimes\sum_{a=1}^d|\sigma(a)\rangle\langle a|,
 \qquad \sigma\in S_d,
\end{equation}
permute the blocks. Every nontrivial permutation induces an outer automorphism of $\mN$, since inner automorphisms fix its center pointwise. Adjoining these unitaries generates $\mM$, and $\mN'\cap\mM=1_B\otimes D_{d\times d}=Z(\mN)$.

To realize this matrix algebra as a crossed product, it suffices to use the cyclic shift $T=1_B\otimes\sum_a|a+1\rangle\langle a|$, with labels understood modulo $d$. Its action on the block labels is free and transitive, and
\begin{equation}
 \mN\rtimes_{\operatorname{Ad}T}\mathbb Z_d\cong\mM.
\end{equation}

For a proper inclusion of finite-dimensional factors, the relative commutant is nontrivial, and every automorphism of either factor is inner. Infinite-dimensional factors can behave differently. If a discrete group $G$ acts on a factor $\mN$ with every nonidentity group element acting outerly, then $\mN\subset\mN\rtimes G$ has trivial relative commutant. Such an inclusion is called \emph{irreducible}.

\subsection{Modular energy and relative entropy}
\label{app:modular-energy-proof}

\paragraph{Proof of Lemma~\ref{lem:modular-energy}:}
\begin{proof}
\emph{Cocycle factorization:}
Write $\omega',\varphi'$ for the states induced on $\mathcal A'$. For cyclic and separating vectors $\Phi,\ket{\Psi}$ the relative modular operators $\Delta_{\Phi|\Psi;\mathcal A}$ of \eqref{eq:relative-modular} satisfy~\cite{Araki1976,Araki1977}
\begin{equation}
 \Delta_{\Phi|\Theta;\mathcal A}^{it}=[D\Phi:D\Psi]^{\mathcal A}_t\,\Delta_{\Psi|\Theta;\mathcal A}^{it},
 \qquad
 \Delta_{\Phi|\Psi;\mathcal A'}=\Delta_{\Psi|\Phi;\mathcal A}^{-1},
 \label{eq:relative-modular-cocycles}
\end{equation}
where $\Theta$ is any cyclic and separating vector (for $\Theta=\Psi$ the first identity reads $\Delta_{\Phi|\Psi}^{it}=[D\Phi:D\Psi]_t\Delta_\Psi^{it}$) and $[D\Phi:D\Psi]^{\mathcal A}_t\in\mathcal A$ is the Connes cocycle of the two states on $\mathcal A$. The first identity changes the state in the first slot. Applied on $\mathcal A'$ and transported by the second identity, it changes the state in the second slot: $\Delta_{\Phi|\Psi;\mathcal A}^{it}=\Delta_{\Phi;\mathcal A}^{it}\,[D\Phi':D\Psi']^{\mathcal A'}_t$. Applying the first identity with $(\Phi,\Psi,\Theta)=(\omega,\varphi,\omega)$ and then the second-slot formula with $(\Phi,\Psi)=(\varphi,\omega)$ to $\Delta_{\omega;\mathcal A}=\Delta_{\omega|\omega;\mathcal A}$ gives
\begin{equation}
 \Delta_{\omega;\mathcal A}^{it}=[D\omega:D\varphi]^{\mathcal A}_t\;\Delta_{\varphi;\mathcal A}^{it}\;[D\varphi':D\omega']^{\mathcal A'}_t .
 \label{eq:three-factor}
\end{equation}
Set $a_t=[D\omega:D\varphi]^{\mathcal A}_t\in\mathcal A$ and $b_t=\Delta_{\varphi;\mathcal A}^{it}[D\varphi':D\omega']^{\mathcal A'}_t\Delta_{\varphi;\mathcal A}^{-it}\in\mathcal A'$, both unitary with $a_0=b_0=1$. Since $\Delta_{\varphi;\mathcal A}^{it}\ket{\Psi}=\ket{\Psi}$, \eqref{eq:three-factor} gives
\begin{align}
 \langle\Psi|\Delta_{\omega;\mathcal A}^{it}|\Psi\rangle&=\langle\Psi|a_tb_t|\Psi\rangle,\nn\\
 \langle\Psi|a_t|\Psi\rangle&=\varphi\bigl([D\omega:D\varphi]_t\bigr),
 \qquad
 \langle\Psi|b_t|\Psi\rangle=\varphi'\bigl([D\varphi':D\omega']_t\bigr).
\end{align}
\smallskip\noindent
\emph{Finite logarithmic moments:}
By the first identity in \eqref{eq:relative-modular-cocycles}, $\varphi([D\omega:D\varphi]_t)=\langle\Psi|\Delta_{\omega|\varphi;\mathcal A}^{it}|\Psi\rangle=\int e^{it\lambda}\,d\mu(\lambda)$, with $\mu$ the spectral measure of $\log\Delta_{\omega|\varphi;\mathcal A}$ in $\ket{\Psi}$. Here $\int e^{\lambda}d\mu=\|\Delta_{\omega|\varphi;\mathcal A}^{1/2}\ket{\Psi}\|^2=\|\ket{\Omega}\|^2=1$ bounds the positive part of $\lambda$, and $S_{\mathcal A}(\varphi\|\omega)=-\int\lambda\,d\mu<\infty$ then bounds the negative part, so $\int|\lambda|\,d\mu<\infty$ and $\varphi([D\omega:D\varphi]_t)$ is differentiable at $0$ with derivative $-iS_{\mathcal A}(\varphi\|\omega)$~\cite{OhyaPetz1993}. The same argument on $\mathcal A'$ gives $\partial_t\varphi'([D\omega':D\varphi']_t)|_{t=0}=-iS_{\mathcal A'}(\varphi\|\omega)$, and since $[D\varphi':D\omega']_t=[D\omega':D\varphi']_t^*$, the derivative of $\langle\Psi|b_t|\Psi\rangle$ at $0$ is $+iS_{\mathcal A'}(\varphi\|\omega)$.

\smallskip\noindent
\emph{The cross term:} Finally,
\begin{equation}
 \langle\Psi|a_tb_t|\Psi\rangle-1
 =\bigl(\langle\Psi|a_t|\Psi\rangle-1\bigr)+\bigl(\langle\Psi|b_t|\Psi\rangle-1\bigr)
 +\langle\Psi|(a_t-1)(b_t-1)|\Psi\rangle,
\end{equation}
and $\|(a_t^*-1)\ket{\Psi}\|^2=\|(a_t-1)\ket{\Psi}\|^2=2\operatorname{Re}\bigl(1-\langle\Psi|a_t|\Psi\rangle\bigr)=o(t)$, because the derivative of $\langle\Psi|a_t|\Psi\rangle$ at $0$ is purely imaginary, and likewise for $b_t$. The cross term is therefore $o(t)$, and
$-i\,\partial_t\langle\Psi|\Delta_{\omega;\mathcal A}^{it}|\Psi\rangle|_{t=0}=-S_{\mathcal A}(\varphi\|\omega)+S_{\mathcal A'}(\varphi\|\omega)$. The last statement is the spectral theorem for $\log\Delta_{\omega;\mathcal A}$ on its form domain. In finite dimensions $\log\Delta_{\omega;\mathcal A}=\log\rho^\omega_{\mathcal A}\otimes1-1\otimes\log\rho^\omega_{\mathcal A'}$ on $\mathcal H_{\mathcal A}\otimes\mathcal H_{\mathcal A'}$, and \eqref{eq:modular-energy} reduces to the equality of the von Neumann entropies of the two marginals of a pure state.
\end{proof}

\subsection{Conditional expectations and product structure}
\label{app:expectation-proofs}

\paragraph{Proof of Theorem~\ref{thm:split-no-expectation}:}
\begin{proof}
Let $\mN\subset\mathcal F\subset\mM$ be an interpolating type I
factor and suppose that $\mE:\mM\to \mN$ is faithful and normal.  Its restriction
\begin{equation}
 \mE_{\mathcal F}:=\mE|_{\mathcal F}:\mathcal F\longrightarrow\mN
\end{equation}
is a faithful normal norm-one projection with range $\mN$.  Tomiyama's type
theorem states that a normally expected von Neumann subalgebra of a type I von
Neumann algebra is type I \cite{Tomiyama1959,Takesaki1972}. This contradicts $\mN$ being a type III factor.
\end{proof}

\paragraph{Proof of Lemma~\ref{lem:expectation-slice}:}
\begin{proof}
For $b\in\mN$ and $c\in\mR$,
\begin{equation}
 b \mE(c)=\mE(b c)=\mE(c b)=\mE(c)b.
\end{equation}
Thus, $\mE(c)\in Z(\mN)=\mathbb C1$, so
$\mE(c)=\chi_{\mR}(c)1$ for a normal state $\chi_{\mR}$.  If $c\geq0$ and
$\chi_{\mR}(c)=0$, faithfulness of $\mE$ implies $c=0$. Hence $\chi_{\mR}$ is faithful.  The
second identity is the bimodule property.
This proves the scalar restriction and bimodule identities. The parametrization of all normal faithful expectations in the conormal case follows from the tensor-product construction below.
\end{proof}

For a faithful normal state $\varphi$ on $\mN$, the state
$\varphi_0:=\varphi\circ \mE$ is faithful and normal and satisfies
\begin{equation}
 \varphi_0(b c)=\varphi_0(b)\chi_{\mR}(c).
\end{equation}

\paragraph{Proof of Lemma~\ref{prop:conormal-factorization}:}
\begin{proof}

Let $\varphi$ be a faithful normal state on $\mN$ and let $\varphi_0=\varphi\circ\mE$. The first part of Lemma~\ref{lem:expectation-slice} gives the faithful normal product state
\begin{equation}
 \varphi_0(bc)=\varphi(b)\chi_\mR(c),\qquad b\in\mN,\ c\in\mR.
\end{equation}
Write $\mathcal A=\mN\vee\mR$ and take the GNS representation of $\varphi_0|_{\mathcal A}$. On the algebraic tensor product of the two GNS cores, the map
\begin{equation}
 b\ket{\Omega_\varphi}\otimes c\ket{\Omega_\chi}
 \longmapsto bc\ket{\Omega_{\varphi_0}}
\end{equation}
preserves inner products by the product identity. Its range is dense, so it extends to a unitary. This unitary intertwines the spatial tensor-product representation of $\mN\overline\otimes\mR$ with multiplication on $\mathcal A$. Since the states and their GNS representations are faithful and normal, multiplication extends to a normal faithful $*$-isomorphism onto $\mathcal A$. Conormality gives $\mathcal A=\mM$ and hence \eqref{eq:conormal-factorization}. The remaining assertion of Lemma~\ref{lem:expectation-slice} follows by transporting the slice maps $\operatorname{id}\otimes\chi$ through this isomorphism.

\end{proof}

\subsection{Proof of the history--core duality (Lemma~\ref{lem:history-memory})}
\label{app:history-core-proof}

\begin{proof}
(a) For von Neumann algebras, the commutant of a join is the intersection of the commutants and the commutant of an intersection is the join of the commutants, and $\operatorname{Ad}U_t$ commutes with taking commutants. Hence $(\bigvee_t\alpha_t(\mN))'=\bigcap_t\alpha_t(\mN')$ and $(\bigcap_t\alpha_t(\mN))'=\bigvee_t\alpha_t(\mN')$.

(b) Intersecting the first identity of (a) with $\mM=\alpha_t(\mM)$ gives $\bigcap_t\alpha_t(\mN')\cap\mM=\bigcap_t\alpha_t(\mN'\cap\mM)=\mR_{\rm core}$. For the inclusion, $\alpha_t(\mN'\cap\mM)\subseteq\alpha_t(\mN')\cap\mM\subseteq(\bigcap_s\alpha_s(\mN))'\cap\mM$ for every $t$. Take the join over $t$. For the equality statement write $\mathcal P^c=\mathcal P'\cap\mM$. The first relation of (b) applied to $\mR$ in place of $\mN$ reads $(\mR_{\rm hist})^c=(\mR^c)_{\rm core}$, and $\mR^c=\mN$ when $\mN=\mN^{cc}$. Taking relative commutants once more, $(\mR_{\rm hist})^{cc}=(\mN_{\rm core})^c$, which equals $\mR_{\rm hist}$ when $\mR_{\rm hist}=(\mR_{\rm hist})^{cc}$. In the tensor-product case $\mN=\mN^{cc}$, and if $\mR$ is autonomous then $\mR_{\rm hist}=\mR=\mR^{cc}$.
\end{proof}

\subsection{Compression identities for subsystem flow}
\label{app:compressed-flow}

The following identity records the effect of compressing the single supported flow. Extend the intrinsic unitary to
$U_\mN(t)=V\Delta_\mN^{it}V^\dagger+(1-e_\mN)$ on $\mH_\mM$.
Then $D(t)=\Delta_\mM^{-it}U_\mN(t)e_\mN$.
Writing $\widetilde\sigma_{-t}^{\omega;\mN}=\operatorname{Ad}U_\mN(-t)$ gives
\begin{align}
 e_\mN D(t+s)e_\mN
 &=(e_\mN D(t)e_\mN)
       \widetilde\sigma_{-t}^{\omega;\mN}(e_\mN D(s)e_\mN)\nonumber\\
 &\quad+\mathcal L(t,s),\nonumber\\
 \mathcal L(t,s)
 &=e_\mN\Delta_\mM^{-it}(1-e_\mN)
       \widetilde\sigma_{-t}^{\omega;\mN}(D(s))e_\mN.
\end{align}
The second term records passage through the orthogonal complement. This algebraic identity does not itself specify an open-system master equation.
On vectors $\psi\in\operatorname{Dom}K_\mN$ with
$V\psi\in\operatorname{Dom}K_\mM$,
\begin{equation}
 -i\left.V^\dagger\partial_tD(t)V\psi\right|_{t=0}
 =(V^\dagger K_\mM V-K_\mN)\psi.
\end{equation}
This compares modular generators on the stated domain.

Put $C_t=e_\mN D(t)e_\mN$. The Heisenberg map
\begin{equation}
 \Phi_t(a)=C_t^\dagger aC_t,\qquad a\in B(e_\mN\mH_\mM),
\end{equation}
is completely positive and subunital. It is unital precisely when $C_t^\dagger C_t=e_\mN$, namely $e_\mN D(t)^\dagger e_\mN D(t)e_\mN=e_\mN$. For the opposite orientation $a\mapsto C_taC_t^\dagger$, the condition is instead $C_tC_t^\dagger=e_\mN$.

\subsection{\texorpdfstring{The two-sided canonical tower in the common-cyclic case}{The two-sided canonical tower in the common-cyclic case}}
\label{subsec:canonical-tower-common-cyclic}

This subsection describes the discrete structure behind the Petz map of Section~\ref{subsec:recovery-Poisson}. When $\ket{\Omega}$ is cyclic and separating for both algebras, the Petz map is the canonical endomorphism and its unitary implementation $\Gamma$ generates a tower of algebras on both sides of $\mN\subset\mM$.

For a common cyclic and separating vector, the unitary $\Gamma$ and canonical endomorphism $\gamma_{\rm can}$ are defined in \eqref{eq:interpolation-Gamma}--\eqref{eq:interpolation-canonical-endomorphism}. Their iterates produce a two-sided tower of von Neumann algebras.
For each $k\in\mathbb Z$, put
\begin{equation}
 \mA_{2k}:=\Gamma^k\mM\Gamma^{-k},\qquad
 \mA_{2k+1}:=\Gamma^k\mN\Gamma^{-k}.
 \label{eq:interpolation-tower-levels}
\end{equation}
The original inclusion is $\mA_1=\mN\subset\mA_0=\mM$,
and $\mA_{j+1}\subseteq\mA_j$ for every integer $j$:
\begin{equation}
 \cdots\mathrel{\supseteq}\Gamma^\dagger\mM\Gamma
 \mathrel{\supseteq}\Gamma^\dagger\mN\Gamma
 \mathrel{\supseteq}\mM\mathrel{\supseteq}\mN
 \mathrel{\supseteq}\Gamma\mM\Gamma^\dagger\mathrel{\supseteq}\Gamma\mN\Gamma^\dagger
 \mathrel{\supseteq}\cdots .
 \label{eq:interpolation-two-sided-tower}
\end{equation}
The left side is a tower of \emph{extensions} of $\mM$,
not merely the commutants of the descending tower. Its first extension
is
\begin{equation}
 \mA_{-1}=\Gamma^\dagger\mN\Gamma
 =J_{\mM}\mN'J_{\mM}\supseteq\mM.
 \label{eq:interpolation-first-extension}
\end{equation}
Equivalently, applying commutants to the descending tower gives the
dual ascending chain
$\mM'\subseteq\mN'\subseteq(\Gamma\mM\Gamma^\dagger)'
\subseteq\cdots$. The inversely shifted original tower gives the
extensions in \eqref{eq:interpolation-two-sided-tower}. This is the
canonical-endomorphism or Jones-tunnel construction familiar from
subfactor theory~\cite{revolutionizing,GuidoLongoWiesbrock1998}.

Every $\mA_j$ has $\ket{\Omega}$ as a cyclic and separating vector.
Consequently, each neighboring pair has its own \emph{modular interpolation tunnel}, generated by the modular-flow translates of $\mA_{j+1}$ within $\mA_j$ over times $|s|<T$,
\begin{equation}
 \mI_{j,T}:=
 \bigvee_{|s|<T}
 \sigma_s^{\omega;\mA_j}(\mA_{j+1}),
 \qquad j\in\mathbb Z,\quad T>0.
 \label{eq:interpolation-tower-net}
\end{equation}
These algebras satisfy
\begin{equation}
 \mA_{j+1}\subseteq\mI_{j,T_1}
 \subseteq\mI_{j,T_2}\subseteq\mA_j,
 \qquad T_1<T_2,
 \label{eq:interpolation-tower-net-order}
\end{equation}
as well as $\mI_{j+1,S}\subseteq\mI_{j,T}$ for all $S,T>0$.
Thus the discrete depth $j$ and the continuous modular-history size $T$
organize a partially ordered net of von Neumann algebras on both
sides of the original inclusion. Since $\Gamma\ket{\Omega}=\ket{\Omega}$,
\begin{equation}
 \Gamma\mA_j\Gamma^\dagger=\mA_{j+2},\qquad
 \Gamma\mI_{j,T}\Gamma^\dagger=\mI_{j+2,T}.
 \label{eq:interpolation-shift-covariance}
\end{equation}
The shift by two is exact, but it is a \emph{discrete} covariance. It does not make the endomorphism the continuous flow of a positive
generator. In half-sided modular inclusions, a continuous geometric
interpolation is available, whereas for a general inclusion none is
implied by the tower alone.

\section{Basic properties of subsystem flow}\label{app:basic-properties}
In this paper, we introduced the notion of a subsystem flow. In this appendix, we clarify some basic mathematical properties of the subsystem flow. These properties inherit directly from the corresponding properties of Borchers' characteristic function \cite{revolutionizing} and should require no motivation.

A subsystem is a pair of von Neumann algebras $\mathcal{N}\subset\mathcal{M}$ along with a reference state $\omega$. The state is represented by the GNS vector $\ket{\Omega}\in\mathcal{H}_\mathcal{M}$ where $\mathcal{H}_\mathcal{M}$ is the GNS Hilbert space, and $\ket{\Omega}$ is cyclic and separating for $\mathcal{M}$. We do not assume $\ket{\Omega}$ is cyclic for $\mathcal{N}$. Hence the canonical isometry $V:\mathcal{H}_\mathcal{N}\to\mathcal{H}_\mathcal{M}$, $Vn\ket{\Omega_\mathcal{N}}=n\ket{\Omega}$, from the GNS Hilbert space of $\mathcal N$ need not be onto. Its range $\overline{\mathcal{N}\ket{\Omega}}$ is the subspace $e_\mathcal{N}\mathcal{H}_\mathcal{M}$, with $VV^\dagger=e_\mathcal{N}$ and $V\ket{\Omega_\mathcal{N}}=\ket{\Omega}$. The subsystem flow is
\begin{equation}
    D(t) = \Delta_\mathcal{M}^{-it}V\Delta_\mathcal{N}^{it}V^\dagger
\end{equation}
If $\ket{\Omega}$ is cyclic for $\mathcal{N}$, the isometry $V$ is the identity and the subsystem flow reduces to Borchers' characteristic function \cite{revolutionizing}. The purpose of this appendix is to extend and restate some basic properties of Borchers' characteristic function to subsystem flow. 

\subsection{Characterization by the subsystem flow}\label{app:characteristic-function}
For a common cyclic and separating vector, Borchers\textquotesingle{} characteristic function has the following properties~\cite{revolutionizing}.

\begin{enumerate}
    \item $D(0)=1$, and $D_{\mN\subset\mM}(t)$ and $D_{\mN\subset\mM}(t +\frac{i}{2})$ are unitary and strongly continuous in $t$,
    \item $D_{\mN\subset\mM}(z)$ is analytic on $0<\operatorname{Im}z<1/2$ and contractive on its closure,
    \item $D_{\mN\subset\mM}(t)\ket{\Omega} = \ket{\Omega}$,
    \item The cocycle condition holds: $D(t+s)=\Delta_{\mM}^{-it}D(s)\Delta_{\mM}^{it}D(t)$. Here conjugation by $\Delta_{\mM}^{it}$ acts on $B(\mH)$, since $D(s)$ need not lie in $\mM$.
    \item The intrinsic boundary symmetry is
    \[
     D(t+i/2)^\dagger J_{\mM} D(t)=D(t)^\dagger J_{\mM} D(t+i/2),
    \]
    independent of $t$. For the given inclusion its value is $J_{\mN}$, equivalently
    $D(t+i/2)=J_{\mM} D(t)J_{\mN}$. In particular, $D_{\mN\subset\mM}(\frac{i}{2}) = J_{\mM}J_{\mN}$. This unitary operator is crucial in subfactor theory, creating a chain of algebras (see Appendix~\ref{subsec:canonical-tower-common-cyclic}):
    \begin{equation}
        \mM\supset\mN\supset D_{\mN\subset\mM}(\frac{i}{2})^\dagger\mM D_{\mN\subset\mM}(\frac{i}{2})\supset D_{\mN\subset\mM}(\frac{i}{2})^\dagger\mN D_{\mN\subset\mM}(\frac{i}{2})\supset\cdots
    \end{equation}
    \item For all $t\in\mathbb{R}$, $D_{\mN\subset\mM}(t)D_{\mN\subset\mM}(\frac{i}{2})^\dagger\mM D_{\mN\subset\mM}(\frac{i}{2})D_{\mN\subset\mM}(t)^\dagger\subset\mM$ holds.
\end{enumerate}
\begin{theorem}
    Given a von Neumann algebra $\mM$ and a cyclic and separating vector $\ket{\Omega}$, each subalgebra $\mN\subset\mM$ for which $\ket{\Omega}$ is cyclic is uniquely determined by its subsystem flow $D_{\mN\subset\mM}$, and every operator-valued function with the properties listed above is the subsystem flow of such a subalgebra. In other words, $\mN\mapsto D_{\mN\subset\mM}$ is a bijection between the von Neumann subalgebras of $\mM$ for which $\ket{\Omega}$ is cyclic and the operator-valued functions with these properties. 
\end{theorem}

The reconstruction starts with $\Delta_\mN^{it}=\Delta_\mM^{it}D(t)$ and $J_\mN=J_\mM D(i/2)$. In particular, the commutant can be recovered as $\mN'=\bigvee_t\operatorname{Ad}(\Delta_\mM^{it}D(t))(\mM')$. The boundary and cocycle conditions ensure that these data define the required standard algebra~\cite[Section~II.3]{revolutionizing}.

\subsection{Subsystem flow beyond Borchers' characteristic function}\label{subsec:beyond-Borchers}
We start with the Tomita operators. Let $S_\mathcal{N}$ be the Tomita operator of $(\mathcal{N},\ket{\Omega_\mathcal{N}})$ on $\mathcal{H}_\mathcal{N}$ and $S_\mathcal{M}$ that of $(\mathcal{M},\ket{\Omega})$ on $\mathcal{H}_\mathcal{M}$. For $n\in\mathcal{N}$, $S_\mathcal{M}Vn\ket{\Omega_\mathcal{N}}=n^*\ket{\Omega}=VS_\mathcal{N}n\ket{\Omega_\mathcal{N}}$. Since $\mathcal{N}\ket{\Omega_\mathcal{N}}$ is a core of $S_\mathcal{N}$ and $S_\mathcal{M}$ is closed, $S_\mathcal{M}V\supseteq VS_\mathcal{N}$. (The operator $VS_\mathcal{N}V^\dagger$ vanishes on $(1-e_\mathcal{N})\mathcal{H}_\mathcal{M}$, where $S_\mathcal{M}$ does not, so $S_\mathcal{M}\not\supseteq VS_\mathcal{N}V^\dagger$ unless $V=1$.) Hence $V\operatorname{Dom}\Delta_\mathcal{N}^{1/2}\subseteq\operatorname{Dom}\Delta_\mathcal{M}^{1/2}$ and
\begin{equation}
    \|\Delta_\mathcal{M}^{1/2}V\xi\|=\|\Delta_\mathcal{N}^{1/2}\xi\|,\qquad \xi\in\operatorname{Dom}\Delta_\mathcal{N}^{1/2},
\end{equation}
that is, $V^\dagger\Delta_\mathcal{M}V\le\Delta_\mathcal{N}$ in the sense of quadratic forms on $\mathcal{H}_\mathcal{N}$. The two forms agree on $\operatorname{Dom}\Delta_\mathcal{N}^{1/2}$, and the form of $V^\dagger\Delta_\mathcal{M}V$ may have a larger domain. The graph of $VS_\mathcal{N}$ (the image of the graph of $S_\mathcal{N}$ under the isometry $V\oplus V$) is contained in the graph of $S_\mathcal{M}$. Comparing the orthogonal projections onto these graphs and compressing to the first summand, as in Borchers~\cite{revolutionizing}, gives
\begin{equation}
    V(1+\Delta_\mathcal{N})^{-1}V^\dagger\le(1+\Delta_\mathcal{M})^{-1}\quad\text{on }\mathcal{H}_\mathcal{M},
\end{equation}
which for $V=1$ is Borchers' $\Delta_\mathcal{M}\le\Delta_\mathcal{N}$. In particular, for $\xi\in\operatorname{Dom}\Delta_\mathcal{N}^{1/2}$, $\|D(\tfrac{i}{2})V\Delta_\mathcal{N}^{1/2}\xi\|=\|\Delta_\mathcal{M}^{1/2}V\xi\|=\|\Delta_\mathcal{N}^{1/2}\xi\|$, so $D(\tfrac{i}{2})^\dagger D(\tfrac{i}{2})=e_\mathcal{N}\le1$.

For $0\le\alpha\le\frac{1}{2}$ the Heinz--Kato inequality~\cite{Kato1952} applied to the norm identity above replaces operator monotonicity of $x\mapsto x^\alpha$: $\|\Delta_\mathcal{M}^\alpha V\xi\|\le\|\Delta_\mathcal{N}^\alpha\xi\|$ for $\xi\in\operatorname{Dom}\Delta_\mathcal{N}^\alpha$.
Hence we have the bounded results:
\begin{align}
    \begin{split}
        \|D(i\alpha)^\dagger D(i\alpha)\| = \|V\Delta^{-\alpha}_\mathcal{N} V^\dagger \Delta_\mathcal{M}^{2\alpha} V\Delta^{-\alpha}_\mathcal{N} V^\dagger\| \leq 1
    \end{split}
\end{align}
where $0\leq \alpha \leq \frac{1}{2}$. Adding unitary components, we have:
\begin{align}
    \begin{split}
        \|D(t + i\alpha)^\dagger D(t + i\alpha)\| &= \|V\Delta_\mathcal{N}^{-it -\alpha}V^\dagger\Delta_\mathcal{M}^{2\alpha}V\Delta_\mathcal{N}^{it - \alpha}V^\dagger\|
        \\&
        \leq \|V\Delta_\mathcal{N}^{-it}V^\dagger\|\cdot\|V\Delta_\mathcal{N}^{-\alpha}V^\dagger\Delta_\mathcal{M}^{2\alpha}V\Delta_\mathcal{N}^{-\alpha}V^\dagger\|\cdot\|V\Delta_\mathcal{N}^{it}V^\dagger\| \leq 1
    \end{split}
\end{align}
where $0\leq \alpha \leq \frac{1}{2}$ and $t\in\mathbb{R}$. 

Finally, to see that the extension $D(z)$ to the strip $0\leq \operatorname{Im}(z) \leq \frac{1}{2}$  is analytic in the interior of the strip, we consider two strongly dense subsets of vectors in $\mathcal{H}_\mathcal{M}$ \cite{Takesaki1970}:
\begin{equation}
    \{\sigma^f_\mathcal{M}(a)\ket{\Omega}:= \int_\mathbb{R} dt f(t)\Delta_\mathcal{M}^{it}a\ket{\Omega}: a\in\mathcal{M}\}
\end{equation}
\begin{equation}
    \{V\sigma_\mathcal{N}^f(a)\ket{\Omega} := V\int_\mathbb{R} dt f(t)\Delta_\mathcal{N}^{it}a\ket{\Omega}:a\in\mathcal{N}\}\oplus \operatorname{ker}(e_\mathcal{N})
\end{equation}
where $f(t)\in L^1(\mathbb{R})$ and has an entire analytic extension. Both sets of vectors are analytic vectors \cite{Takesaki1970} in the sense that the functions:
\begin{equation}
    t\mapsto \Delta_\mathcal{M}^{it}\sigma^f_\mathcal{M}(a)\ket{\Omega}\text{ , }t\mapsto\Delta_\mathcal{N}^{it}\sigma_\mathcal{N}^f(b)\ket{\Omega}
\end{equation}
can be extended to entire functions on $\mathbb{C}$. Then the following function of $z$ can be analytically continued to domains where the function is bounded:
\begin{align}
    \begin{split}
        \langle \sigma_\mathcal{M}^f(a)\Omega, D(z)V\sigma_\mathcal{N}^f(b)\Omega\rangle &= \langle\Delta_\mathcal{M}^{i\bar{z}}\sigma_\mathcal{M}^f(a)\Omega, V \Delta_\mathcal{N}^{iz}\sigma_\mathcal{N}^f(b)\Omega\rangle
    \end{split}
\end{align}
Combined with the fact that $D(z)$ is bounded within the strip, we have a proof that the extension is analytic.

\section{Additional quantum error-correction examples}\label{app:extra-qec-examples}

\subsection{Knill--Laflamme and the complementary channel}
\label{app:finite-KL}

\paragraph{The finite-dimensional dictionary:}
Let $\mathcal F_*(\rho)=\sum_kE_k\rho E_k^\dagger$ be a channel on $B(\mathcal H_S)$ with Stinespring isometry $W\ket{\xi}=\sum_kE_k\ket{\xi}\otimes\ket{k}_E$, so that $\mathcal F_*=\Tr_E\circ\Ad W$. The complementary channel is $\mathcal F^c_*=\Tr_S\circ\Ad W$, explicitly
\begin{equation}
 \mathcal F^c_*(\rho)=\sum_{k,l}\Tr\!\bigl(E_l^\dagger E_k\,\rho\bigr)\,\ket{k}\!\bra{l}.
 \label{eq:complementary-channel}
\end{equation}
A code subspace $P\mathcal H_S$ is exactly correctable for $\mathcal F_*$ if and only if~\cite{KnillLaflamme1997}
\begin{equation}
 P\,E_l^\dagger E_k\,P=\lambda_{lk}\,P\qquad\hbox{for all }k,l,
 \label{eq:KL}
\end{equation}
for a positive matrix $\lambda$ with $\sum_k\lambda_{kk}=1$. For a state $\rho$ supported on $P$ the matrix elements of \eqref{eq:complementary-channel} are $\mathcal F^c_*(\rho)_{kl}=\Tr(PE_l^\dagger E_kP\,\rho)$. Hence \eqref{eq:KL} holds if and only if
\begin{equation}
 \mathcal F^c_*(\rho)=\Lambda\quad\hbox{for every state $\rho$ supported on }P,\qquad\Lambda_{kl}=\lambda_{lk},
 \label{eq:forgetful}
\end{equation}
because the code states span all operators on $P\mathcal H_S$. Condition \eqref{eq:forgetful} says that the complementary channel is constant, or forgetful, on the code: the environment ends in the same state whatever logical state was encoded. Conditions \eqref{eq:KL} and \eqref{eq:forgetful} are therefore one statement read from two sides. From the system's side, no pair of errors can be told apart on the code except through the code-independent Gram matrix $\lambda$. From the environment's side, the environment, whose state records which error occurred, carries no information about the logical input. The content of the Knill--Laflamme theorem is the further implication that \eqref{eq:forgetful} suffices for recovery by a channel acting on the system alone. Its approximate version is the information--disturbance trade-off~\cite{SchumacherWestmoreland2002,KretschmannSchlingemannWerner2008}, and for a logical \emph{algebra} rather than a subspace the scalar $\lambda_{lk}$ is replaced by an element of the commutant of the logical algebra on the code~\cite{BenyKempfKribs2007}.

\paragraph{The finite-dimensional Petz map:}
With $\mathcal F_*$ now the encoded channel \eqref{eq:two-cut-noise}, and $\tau_B$, $Q$ and $\mM_{\rm ret}$ as in \eqref{eq:retained-reference-support} of Section~\ref{subsec:finite-QEC}, the Heisenberg Petz map is
\begin{equation}
 \mathcal D_\sigma(b)
 =\tau_{B}^{-1/2}
   \mathcal F_*\bigl(\sigma_L^{1/2}b\,\sigma_L^{1/2}\bigr)
   \tau_{B}^{-1/2},\qquad b\in\cL,
 \label{eq:finite-Heisenberg-Petz}
\end{equation}
where inverse powers are taken on $Q$. It is unital with respect to the unit $Q$ of $\mM_{\rm ret}$, and if \eqref{eq:Schrodinger-exact-QEC} holds, then $\Phi\mathcal D_\sigma=\operatorname{id}$. For a general exactly correctable subsystem, the transpose map associated with the maximally mixed state on the code similarly corrects its logical algebra~\cite{MandayamNg2012}\footnote{No unitality assumption on the original Schr\"odinger noise channel is required. A unitary decoding may reveal the output logical subsystem~\cite{KribsSpekkens2006}, but a unitary implementation is not needed for the argument below.}. With $\rho_{\rm enc}=V_{\rm enc}\sigma_LV_{\rm enc}^\dagger$ and $\tau_B=\Tr_R\rho_{\rm enc}$, $\mathcal D_\sigma(b)\otimes1_R=\mathsf E_{\rho_{\rm enc}}(V_{\rm enc}bV_{\rm enc}^\dagger)$, where $\mathsf E_\rho(a)=\tau_B^{-1/2}\Tr_R(\rho^{1/2}a\rho^{1/2})\tau_B^{-1/2}\otimes1_R$ is the matrix form of \eqref{eq:background-Petz-dual}. The encoding isometry $V_{\rm enc}$ is not the canonical GNS isometry $V$ of Section~\ref{subsec:subsystem-flow-background}: $\rho_{\rm enc}$ is not faithful on $B(\mH_B\otimes\mH_R)$, so $\Gamma=J_\mN V^\dagger J_\mM$ is not defined for it, and the construction lives on the occupied algebra $\mM_{\rm ret}$.

\section{Properties of inclusion parent Hamiltonians}
\label{app:inclusion-Hamiltonian-details}

This appendix supplies the reflection, commutant-dual, restriction and spectral arguments used in Section~\ref{sec:probing-outside-M}, together with the full ground-space and recurrence proofs. The GNS spaces and the full-support convention are those of that section.

\subsection{Reflection and modular conjugations}
\label{app:Q-reflection}
Since $X_\alpha=D_\alpha V$ is bounded for $0\leq\alpha\leq\tfrac12$ by \eqref{eq:W-strip-bound}, it suffices to check $X_{1/2-\alpha}=J_\mM X_\alpha J_\mN$ on the dense set $\Delta_\mN^{1/2-\alpha}n\ket{\Omega_\mN}$, $n\in\mN$. We have $J_\mN\Delta_\mN^{1/2-\alpha}n\ket{\Omega_\mN}=\Delta_\mN^{\alpha}n^*\ket{\Omega_\mN}$ and
$J_\mM X_\alpha\Delta_\mN^{\alpha}n^*\ket{\Omega_\mN}=J_\mM\Delta_\mM^{\alpha}n^*\ket{\Omega}=\Delta_\mM^{-\alpha}\Delta_\mM^{1/2}n\ket{\Omega}=X_{1/2-\alpha}\Delta_\mN^{1/2-\alpha}n\ket{\Omega_\mN}$. Multiplying by $V^\dagger$ on the right gives
\begin{equation}
 J_\mM D_\alpha VJ_\mN V^\dagger=D_{\frac12-\alpha}.
 \label{eq:W-reflection}
\end{equation}

Consequently,
\begin{equation}
    J_{\mN}\rho_{\alpha,\mN\subset\mM}J_{\mN}
    =\rho_{\frac12-\alpha,\mN\subset\mM},
    \label{eq:rho-reflection}
\end{equation}
and, because $\sin(2\pi(1/2-\alpha))=\sin(2\pi\alpha)$,
\begin{equation}
    J_{\mN}Q_{\alpha,\mN\subset\mM}J_{\mN}
    =Q_{\frac12-\alpha,\mN\subset\mM}.
    \label{eq:H-reflection}
\end{equation}
At the midpoint,
\begin{equation}
    J_{\mN}Q_{\frac14,\mN\subset\mM}J_{\mN}
    =Q_{\frac14,\mN\subset\mM}.
    \label{eq:H-quarter-real}
\end{equation}
This is an antiunitary time-reflection symmetry of the midpoint flow,
not by itself a statement that the flow is a geometric null translation.

\paragraph{Action of $J_{\mM}$:}
The modular conjugation of the ambient algebra acts naturally on
$\mH_{\mM}$, whereas $\rho_\alpha$ and $Q_\alpha$ act on
$\mH_{\mN}$. For a general compressed inclusion, the well-defined
ambient identity following from \eqref{eq:W-reflection} is
\begin{equation}
    J_{\mM}(D_\alpha D_\alpha^\dagger)J_{\mM}
    =D_{\frac12-\alpha}D_{\frac12-\alpha}^*.
    \label{eq:Q-JM-ambient}
\end{equation}
If $\ket{\Omega}$ is cyclic for both algebras, so that
$\mH_{\mN}=\mH_{\mM}$, put
$\widetilde\mN=J_{\mM}\mN J_{\mM}\subset\mM'$. Then
\begin{equation}
    J_{\mM}Q_{\alpha,\mN\subset\mM}J_{\mM}
    =Q_{\alpha,\widetilde\mN\subset\mM'}.
    \label{eq:Q-JM-mirror}
\end{equation}
This identifies the parent Hamiltonian with that of the reflected
inclusion, rather than asserting $J_{\mM}Q_\alpha J_{\mM}=Q_\alpha$.
With the convention $\Gamma=J_\mN J_\mM$ of \eqref{eq:interpolation-Gamma}, the common-cyclic identity is $J_\mM Q_\alpha J_\mM=\Gamma^\dagger Q_{1/2-\alpha}\Gamma$.

\subsection{\texorpdfstring{Commutant-dual inclusion and the two orderings}{Commutant-dual inclusion and the two orderings}}
\label{app:Q-dual-inclusion}

Let $\ket{\Omega}$ be cyclic and separating for both $\mN$ and $\mM$, so that $V=1$ and $D(z)=\Delta_\mM^{-iz}\Delta_\mN^{iz}$. Taking commutants reverses the inclusion, $\mM'\subset\mN'$, and $\Delta_{\mM'}=\Delta_\mM^{-1}$, $\Delta_{\mN'}=\Delta_\mN^{-1}$. The characteristic function of the dual inclusion is $D^\vee(z):=D_{\mM'\subset\mN'}(z)$, which is formally $\Delta_\mN^{iz}\Delta_\mM^{-iz}$, so that $D^\vee(t)=D(-t)^{-1}$ on the real axis. The functions $z\mapsto D^\vee(z)$ and $z\mapsto D(-\bar z)^\dagger$ are bounded and holomorphic on $0<\operatorname{Im}z<\tfrac12$ and weakly continuous up to $\operatorname{Im}z=0$, where they agree. By the Schwarz reflection principle applied to matrix elements,
\begin{equation}
 D^\vee(z)=D(-\bar z)^\dagger,\qquad D^\vee(i\alpha)=X_\alpha^\dagger,\qquad 0<\alpha<\tfrac12 .
 \label{eq:D-dual-reflection}
\end{equation}
The two orderings of the product, $D^\vee(z)D(z)$ for $\mN\subset\mM$ and $D(z)D^\vee(z)$ for $\mM'\subset\mN'$, are unitary on the real axis, where their logarithms are defined only up to winding ambiguities. At $z=i\alpha$ they are the positive contractions
\begin{equation}
 D^\vee(i\alpha)D(i\alpha)=X_\alpha^\dagger X_\alpha=\rho_{\alpha,\mN\subset\mM},\qquad
 D(i\alpha)D^\vee(i\alpha)=X_\alpha X_\alpha^\dagger=\rho_{\alpha,\mM'\subset\mN'},
 \label{eq:rho-dual-orderings}
\end{equation}
whose logarithms have a unique real spectral branch and give $Q_{\alpha,\mN\subset\mM}$ and $Q_{\alpha,\mM'\subset\mN'}$ through \eqref{eq:H-alpha-general}. If $X_\alpha=U_\alpha|X_\alpha|$ is the polar decomposition, then $X_\alpha X_\alpha^\dagger=U_\alpha(X_\alpha^\dagger X_\alpha)U_\alpha^\dagger$ on the supports, and hence
\begin{equation}
 Q_{\alpha,\mM'\subset\mN'}=U_\alpha\,Q_{\alpha,\mN\subset\mM}\,U_\alpha^\dagger .
 \label{eq:Q-dual-polar}
\end{equation}
The inclusion and its commutant dual therefore have identical inclusion-energy spectra on their supports, including their zero-energy eigenspaces.

\subsection{Further restriction and the small-exponent limit}
\label{app:Q-restriction}

Consider $\mN_2\subset\mN_1\subset\mM$ with the same reference vector.
Let $V_1:\mH_{\mN_1}\to\mH_\mM$ and
$V_{12}:\mH_{\mN_2}\to\mH_{\mN_1}$ be the canonical inclusions.
We write $D_\alpha^{\mM\leftarrow\mN}:=D(i\alpha)$ for the continued subsystem flow of the inclusion $\mN\subset\mM$, an operator on $\mH_\mM$, so that $X_\alpha=D_\alpha^{\mM\leftarrow\mN}V$, and similarly for the other inclusions. The canonical isometry of $\mN_2$ is $V_1V_{12}$ and $V_1^\dagger V_1=1$, so the composition law is
\begin{equation}
 D_\alpha^{\mM\leftarrow\mN_2}
 =D_\alpha^{\mM\leftarrow\mN_1}V_1
      D_\alpha^{\mN_1\leftarrow\mN_2}V_1^\dagger.
 \label{eq:Q-inclusion-composition}
\end{equation}
With $T=D_\alpha^{\mN_1\leftarrow\mN_2}V_{12}$, a contraction,
\begin{equation}
 \rho_{\alpha,\mN_2\subset\mM}
 =T^\dagger\rho_{\alpha,\mN_1\subset\mM}T
 \leq T^\dagger T\leq1.
 \label{eq:Q-inclusion-restriction}
\end{equation}

In particular, when the characteristic operators are compact, their
ordered singular values obey
\begin{equation}
    s_k(D_\alpha^{\mM\leftarrow\mN_2})
    \leq s_k(D_\alpha^{\mM\leftarrow\mN_1}).
    \label{eq:Q-spectral-monotonicity}
\end{equation}
Thus the ordered finite inclusion-energy levels cannot decrease when the retained algebra shrinks. The Schatten norms of the characteristic contractions $D_\alpha$ and traces of their positive powers $\rho_\alpha^p$, $p>0$, are nonincreasing when finite. This is not a claim that traces of $Q_\alpha$ decrease. This is a \emph{spectral}
monotonicity statement, not in general the Loewner-order inequality
$Q_{\alpha,\mN_2\subset\mM}\geq
Q_{\alpha,\mN_1\subset\mM}$: the two operators act on different
GNS subspaces, and even with common cyclicity the intervening
contraction need not preserve operator order.

\paragraph{The limit $\alpha\to0$:}
If both modular Hamiltonians are bounded, differentiation is valid in operator norm. Put $G=V^\dagger K_\mM V-K_\mN$. Then
\begin{align}
 D_\alpha V&=V-\alpha(K_\mM V-VK_\mN)+O(\alpha^2),\nonumber\\
 \rho_\alpha&=1-2\alpha G+O(\alpha^2),\nonumber\\
 -\log\rho_\alpha&=2\alpha G+O(\alpha^2).
\end{align}
Since $2\sin(2\pi\alpha)=4\pi\alpha+O(\alpha^3)$, this proves
\begin{equation}
 \lim_{\alpha\downarrow0}Q_{\alpha,\mN\subset\mM}
 =\frac{1}{2\pi}(V^\dagger K_\mM V-K_\mN)
 \label{eq:Q-alpha-zero}
\end{equation}
in operator norm in the bounded case, including finite-dimensional examples. Both sides act on $\mH_\mN$: $Q_\alpha$ is defined there, and $V^\dagger K_\mM V$ is the compression of the ambient modular Hamiltonian to the subsystem GNS space. For common cyclicity the limit is $(K_\mM-K_\mN)/2\pi$ on $\mH_\mM$. For unbounded modular Hamiltonians, the same coefficient is a formal quadratic-form expansion. A form-limit statement additionally requires a specified common domain and control allowing the logarithm to be differentiated. The Taylor expansion alone does not prove such convergence. We make no unconditional form- or norm-convergence assertion for QFT modular operators.

\subsection{Proof of the ground-space theorem}
\label{app:Q-ground-proof}

We prove Theorem~\ref{thm:Q-kernel-core}.
\begin{proof}

First, the contraction $D(z)$ is analytic in the strip, and
$D(0)=e_\mN$. For $\ket{\xi}\in\mH_\mN$, equality at one interior point,
$\|D_\alpha V\ket{\xi}\|=\|\ket{\xi}\|$, implies by the vector-valued
maximum-modulus principle that $D(z)V\ket{\xi}=V\ket{\xi}$ throughout the strip.
For $\ket{\xi}\ne0$, test $D(z)V\ket{\xi}/\|\ket{\xi}\|$ against the unit vector $D_\alpha V\ket{\xi}/\|\ket{\xi}\|$. The resulting scalar analytic contraction attains the value $1$ at $i\alpha$, so it is constant. Equality in Cauchy--Schwarz makes the vector constant as well, and continuity at $z=0$ identifies it with $V\ket{\xi}$.

Thus
\begin{equation}
 \ker(1-\rho_\alpha)
 =\{\ket{\xi}\in\mH_\mN:\ 
     \Delta_\mM^{it}V\ket{\xi}=V\Delta_\mN^{it}\ket{\xi}
       \text{ for all }t\in\mathbb R\},
 \label{eq:Q-kernel-modular-intertwining}
\end{equation}
independently of $\alpha$. The inclusion ``$\subseteq$'' is the case $z=t$ of $D(z)V\ket{\xi}=V\ket{\xi}$ proved above. Conversely, let $\ket{\xi}$ satisfy the intertwining relation for all $t$. Then $V$ maps the cyclic subspace of $\Delta_\mN^{it}$ generated by $\ket{\xi}$ unitarily onto the cyclic subspace of $\Delta_\mM^{it}$ generated by $V\ket{\xi}$ and intertwines the two groups, so $f(\Delta_\mM)V\ket{\xi}=Vf(\Delta_\mN)\ket{\xi}$ for every bounded Borel function $f$. For $\ket{\xi_n}:=\mathbf 1_{[1/n,n]}(\Delta_\mN)\ket{\xi}\in\operatorname{Ran}\Delta_\mN^{\alpha}$ this gives $X_\alpha\ket{\xi_n}=\Delta_\mM^{\alpha}V\Delta_\mN^{-\alpha}\ket{\xi_n}=V\ket{\xi_n}$, hence $\langle\xi_n,\rho_\alpha\xi_n\rangle=\|X_\alpha\xi_n\|^2=\|\xi_n\|^2$ and, because $0\leq\rho_\alpha\leq1$, $\rho_\alpha\xi_n=\xi_n$. Letting $n\to\infty$, and using that $\ker(1-\rho_\alpha)$ is closed, we get $\rho_\alpha\ket{\xi}=\ket{\xi}$, so \eqref{eq:Q-kernel-modular-intertwining} is an equality. Put $K:=\ker(1-\rho_\alpha)$, the space of zero modes. It reduces $\Delta_\mN$: for $\ket{\xi}\in K$ and $s,t\in\mathbb R$,
\begin{equation*}
 \Delta_\mM^{it}V\Delta_\mN^{is}\ket{\xi}=\Delta_\mM^{i(t+s)}V\ket{\xi}=V\Delta_\mN^{i(t+s)}\ket{\xi}=V\Delta_\mN^{it}(\Delta_\mN^{is}\ket{\xi}),
\end{equation*}
so $\Delta_\mN^{is}\ket{\xi}$ lies in the intertwining set, which is $K$, for every $s\in\mathbb R$. In particular, if $a\in\mN$ and $a\ket{\Omega}$ is a zero mode, the
intertwining identity and separatingness of $\ket{\Omega}$ for $\mM$
show that $a$ belongs to the persistent core. Conversely, let $b\in\mN_{\rm core}$. The core is $\sigma^{\omega;\mM}$-invariant, so Takesaki's theorem gives an $\omega$-preserving normal expectation $\mM\to\mN_{\rm core}$. Its restriction to $\mN$ shows that $\mN_{\rm core}$ is also $\sigma^{\omega;\mN}$-invariant, and both modular groups restrict to the modular group of $\mN_{\rm core}$. Hence $\Delta_\mM^{it}b\ket{\Omega}=\sigma_t^{\omega;\mM}(b)\ket{\Omega}=\sigma_t^{\omega;\mN}(b)\ket{\Omega}=V\Delta_\mN^{it}b\ket{\Omega}$ for all $t$, that is, $b\ket{\Omega}\in K$. Thus $\{a\in\mN:a\ket{\Omega}\in K\}=\mN_{\rm core}$.

To obtain the closure of the bounded-observable zero modes, use the vectors $\Delta_\mN^{1/4}a\ket{\Omega}$ with $a\in\mN$, which span a dense subspace of $\mH_\mN$. Indeed, $\mN\ket{\Omega}$ is a core for $\Delta_\mN^{1/2}$, hence for $\Delta_\mN^{1/4}$, which is injective with dense range.
The characteristic map obeys
$D_{1/4}V\Delta_\mN^{1/4}a\ket{\Omega}=\Delta_\mM^{1/4}a\ket{\Omega}$ for $a\in\mN$.
Let $\mathsf E_\omega:\mM\to\mN$ be the
Petz map $\mathsf E_\omega(m)=\Gamma m\Gamma^\dagger$ of \eqref{eq:QEC-Petz-dual}, and put
$\Phi=\mathsf E_\omega|_\mN$. It is a normal unital completely
positive map preserving $\omega|_\mN$. Its symmetric $L^2$ implementation is \eqref{eq:Q-AC-L2}.
Indeed, for $a,b\in\mN$,
\begin{align*}
\langle\Delta_\mN^{1/4}b\Omega,\rho_{1/4}\Delta_\mN^{1/4}a\Omega\rangle
&=\langle\Delta_\mM^{1/4}b\Omega,\Delta_\mM^{1/4}a\Omega\rangle=\langle b\Omega,J_\mM a^*\Omega\rangle
=\langle b\Omega_\mN,J_\mN\mathsf E_\omega(a^*)\Omega_\mN\rangle\\
&=\langle b\Omega_\mN,\Delta_\mN^{1/2}\Phi(a)\Omega_\mN\rangle=\langle\Delta_\mN^{1/4}b\Omega,\Delta_\mN^{1/4}\Phi(a)\Omega\rangle,
\end{align*}
using $\rho_{1/4}=X_{1/4}^\dagger X_{1/4}$ with $X_{1/4}\Delta_\mN^{1/4}b\Omega=\Delta_\mM^{1/4}b\Omega$, then $\mathsf E_\omega(c)\Omega_\mN=J_\mN V^\dagger J_\mM c\Omega$ from \eqref{eq:background-Petz-dual} and $\mathsf E_\omega(a^*)=\Phi(a)^*$, and finally $J_\mN c^*\Omega_\mN=\Delta_\mN^{1/2}c\Omega_\mN$. By the density of $\Delta_\mN^{1/4}\mN\ket{\Omega}$ this proves \eqref{eq:Q-AC-L2}. Now let $b\in\mN$. By \eqref{eq:Q-AC-L2}, $\Delta_\mN^{1/4}b\ket{\Omega}\in K\iff\Delta_\mN^{1/4}\Phi(b)\ket{\Omega}=\Delta_\mN^{1/4}b\ket{\Omega}\iff\Phi(b)=b$, because $\Delta_\mN^{1/4}$ is injective and $\ket{\Omega}$ is separating for $\mN$. Since $K$ reduces $\Delta_\mN$ and $b\ket{\Omega}\in\operatorname{Dom}\Delta_\mN^{1/2}$, also $\Delta_\mN^{1/4}b\ket{\Omega}\in K\iff b\ket{\Omega}\in K\iff b\in\mN_{\rm core}$ by the zero-mode statement above. Hence
\begin{equation}
 \operatorname{Fix}(\Phi)
 :=\{a\in\mN:\Phi(a)=a\}
 =\mN_{\rm core}.
 \label{eq:Q-AC-fixed-point}
\end{equation}

Let $P_0$ denote the spectral projection of $\rho_{1/4}$
onto its eigenvalue $1$. Since $0\leq\rho_{1/4}\leq1$,
\begin{equation}
 P_0=\operatorname*{s-lim}_{n\to\infty}\rho_{1/4}^{\,n}.
\end{equation}
For $a\in\mN$ we have
$\rho_{1/4}^{\,n}\Delta_\mN^{1/4}a\ket{\Omega}=\Delta_\mN^{1/4}\Phi^n(a)\ket{\Omega}$.
The iterates satisfy $\|\Phi^n(a)\|\leq\|a\|$. Hence a subnet
converges ultraweakly to some $b\in\mN$. The map $a\mapsto\Delta_\mN^{1/4}a\ket{\Omega}$
is bounded, $\|\Delta_\mN^{1/4}a\ket{\Omega}\|\leq\|a\|$, and weakly continuous on norm-bounded sets, so
$P_0\Delta_\mN^{1/4}a\ket{\Omega}=\Delta_\mN^{1/4}b\ket{\Omega}$. Applying $\rho_{1/4}$ to this identity
and using \eqref{eq:Q-AC-L2} gives $\Phi(b)=b$, so $b\in\mN_{\rm core}$ by \eqref{eq:Q-AC-fixed-point}. Thus
\begin{equation}
 P_0\Delta_\mN^{1/4}\mN\ket{\Omega}\subseteq
 \Delta_\mN^{1/4}\mN_{\rm core}\ket{\Omega}.
\end{equation}
Conversely every fixed point of $\Phi$ gives an eigenvalue-one
vector by \eqref{eq:Q-AC-L2}. The density of $\Delta_\mN^{1/4}\mN\ket{\Omega}$
then implies
\begin{equation}
 \operatorname{Ran}P_0
 =\overline{\Delta_\mN^{1/4}\mN_{\rm core}\ket{\Omega}}.
\end{equation}
The core is globally invariant under both modular groups, whose
restrictions to the core agree. Its $L^2$ closure therefore equals
its GNS closure:
$\overline{\Delta_\mN^{1/4}\mN_{\rm core}\ket{\Omega}}
 =\overline{\mN_{\rm core}\ket{\Omega}}$. Indeed, the core has an $\omega$-preserving expectation from $\mM$, whose restriction to $\mN$ is also an expectation. Its GNS sector therefore reduces $\Delta_\mN$, and the restricted modular operator is the intrinsic operator of the core. The dense symmetric embedding of the core has exactly that sector as its closure. The expectation $\mN\to\mN_{\rm core}$ commutes with the Tomita operator of $\mN$. Moreover $\mN_{\rm core}\ket{\Omega}$ is a core for $\Delta_{\mN_{\rm core}}^{1/2}$, hence for $\Delta_{\mN_{\rm core}}^{1/4}$, which is injective with dense range, so $\Delta_\mN^{1/4}\mN_{\rm core}\ket{\Omega}$ is dense in the sector.
This proves the theorem for $\alpha=1/4$. The first step
extends it to all interior $\alpha$.
\end{proof}

\subsection{Positive energy and invariant algebras}
\label{app:Q-positive-energy}

\paragraph{Positive-energy invariant algebras:}
We will also use the following elementary consequence of Tomita theory.
Let $\mathcal Y\subset B(\mH_{\mN})$ be a von Neumann subalgebra satisfying
\begin{equation}
    \gamma_t^\alpha(\mathcal Y)
    =
    \mathcal Y,
    \qquad
    t\in\mathbb R,
    \label{eq:Q-invariant-Halpha}
\end{equation}
and suppose $\ket{\Omega}$ is separating for $\mathcal Y$. On
\begin{equation}
    \mathcal K_{\mathcal Y}
    :=
    \overline{\mathcal Y\ket{\Omega}}
\end{equation}
the vector $\ket{\Omega}$ is cyclic and separating for the restricted algebra.
Since $U_t^{(\alpha)}\ket{\Omega}=\ket{\Omega}$ and
$U_t^{(\alpha)}$ preserves $\mathcal Y$, it commutes on
$\mathcal K_{\mathcal Y}$ with the Tomita operator and hence with the modular
conjugation $J_{\mathcal Y}$. Antiunitarity then gives
\begin{equation}
    J_{\mathcal Y}
    Q_{\alpha,\mN\subset\mM}\big|_{\mathcal K_{\mathcal Y}}
    J_{\mathcal Y}
    =
    -
    Q_{\alpha,\mN\subset\mM}\big|_{\mathcal K_{\mathcal Y}}.
\end{equation}
Since the restriction of $Q_{\alpha,\mN\subset\mM}$ is positive, it follows that
\begin{equation}
    Q_{\alpha,\mN\subset\mM}\big|_{\mathcal K_{\mathcal Y}}=0.
    \label{eq:positive-H-invariant-Q}
\end{equation}
Thus, a positive-energy flow fixing the reference vector can preserve a
standard autonomous subalgebra only by acting trivially on the cyclic sector generated by
that subalgebra.

\subsection{One-sided histories and recurrence}
\label{app:Q-recurrence}

To distinguish the two time orientations, define the future and past
inclusion parent Hamiltonian histories by
\begin{align}
    \mathcal A_\alpha^\pm
    &:=\bigvee_{\pm t>0}\gamma_t^\alpha(\mN).
    \label{eq:Halpha-one-sided-histories-cores}
\end{align}
We call the flow \emph{future generating} or \emph{past generating} on
$\mH_{\mN}$ when $\mathcal A_\alpha^+=B(\mH_{\mN})$ or
$\mathcal A_\alpha^-=B(\mH_{\mN})$, respectively.

We now prove Proposition~\ref{prop:Halpha-pure-point-one-sided-generation}.
\begin{proof}
Choose an orthonormal eigenbasis $\{\psi_j\}$ of $Q_\alpha$, with
eigenvalues $E_j$. Simultaneous approximation of finitely many phases,
followed by a diagonal choice, gives times $t_n\to+\infty$ such that
$e^{it_nE_j}\to1$ for every $j$. Consequently
\begin{equation}
    U_{t_n}^{(\alpha)}\longrightarrow1,
    \qquad U_{-t_n}^{(\alpha)}\longrightarrow1
    \quad\text{strongly}.
    \label{eq:Halpha-pure-point-recurrence}
\end{equation}
For $a\in\mN$, $r\in\mathbb R$, and
$\varepsilon\in\{+1,-1\}$, it follows that
\begin{equation}
    \gamma_{\varepsilon t_n+r}^\alpha(a)
    \longrightarrow\gamma_r^\alpha(a)
    \quad\text{strongly}.
    \label{eq:Halpha-one-sided-recurrence-approximation}
\end{equation}
For large $n$, the times $\varepsilon t_n+r$ have sign $\varepsilon$.
The approximating operators therefore belong to
$\mathcal A_\alpha^+$ if $\varepsilon=+1$, and to
$\mathcal A_\alpha^-$ if $\varepsilon=-1$. Since von Neumann algebras
are strongly closed, each half-history contains
$\gamma_r^\alpha(\mN)$ for every real $r$. Hence
$\mathcal A_\alpha^+=\mathcal A_\alpha^-=\mathcal A_\alpha$, and
Proposition~\ref{prop:Halpha-full-generation} proves
\eqref{eq:Halpha-pure-point-one-sided-generation}.

If $\rho_\alpha$ is trace class with full support, it is a compact
positive operator with an orthonormal eigenbasis and positive
eigenvalues $\lambda_j$. The same basis diagonalizes $Q_\alpha$, with
$E_j=-[2\sin(2\pi\alpha)]^{-1}\log\lambda_j$, so the preceding argument
applies. Full support is essential here to define the unitary group
on all of $\mH_{\mN}$.
\end{proof}

\subsection{Three-lines majorization and alternative normalizations}
\label{subsec:Q-three-lines}

To compare attenuation across the analytic strip, it is convenient to work with the unrescaled logarithm $-\tfrac12\log\rho_\alpha=\sin(2\pi\alpha)\,Q_\alpha$, which agrees with $Q_\alpha$ at the midpoint. We denote it $Q_\alpha^{\rm gap}$ and write $F_\alpha:=-\log\rho_\alpha=2Q_\alpha^{\rm gap}$.

The reflection-symmetric midpoint has a second distinction besides
$J_\mN Q_{1/4}J_\mN=Q_{1/4}$. It maximizes the spectral
attenuation of the \emph{unnormalized} inclusion operator, as follows.

\begin{theorem}[Three-lines weak log-majorization]
\label{thm:Q-three-lines}
For each integer $1\leq k\leq\dim\mH_{\mN}$, define
\begin{equation}
 f_k(\alpha)
 :=\log\left\|\bigwedge^kD_\alpha\right\|,
 \qquad 0\leq\alpha\leq\tfrac12.
 \label{eq:Q-fk-definition}
\end{equation}
where $\bigwedge^k$ is the exterior-power map\footnote{The theorem uses exterior powers. For a bounded operator $A$ on a Hilbert space $\mH$ and an integer $k\geq1$, the exterior power $\bigwedge^kA$ is the operator on the antisymmetric subspace $\bigwedge^k\mH\subset\mH^{\otimes k}$ defined by $\bigwedge^kA\,(v_1\wedge\cdots\wedge v_k)=Av_1\wedge\cdots\wedge Av_k$. Physically, it is the one-particle operator $A$ acting on each of $k$ identical fermions. If $A$ is compact with singular values $s_1\geq s_2\geq\cdots$, then $\|\bigwedge^kA\|=s_1\cdots s_k$. This is why exterior powers are useful here. A product of the $k$ largest singular values is the norm of an operator, so it obeys the three-lines convexity, whereas a single singular value other than the largest is not a norm and in general does not. For $k=2$ and $s_1=1$, as for $D_\alpha$ below, the norm is the second singular value, whose negative logarithm is the attenuation gap.}.
Then $f_k$ is convex, $f_k(0)=f_k(1/2)=0$, and
$f_k(1/2-\alpha)=f_k(\alpha)$. In particular,
\begin{equation}
 f_k(\beta)\leq f_k(\alpha),
 \qquad 0<\alpha<\beta\leq\tfrac14.
 \label{eq:Q-fk-monotone}
\end{equation}
When $D_\alpha$ is compact and has at least $k$ nonzero singular
values $s_1(\alpha)\geq\cdots\geq s_k(\alpha)$,
\begin{equation}
 f_k(\alpha)=\sum_{j=1}^k\log s_j(\alpha).
\end{equation}
Consequently, the sums of the $k$ lowest finite energies of
$F_\alpha$ and
$Q_\alpha^{\rm gap}$ do not decrease on $(0,1/4)$.
On the other hand, the alternative normalization
\begin{equation}
 \widehat Q_\alpha:=-\frac{1}{4\pi\alpha}\log\rho_\alpha
 =\frac{\sin(2\pi\alpha)}{2\pi\alpha}Q_\alpha
 \label{eq:Q-alpha-linear-normalization}
\end{equation}
has \emph{nonincreasing} sums of its lowest $k$ finite energies on
$0<\alpha<1/2$, whenever the relevant spectral levels are defined.
These are spectral statements. They do not imply a Loewner-order
inequality for the operators at different $\alpha$.
Moreover $f_1\equiv0$ and, for all $0\leq\alpha\leq\tfrac12$ and without compactness, $f_2(\alpha)=\log\|X_\alpha|_{\ket{\Omega}^\perp}\|=-\lambda_{\rm gap}(\alpha)$ (identically zero if $\mN_{\rm core}\neq\mathbb C1$), so $\lambda_{\rm gap}$ of \eqref{eq:Q-midpoint-gap} is concave, symmetric about $\tfrac14$ and maximal at $\alpha=\tfrac14$.
\end{theorem}

\begin{proof}
The exterior-power map $z\mapsto\bigwedge^k D(z)$ is bounded
and analytic on the modular strip. The relation
$D(t+i\alpha)=\Delta_\mM^{-it}D_\alpha V\Delta_\mN^{it}V^\dagger$
shows that its norm is independent of the real part of $z$.
Hadamard's three-lines theorem \cite{BerghLofstrom1976}
therefore gives the convexity of $f_k$.
Both boundary maps are isometries on the code subspace and vanish on its orthogonal complement, so $f_k(0)=f_k(1/2)=0$. The strip-reflection identity
$D_{1/2-\alpha}=J_\mM D_\alpha VJ_\mN V^\dagger$
preserves exterior-power norms and implies the stated symmetry.
Convexity and midpoint symmetry imply \eqref{eq:Q-fk-monotone}.
Since $D_\alpha\ket{\Omega}=D_\alpha^\dagger\ket{\Omega}=\ket{\Omega}$ and $\|D_\alpha\|\leq1$, $D_\alpha=1\oplus D_\alpha|_{\ket{\Omega}^\perp}$ on $\mathbb C\ket{\Omega}\oplus\ket{\Omega}^\perp$, hence $f_1\equiv0$. Write $B=D_\alpha|_{\ket{\Omega}^\perp}$. The operator $\bigwedge^2D_\alpha$ acts as $B$ on $\ket{\Omega}\wedge\ket{\Omega}^\perp$ and as $\bigwedge^2B$ on $\bigwedge^2\ket{\Omega}^\perp$, so $\|\bigwedge^2D_\alpha\|=\max(\|B\|,\|\bigwedge^2B\|)=\|B\|$, because $\|\bigwedge^2B\|\leq\|B\|^2\leq\|B\|$. Finally $\|B\|=\|X_\alpha|_{\ket{\Omega}^\perp}\|$.

For compact $D_\alpha$, the exterior-power norm equals the
product of its $k$ largest singular values. The corresponding
eigenvalues of $\rho_\alpha$ are $s_j(\alpha)^2$, hence the
lowest-energy partial sum of $F_\alpha$ is $-2f_k(\alpha)$,
and that of $Q_\alpha^{\rm gap}$ is $-f_k(\alpha)$.
Finally, convexity and $f_k(0)=0$ imply that
$f_k(\alpha)/\alpha$ is nondecreasing. The lowest-energy
partial sum of $\widehat Q_\alpha$ is
$-f_k(\alpha)/(2\pi\alpha)$, and is therefore nonincreasing.
\end{proof}

\section{Classical subsystem flows and ergodic theory}\label{sec:classical}

In this appendix, we review established constructions from three related traditions to draw an analogy between quantum and classical dynamics and ergodicity. The first is conditional expectation and statistical sufficiency in probability and information theory~\cite{Doob1953,KullbackLeibler1951,Csiszar1975}. The second is factors, generating algebras and Kolmogorov systems in ergodic theory~\cite{CornfeldFominSinai1982}. The third is state-dependent Poisson dynamics in classical KMS theory and Poisson modular geometry~\cite{GallavottiVerboven1975,GallavottiPulvirenti1976,Weinstein1997}. Our purpose is to organize these constructions for comparison with the quantum subsystem flows of the main text. Table~\ref{tab:classical-quantum-dictionary} summarizes the dictionary between the two settings.

\begin{table}[!t]
\captionsetup{font=footnotesize}
\centering
\begingroup
\footnotesize
\setlength{\tabcolsep}{4pt}
\renewcommand{\arraystretch}{1.08}
\begin{tabular}{@{}>{\raggedright\arraybackslash}p{0.19\linewidth}
                >{\raggedright\arraybackslash}p{0.365\linewidth}
                >{\raggedright\arraybackslash}p{0.365\linewidth}@{}}
\toprule
\textbf{Concept} & \textbf{Classical} & \textbf{Quantum} \\
\midrule
Observable algebra &
$\mM=L^\infty(X,\Sigma,\mu)$, commutative. &
$\mM\subset B(\mH_{\mM})$, generally noncommutative \\
\addlinespace[2pt]
Coarse-graining &
Sub-$\sigma$-algebra $\Sigma_0\subset\Sigma$, or for a subsystem a projection $\pi:X\to Y$. &
Inclusion $\mN\subset\mM$, with no tensor factorization required. \\
\addlinespace[2pt]
Reduced state &
$\rho_0=\mE_\mu(\rho)$ (in a product, $\rho_0(y,z)=\rho_Y(y)$). &
Restriction $\omega|_{\mN}$, with intrinsic modular operator $\Delta_{\mN}$. \\
\addlinespace[2pt]
State-generated flow &
Poisson flow of $K_\rho=-\log\rho$:
$\delta_\rho(f)=\{K_\rho,f\}$. &
Modular flow $\sigma_t^{\omega;\mM}=\operatorname{Ad}\Delta_{\mM}^{it}$,
$K_{\mM}=-\log\Delta_{\mM}$. \\
\addlinespace[2pt]
Subsystem flow &
$\mathcal V_t=\alpha_t^\rho\circ\alpha_{-t}^{0}$,
comparing full and lifted reduced motions. &
$D(t)=\Delta_\mM^{-it}V\Delta_\mN^{it}V^\dagger$,
$V:\mH_{\mN}\hookrightarrow\mH_{\mM}$. \\
\addlinespace[2pt]
Infinitesimal comparison &
$K_{\mathrm{rel}}=K_\rho-K_0
=-\log(\rho/\rho_0)$ generates the relative Poisson motion. &
$V^\dagger K_{\mM}V-K_{\mN}$ is the comparison quadratic form on $\mH_{\mN}$. \\
\addlinespace[2pt]
Information loss &
$S(\rho\|\rho_0)=S(\rho\|1)-S(\rho_0\|1)=-\langle K_{\mathrm{rel}}\rangle_\rho\ge0$ (probe $\rho$ relative to $\mu$, where $K_{\rm rel}$ itself is indefinite: $\langle K_{\rm rel}\rangle_\mu\ge0\ge\langle K_{\rm rel}\rangle_\rho$). &
$\langle V^\dagger K_{\mM}V-K_{\mN}\rangle_\Psi
\ge0$, the sum of the retained and complementary gaps of Proposition~\ref{prop:sufficiency}. It vanishes at $\Psi=\Omega_\mN$. \\
\addlinespace[2pt]
Conditional expectation &
A state-preserving expectation onto every measurable coarse algebra exists. &
An $\omega$-preserving normal expectation exists iff
$\sigma_t^{\omega;\mM}(\mN)=\mN$ (Takesaki). \\
\addlinespace[2pt]
Autonomous subsystem &
$\alpha_t(\mN)=\mN$. For the product Poisson model, $\rho=\rho_Y\rho_Z$ ensures agreement of full and reduced flows. &
$\sigma_t^{\omega;\mM}(\mN)=\mN$ iff
$D_{\mN\subset\mM}(t)=e_\mN$, equivalently reference-preserving QEC. \\
\addlinespace[2pt]
Ergodicity and mixing &
$\mM^\alpha=\mathbb C1$. Clustering of correlations is stronger. &
$\mM^\omega=\mathbb C1$ (trivial centralizer). Modular mixing implies this and requires type $\mathrm{III}_1$. \\
\addlinespace[2pt]
History and persistent core &
$\mN_{\rm hist}=\bigvee_t\alpha_t(\mN)$,
$\mN_{\rm core}=\bigcap_t\alpha_t(\mN)$. &
$\mN_{\rm hist}=\bigvee_t\sigma_t^{\omega;\mM}(\mN)$,
$\mN_{\rm core}=\bigcap_t\sigma_t^{\omega;\mM}(\mN)$. \\
\addlinespace[2pt]
One-sided nesting &
$\alpha_t(\mN_+)\subseteq\mN_+$ for $t\ge0$. Full history and trivial tail give the $K$-property. &
$\sigma_t^{\omega;\mM}(\mN)\subseteq\mN$ for $t\le0$ (the opposite time orientation to the classical row). 
For a common-cyclic half-sided modular inclusion with this orientation, $K_{\mM}-K_{\mN}=2\pi P\ge0$. \\
\bottomrule
\end{tabular}
\endgroup
\vspace{0.3em}

\caption{Dictionary between the classical constructions of this appendix and the
quantum subsystem flows studied in the main text. Note that the classical state-generated
Poisson flow is an analogy, not the Tomita--Takesaki modular flow of an
abelian von Neumann algebra.}
\label{tab:classical-quantum-dictionary}
\end{table}

This appendix uses three distinct constructions. First, a sub-$\sigma$-algebra defines a measure-theoretic coarse-graining and its conditional expectation. Second, a specified measure-preserving dynamics transports that coarse algebra and defines its history and tail. Third, a Poisson bracket and a smooth positive density define a state Hamiltonian $-\log\rho$. Only the third construction turns a density into a motion of phase-space points.

The quantum comparison uses modular dynamics, whereas modular flow of an abelian von Neumann algebra is trivial. The Poisson state flow is therefore an additional geometric analogy, not its literal commutative specialization. It preserves the reference state and obeys a classical KMS identity, but its relative Hamiltonian need not be positive under coarse-graining. Appendices~\ref{subsec:classical-definitions}--\ref{subsec:classical-ergodic-hierarchy} cover the measure-theoretic constructions. Appendices~\ref{subsec:classical-state-flow}--\ref{subsec:classical-subalgebra-Hamiltonian} develop the Poisson comparison and examples. The appendix can be read independently of the quantum proofs.

\subsection{Review of classical dynamical systems}\label{subsec:classical-definitions}

\paragraph{Algebra of observables:} In classical mechanics, one often begins with a phase space that we take to be an orientable smooth manifold \(X\) that admits a strictly positive volume measure. Points \(x\in X\) represent microstates of the system. But from the algebraic viewpoint, the primary object is not the set of points but the commutative von Neumann algebra of observables (complex measurable functions)
\[
\mM=L^\infty(X,\Sigma,\mu),
\]
where \((X,\Sigma,\mu)\) is a standard probability space. When $X$ is a smooth phase space of finite volume, we take $\mu$ to be the normalized volume measure\footnote{$\Sigma$ is the completion of the Borel $\sigma$-algebra with respect to $\mu$. Elements of $L^\infty(X,\Sigma,\mu)$ are equivalence classes of essentially bounded measurable functions that agree $\mu$-almost everywhere.}.

\paragraph{States:} The normalized measure $\mu$ above is interpreted as a distinguished reference state\footnote{A $\sigma$-finite volume measure of infinite total volume instead defines a distinguished weight. Weights are introduced in Appendix~\ref{sec:modular-vs-dynamics}.}. General normal states correspond to probability measures $\nu$ on $X$ that are absolutely continuous with respect to $\mu$ (denoted $\nu\ll \mu$), meaning that for every measurable set $E\subseteq X$, if $\mu(E)=0$ then $\nu(E)=0$.
There exists a $\mu$-measurable function $\rho_\nu:X\to [0,\infty)$ such that $\nu(E)=\int_E \rho_\nu d\mu$\footnote{In mathematical terminology, $\rho_\nu$ is called a Radon-Nikodym derivative.}. In other words, every normal state can be viewed as a probability density function $\rho_\nu$ on $X$, with expectation values of an observable $f$ in this state captured by integrals on $X$:
\begin{align}
\braket{f}_\nu=\int_{x\in X}f(x) d\nu(x)=\int_{x\in X} \rho_\nu(x) f(x) d\mu(x)\ .
\end{align}

\paragraph{GNS Hilbert space:} The GNS Hilbert space of the reference state $\mu$ is
\[
  \cH=L^2(X,\Sigma,\mu),
  \qquad
  \mM=L^\infty(X,\Sigma,\mu).
\]
Observables act by multiplication, and the reference state $\mu$ is represented by the constant unit vector $\ket{1}$. A general normal state with density $\rho$ is represented by
\[
  \ket{\sqrt{\rho}}\in\cH,
  \qquad
  \|\ket{\sqrt{\rho}}\|_2^2=\int_X\rho\,\dd\mu=1.
\]
The GNS construction builds a connection between observables and states.

\paragraph{Dynamics:} Dynamics will be described by a measurable one-parameter group of invertible, measure-preserving transformations $\Phi_t:X\to X$. General dynamics need not be smooth. In physics, we are often interested in smooth flows described by a one-parameter family of volume-preserving diffeomorphisms of $X$. Taking a $t$-derivative of how a point $x\in X$ moves under this flow defines a vector field $v$, which preserves the volume measure if and only if $\mathcal{L}_{v}\mu=0$, i.e., it is divergence-free with respect to $\mu$.

\paragraph{Coarse-grainings:}

The measure-theoretic description used here is the standard encoding of retained information by a sub-$\sigma$-algebra. Conditional expectation is both an averaging operation and, on $L^2$, an orthogonal projection~\cite{Doob1953}. The region and variable-selection constructions below are two realizations of this common framework. Classical coarse-graining retains a subalgebra of observables and discards the information needed to distinguish states that agree on this subalgebra. Two useful constructions arise from a subregion of the space of microstates and from a subsystem of observed variables. Both can be formulated in measure theory, before introducing a Hamiltonian or a Poisson bracket. We first describe these constructions and then ask when the retained observables evolve autonomously.

In the measure-theoretic setting above, a coarse description is specified by a sub-$\sigma$-algebra $\Sigma_0\subseteq\Sigma$, with associated unital von Neumann subalgebra
\begin{align}
    \mN=L^\infty(X,\Sigma_0,\mu)\subseteq\mM\ .
\end{align}
The reference measure $\mu$ determines the coarse-graining map
\begin{align}
    \mE_{\mN}(f)=\mathbb E_\mu[f\mid\Sigma_0]\ .
\end{align}
This is the unique $\mu$-preserving normal conditional expectation onto $\mN$. On $L^2(X,\mu)$ it acts as the orthogonal projection onto $L^2(X,\Sigma_0,\mu)$. A general normal state $\omega_\rho(f)=\int_X\rho f\,d\mu$ is coarse-grained to $\omega_0=\omega_\rho\circ \mE_{\mN}$, whose density is $\rho_0=\mE_{\mN}(\rho)$, using the $L^1$ extension of conditional expectation. The two states agree on $\mN$. Thus the subalgebra specifies the retained information, while the chosen expectation specifies how a state on this subalgebra is extended back to the full algebra.

\paragraph{Entropy of a coarse-graining:}

Relative entropy provides the classical information-theoretic measure of loss under an observation, and equality in its monotonicity is tied to statistical sufficiency~\cite{KullbackLeibler1951}. The identity below is a conditional-expectation chain rule. It also fits the information-geometric study of divergence and projections~\cite{Csiszar1975}. Consider a $\mu$-preserving coarse-graining $\mE$, a normal state with density $\rho$, and its coarse-grained density $\rho_0=\mE(\rho)$. Then, the relative entropy (KL-divergence) of these two states associates an entropy to this coarse-graining:
\begin{align}
    \label{eq:KL-classical}
    S(\rho\|\rho_0)=\int_X \rho \log(\rho/\rho_0)d\mu=H(\rho_0)-H(\rho)
\end{align}
where $H(\rho)=-\int_X\rho\log\rho\,d\mu$ is the entropy relative to the chosen reference measure, and the equality assumes that the two entropies are finite. The nonnegativity of relative entropy implies that coarse-graining cannot decrease this entropy. This compares a state with its coarse-grained version. It does not by itself imply monotonic entropy growth under time evolution. For a distinct quantum-field-theoretic application of relative-entropy losses and recovery, see~\cite{Lashkari2019EntanglementScale}.

\paragraph{Generating function for entropy:} For faithful densities $\rho,\rho_0>0$ almost everywhere, the multiplicative relative phase
\[
u_t
=
\left(\frac{\rho}{\rho_0}\right)^{it}
\]
is the generating function for coarse-graining entropy because
\begin{align}
&-i\left.\frac{\dd}{\dd t}u_t\right|_{t=0}
=
\log\frac{\rho}{\rho_0},\nn\\
&S(\rho\|\rho_0)=-i \int_X \rho (\partial_t u_t\big|_{t=0})d\mu\ .
\end{align}
Here the differentiation under the integral is justified when $\rho\log(\rho/\rho_0)$ is integrable. More generally, for every faithful density $\rho$ we can define the strongly continuous unitary multiplication group
\[
  U_t^\rho:=\Mop_{\rho^{it}}
  =\exp\!\bigl(it\Mop_{\log\rho}\bigr),
  \qquad
  (U_t^\rho\psi)(x)=e^{it\log\rho(x)}\psi(x).
\]
Similarly, the coarse-grained density defines
\[
  U_t^{\rho_0}:=\Mop_{\rho_0^{it}}.
\]
The parameter $t$ is an auxiliary phase parameter. Nothing in this construction identifies it with laboratory time. In particular, $U_t^\rho$ does not transport points of $X$. It only rotates the phase of a Hilbert space vector pointwise\footnote{For faithful normal states on the same abelian algebra, $u_t=(\rho/\rho_0)^{it}$ is the Connes Radon-Nikodym cocycle. Relative modular operators remain nontrivial, even though the modular automorphism group of each state is trivial.}.

\subsection[Subspace (single-particle) and subsystem (multi-particle) coarse-grainings]{Subspace (single-particle) and subsystem (multi-particle)\\coarse-grainings}
\label{subsec:classical-region-subsystem}

A region $O\subset X$ selects alternatives among whole microstates: a point lies in $O$ or in its complement. The corresponding decomposition of observable algebras is a direct sum. For example, one can resolve a pendulum's phase-space points with $|\theta|<\theta_0$ and identify all points outside that angular range.

A subsystem instead retains selected variables at every microstate. For two particles, projection from $(q_1,p_1,q_2,p_2)$ to $(q_1,p_1)$ keeps the first particle and discards the second. In a product phase space, the observable algebras form tensor factors. More generally, a subsystem is specified by a map $\pi:X\to Y$. Its coarse space need not be an embedded region of $X$. The distinction concerns which information is retained, rather than the dimension of the selected space.

\paragraph{Subspace coarse-grainings:}
For a measurable region $O\subset X$, define the algebra generated by observables that are constant outside of $O$:
\begin{align}
    \mM_O=\{f\in\mM:\ f=f \cdot 1_O +c 1_{O^c}\text{ for some }c\in\mathbb C\}
\end{align}
where $1_O$ is the indicator function on $O$, and $(f\cdot 1_O)(x)=f(x)1_O(x)$ is point by point multiplication.
This algebra resolves the microstate completely within $O$ and retains only the event ``outside $O$'' on its complement. Assuming $0<\mu(O)<1$, its conditional expectation is
\begin{align}
    \mE_O(f)=1_O f+
    1_{O^c}\mathbb{E}_{O^c}(f)
\end{align}
where
\begin{align}
    \mathbb{E}_{O^c}(f)=\frac{1}{\mu(O^c)}\int_{O^c} f d\mu
\end{align}
is the average of $f$ on $O^c$.
For $O_1\subseteq O_2$, we have $\mM_{O_1}\subseteq\mM_{O_2}$, so this construction gives a net of coarse observable algebras. Since a microstate $x\in X$ is either in $O$ or $O^c$ we have
\begin{align}
    \mM=1_O\mM\oplus 1_{O^c}\mM
    \simeq L^\infty(O,\mu|_O)\oplus L^\infty(O^c,\mu|_{O^c})\ .
\end{align}
Our unital subspace algebras have the form $\mM_O=1_O\mM\oplus\mathbb C1_{O^c}$, so $\mM_O$ and $\mM_{O^c}$ are not themselves the two direct summands. In fact,
\begin{align}
    \mM_O\cap\mM_{O^c}=\operatorname{span}\{1_O,1_{O^c}\}\ .
\end{align}

\paragraph{Subsystem coarse-grainings:}
Suppose instead that the probability space has a product decomposition
\begin{align}
    (X,\Sigma,\mu)=(Y\times Z,\Sigma_Y\otimes\Sigma_Z,\mu_Y\otimes\mu_Z)\ .
\end{align}
A microstate now specifies both an observed variable $y$ and a hidden variable $z$. The observable algebra factorizes as
\begin{align}
    \mM\simeq\mM_Y\,\bar\otimes\,\mM_Z,
    \qquad \mM_Y=L^\infty(Y,\mu_Y),\quad
    \mM_Z=L^\infty(Z,\mu_Z)\ .
\end{align}
Retaining the subsystem $Y$ means retaining $\mN_Y=\mM_Y\otimes 1$. The corresponding conditional expectation averages over the hidden variables:
\begin{align}
    (\mE_Y f)(y,z)=\int_Z f(y,z')\,d\mu_Z(z')\ .
\end{align}
This operation forgets the hidden variables at every microstate. The tensor factors describe simultaneously specified degrees of freedom. Although the reference measure is a product, a general state with density $\rho(y,z)$ can have correlations between them.

More generally, a measurable map $\pi:X\to Y$ between standard probability spaces specifies coarse observables through
\begin{align}
    &\mN_Y=\pi^*\mM_Y\subset \mM,
    \qquad \mu_Y=\mu\circ \pi^{-1}\nn\\
    &\mM_Y=L^\infty(Y,\mu_Y)
\end{align}

Here $\mN_Y=\pi^*\mM_Y$ is the pullback of $\mM_Y$, and $\mu_Y=\pi_*\mu$ is the
push-forward measure. Under the standard Borel assumptions, disintegration gives
conditional probability measures $\mu_y$ concentrated on the fibers $\pi^{-1}(y)$.

The conditional expectation is
\begin{align}
    (\mE_\pi f)(x)=\int_{\pi^{-1}(\pi(x))}f(x')\,d\mu_{\pi(x)}(x')\ .
\end{align}
The fiber measures may depend on $y$, and no global product decomposition is required.

Note that both subspace and subsystem conditional expectations are special cases of a common measure-theoretic notion of coarse-graining.

\paragraph{Dynamically closed and invariant coarse-grainings:}

In standard ergodic-theory terminology, an invariant coarse algebra describes a factor system~\cite{CornfeldFominSinai1982}. The word ``autonomous'' below refers to this exact closure of the retained observables under the specified reversible dynamics. It is stronger than merely specifying their equilibrium marginal.
Let $\Phi_t:X\to X$ be a measurable one-parameter group of invertible, measure-preserving transformations. It evolves observables according to
\begin{align}
    \alpha_t(f)=f\circ\Phi_{-t}\ .
\end{align}
For any coarse algebra $\mN\subseteq\mM$, we can follow the family $\alpha_t(\mN)$. We call the coarse description dynamically closed when
\begin{align}\label{eq:classical-coarse-invariance}
    \alpha_t(\mN)=\mN,\qquad t\in\mathbb R\ .
\end{align}
For nonatomic $\mu$ and $0<\mu(O)<1$, invariance of the subspace algebra $\mM_O$ violates ergodicity, as we show below.
For the subsystem $Y$ in a product space $Y\times Z$, dynamical closure means that the $Y$-component of $\Phi_t(y,z)$ depends only on $y$\footnote{Note that this does not follow from preservation of the product measure.}. For a coarse-graining specified by $\pi$, this means that there is a measure-preserving evolution $\Psi_t$ on the coarse space such that
\begin{align}\label{eq:classical-factor-intertwining}
    \pi\circ\Phi_t=\Psi_t\circ\pi\ ,
\end{align}
understood modulo null sets. Evolving the microstate and then retaining its coarse variables gives the same result as evolving the coarse variables themselves. In ergodic theory such a coarse description is called a {\it factor system}, and we say the coarse-grained variables evolve autonomously. Intuitively, this means that the future of the macroscopic (or observed) variables can be determined from the present macroscopic state, without the need to know the hidden microscopic details. In other words, there is an effective theory that describes the coarse-grained variables.

\paragraph{Effective theory (Compatibility of coarse-graining and evolution):}
For the measure-preserving dynamics above, invariance of $\mN$ is equivalent to
\begin{align}\label{eq:classical-expectation-equivariance}
    \mE_{\pi}\circ\alpha_t=\alpha_t\circ \mE_{\pi},
    \qquad t\in\mathbb R\ .
\end{align}
Indeed, when $\mN_Y$ is invariant, $\alpha_t\circ \mE_{\pi}\circ\alpha_{-t}$ is another $\mu$-preserving conditional expectation onto $\mN_Y$, so uniqueness gives the equality. Conversely, this equality ensures that $\alpha_t$ maps the range of $\mE_{\pi}$ onto itself. The physical interpretation is macroscopic predictability: the future observed variables can be determined from their present values. This compatibility supplies an effective dynamics on the coarse algebra.

\paragraph{Correlation functions of observables and subalgebras:}
For complex functions $f,g\in\mM$, define the correlation function and its connected part as
\begin{align}
    &C_{f,g}(t)=\mu\bigl(f^*\alpha_t(g)\bigr)\nn\\
    &C^{con}_{f,g}(t)
    :=\mu\bigl(f^*\alpha_t(g)\bigr)-\mu(f^*)\mu(g),
    \qquad \mu(f)=\int_X f\,d\mu\ .
\end{align}
Ergodicity is equivalent to the vanishing of the time-averaged connected correlation for every pair of observables:
\begin{align}
    \lim_{T\to\infty}\frac{1}{T}\int_0^T C^{con}_{f,g}(t)\,dt=0\ .
\end{align}
A stronger condition, {\it weak mixing}, requires
\begin{align}
    \lim_{T\to\infty}\frac{1}{T}\int_0^T |C^{con}_{f,g}(t)|^2\,dt=0\ .
\end{align}
The flow is {\it mixing} if $C^{con}_{f,g}(t)\to0$ as $t\to+\infty$. These conditions distinguish cancellation after time averaging from actual loss of correlation at late times. Mixing does not specify a decay rate and does not imply that there are no autonomous subsystems.

Quantitative measures of the correlation between two subspace algebras, or between two subsystems, and their relation to the total variation distance from the product measure, are collected in Appendix~\ref{app:classical-correlations}.

\paragraph{Subsystem coarse-grainings of a free relativistic field theory:} Classical field theory provides a natural setting for a net of subsystem algebras with infinitely many degrees of freedom. For an unconstrained real scalar field on a Cauchy surface $\Sigma$, a phase-space point is a pair of initial data $(\phi,\pi)$. An open spatial region $U\subset\Sigma$ selects the unital polynomial algebra
\begin{align}
    \mN_U
    :=\operatorname{Alg}_{\mathbb R}\{1,\phi(f),\pi(g):f,g\in C_c^\infty(U)\},
    \qquad
    \{\phi(f),\pi(g)\}=\int_\Sigma fg\,d\Sigma\ .
\end{align}
Here $\phi(f)=\int_\Sigma f\phi\,d\Sigma$, and similarly for $\pi(g)$. The other brackets between generators vanish. Thus $\mN_U$ is a Poisson subalgebra, and algebras associated with disjoint spatial regions Poisson commute. Retaining $\mN_U$ resolves the field data in $U$ while discarding the exterior data. Although labeled by a spatial region, this is a subsystem coarse-graining: $U$ selects degrees of freedom, whereas a region of phase space selects entire field configurations.

Choose an ambient Poisson algebra $\mM$ containing these observables and stable under a complete Hamiltonian evolution. With our convention for the pullback, the transported subsystem is
\begin{align}
    \alpha_t^H(F)=F\circ\Phi_{-t}^H,
    \qquad \partial_t\alpha_t^H(F)=\{H,\alpha_t^H(F)\},
    \qquad \alpha_t^H(\mN_U)\subseteq\mM\ .
\end{align}
Hamiltonian evolution preserves Poisson brackets, so every $\alpha_t^H(\mN_U)$ remains a Poisson subalgebra. Invariance for all $t$ gives an autonomous reversible subsystem. More generally, propagation can make the transported observables depend on initially discarded data. Whether their time translates generate $\mM$, or instead form a nested family, is a property of the chosen subsystem and dynamics together.

\subsection{Correlation coefficients for classical subspaces and subsystems}\label{app:classical-correlations}

This subsection records the measures of correlation between two coarse algebras of Appendix~\ref{subsec:classical-region-subsystem}, for a subspace pair and for a subsystem pair, and their relation to the total variation distance from the product measure. They quantify the distinction between the two kinds of coarse-graining but are not used in the main text.

\paragraph{Correlations between subspaces:}
Write $C(f,g)=\mu(f^*g)$. For measurable regions $O_1,O_2$, define
\begin{equation}
 C(O_1,O_2):=\sup_{\substack{f_i=0\ {\rm a.e.\ on}\ O_i^c\\\|f_i\|_\infty\leq1}}
 |C(f_1,f_2)|,
\end{equation}
where $\|\cdot\|_\infty$ is the essential-supremum norm. The integrand is bounded by $1_{O_1\cap O_2}$, and the indicator functions attain the bound. Hence
\begin{equation}
 C(O_1,O_2)=\mu(O_1\cap O_2)=C(1_{O_1},1_{O_2}).
\end{equation}

For connected correlations, define the event coefficient
\begin{equation}
 \alpha(O_1,O_2)=
 \sup_{\substack{A\subset O_1\\B\subset O_2}}
 |\mu(A\cap B)-\mu(A)\mu(B)|.
\end{equation}
In particular,
\begin{equation}
 C^{con}_{1_{O_1},1_{O_2}}(t)
 =\mu\bigl(O_1\cap\Phi_t(O_2)\bigr)-\mu(O_1)\mu(O_2).
\end{equation}
This measures the correlation of an event with a transported event.

\paragraph{Correlations between subsystems:}

The coefficient $\alpha$ below is the strong-mixing coefficient of the two coordinate $\sigma$-algebras~\cite{Bradley2005}. With our convention for total variation, the total variation distance between the joint law and the product of its marginals is their absolute-regularity coefficient $\beta$ (on the standard spaces considered here). These coefficients quantify dependence between two algebras. Uniform decay of such coefficients is a stronger requirement than mixing tested separately against each fixed pair of observables. No such uniform decay assumption is imposed here.
Let $X=Y\times Z$ carry a joint probability measure $\mu_{YZ}$ with marginals
$\mu_Y,\mu_Z$. The unconnected correlation has supremum one over functions bounded
by one, because constant functions attain it. For connected correlations, distinguish
\begin{align}
 \alpha(Y,Z)&=\sup_{\substack{A\subset Y\\B\subset Z}}
 |\mu_{YZ}(A\times B)-\mu_Y(A)\mu_Z(B)|,\nn\\
 \Gamma_{\rm cl}(Y,Z)&=\sup_{\substack{f,g\ {\rm real}\\\|f\|_\infty,\|g\|_\infty\leq1}}
 |\mu_{YZ}(fg)-\mu_Y(f)\mu_Z(g)|=4\alpha(Y,Z).
\end{align}
The factor four follows by rescaling real functions from $[-1,1]$ to $[0,1]$
and writing the latter as integrals of indicator functions. With
$\nu=\mu_{YZ}-\mu_Y\otimes\mu_Z$, total variation satisfies
\begin{equation}
 \alpha(Y,Z)\leq\tfrac12\operatorname{TV},
 \qquad \Gamma_{\rm cl}(Y,Z)\leq2\operatorname{TV},
 \qquad \operatorname{TV}:=\sup_E|\nu(E)|=\tfrac12\|\nu\|_1.
\end{equation}
Here $\|\nu\|_1$ is the total variation norm of the signed measure. The first bound is sharp: two perfectly correlated fair bits give $\alpha=\tfrac14=\tfrac12\operatorname{TV}$. All these
coefficients vanish precisely when the joint measure is a product.

\subsection{Ergodic hierarchy, subspaces and subsystems}
\label{subsec:classical-ergodic-hierarchy}

The terminology in this subsection is that of classical factor theory and $K$-systems~\cite{CornfeldFominSinai1982}. Our ``history'' is the algebra generated by the time translates of an observation algebra. The one-sided ``tail'' is the intersection of a decreasing family. These constructions concern the chosen dynamics and observation algebra, independently of the later Poisson state construction.

The ergodic hierarchy describes how much information about a state remains accessible at late times~\cite{CornfeldFominSinai1982}. In the algebraic language, it can be studied both through correlations of observables and through the evolution of coarse observable algebras. For our purposes, the relevant distinction is between conserved events, autonomous subsystems, and subalgebras that are preserved only in one time direction.

\paragraph{Ergodicity:}
Let $(\mM,\mu,\alpha_t)$ be the measure-preserving dynamical system introduced above. The flow is {\it ergodic} if every measurable set invariant under all $\Phi_t$ has measure zero or one. Equivalently, its fixed-point algebra is trivial:
\begin{align}
    \mM^\alpha
    :=\{f\in\mM:\alpha_t(f)=f\text{ for all }t\in\mathbb R\}
    =\mathbb C1\ .
\end{align}
For nonatomic $\mu$ and $0<\mu(O)<1$, ergodicity forbids invariant subspace algebras. If
\begin{align}
    \alpha_t(\mM_O)=\mM_O,\qquad t\in\mathbb R,
\end{align}
then $\Phi_t(O)=O$ modulo null sets. Indeed, if $E:=O^c\cap\Phi_t(O)$ had positive measure, non-atomicity gives $E_1\subset E$ with $0<\mu(E_1)<\mu(E)$. Then $1_B\in\mM_O$ for $B=\Phi_{-t}(E_1)\subset O$, while $\alpha_t(1_B)=1_{E_1}$ is not a.e. constant on $O^c$. Hence $\Phi_t(O)\subseteq O$, and equality follows from measure preservation. Thus $1_O\in\mM^\alpha$ (it is a {\it conserved event}), which contradicts ergodicity. However, ergodicity does not exclude an invariant subsystem $\mN\subset\mM$: the condition $\alpha_t(\mN)=\mN$ permits observables in $\mN$ to evolve into one another. Such a subsystem is an autonomous factor of the dynamics (conserved subsystem). Taking two identical independent copies of a classical dynamical system always results in autonomous factors. However, a standard result in ergodic theory implies that if the original system is weakly mixing, the combined dynamical system is ergodic\footnote{For an invertible discrete-time example, take two independent bilateral Bernoulli shifts on $\{0,1\}^{\mathbb Z}$ with their product probability measures. Each shift, and their product, is mixing.}.

\paragraph{Generating property:}
Given a coarse algebra $\mN$, the family $\alpha_t(\mN)$ describes the observables accessible as this coarse description is transported by the dynamics. If $\alpha_t(\mN)=\mN$, evolution remains within the original description. Otherwise, the transported observables can depend on variables that were initially hidden. The algebra
\begin{align}
    \mN_{\rm hist}:=\bigvee_{t\in\mathbb R}\alpha_t(\mN)
\end{align}
contains the information accessible from all time translates of the coarse observables. Here $\bigvee$ denotes the generated von Neumann algebra. The condition $\mN_{\rm hist}=\mM$ says that these translates generate the full observable algebra. This is called the {\it generating property} and is distinct from ergodicity of the dynamics.

The dual notion is the persistent core $\mN_{\rm core}:=\bigcap_{t}\alpha_t(\mN)$, the observables of $\mN$ that remain in $\alpha_t(\mN)$ at every time. The core is trivial when $\mN_{\rm core}=\mathbb C1$. For discrete time the joins and intersections run over $t\in\mathbb Z$.

Figure~\ref{fig:ink-history-core} in Section~\ref{subsec:QEC-history-memory} illustrates generation and a trivial core for the baker map.

\paragraph{Kolmogorov property:}
Of particular interest are subalgebras satisfying
\begin{align}\label{eq:classical-one-sided-inclusion}
    \alpha_t(\mN_+)\subseteq\mN_+,\qquad t\geq0\ .
\end{align}
The family $\alpha_t(\mN_+)$ is then decreasing: $\alpha_{t+s}(\mN_+)\subseteq\alpha_t(\mN_+)$ for $s\geq0$. To interpret this inclusion, one can take $\mN_+$ to be generated by observations on a half-infinite time interval, with the orientation chosen so that positive $t$ removes the earliest part of the record.

The Kolmogorov, or $K$, property can be formulated through the existence of such a one-sided algebra whose translates generate the full algebra and whose tail is trivial:
\begin{align}\label{eq:classical-K-algebra}
    \bigvee_{t\in\mathbb R}\alpha_t(\mN_+)=\mM,
    \qquad
    \bigcap_{t\geq0}\alpha_t(\mN_+)=\mathbb C1\ .
\end{align}
The first condition says that the complete record resolves all observables. The second says that no nonconstant observable remains accessible arbitrarily far into the retained half of the record. The analogous conditions with integer time describe a $K$-automorphism. These are the standard generating and trivial-tail conditions of a $K$-system~\cite{CornfeldFominSinai1982}. Their use in the modular future/past-algebra setting is discussed in~\cite{ouseph2024local}. The classical conditions do not by themselves supply modular analyticity or positivity of a quantum translation generator.

\paragraph{Example: a Bernoulli shift:}
For independent uniform bits $x_n$, $n\in\mathbb Z$, with the shift $\alpha(x_n)=x_{n+1}$ and $\mN_+$ the algebra generated by $\{x_n:n\geq0\}$, one has $\alpha^k(\mN_+)=\operatorname{vN}\{x_n:n\geq k\}\subsetneq\mN_+$ for $k>0$, the translates generate $\mM$, and the tail is trivial by the Kolmogorov zero--one law. The retained description forgets one coordinate per step although the shift is invertible and measure preserving. This is a subsystem coarse-graining, the retained and discarded coordinates being tensor factors. Within the finite-measure invertible measure-preserving setting used here, strict one-sided nesting of this kind is possible for subsystem algebras but not for the region-type subspace algebras above. In a finite measure space $\Phi_t(O)\subseteq O$ and measure preservation force $\mu(O\setminus\Phi_t(O))=0$, so a region, and with it a subspace algebra $\mM_O$ in the nonatomic setting of Appendix~\ref{subsec:classical-region-subsystem}, cannot shrink strictly under a measure-preserving flow.

\paragraph{Coarse-graining along a decreasing family:}
Each $\alpha_t(\mN_+)$ has a $\mu$-preserving conditional expectation $\mE_t:\mM\to\alpha_t(\mN_+)$. For $s\geq t$, the tower property gives
\begin{align}
    \mE_s\mE_t=\mE_s\ .
\end{align}
For a fixed state with density $\rho$, its coarse-grained densities $\rho_t=\mE_t(\rho)$ therefore satisfy
\begin{align}
    H(\rho_s)\geq H(\rho_t),
    \qquad
    S(\rho\|\rho_s)\geq S(\rho\|\rho_t),
    \qquad s\geq t,
\end{align}
whenever the entropies are finite. Here the loss of information follows from the nesting of the retained algebras. It does not require irreversible microscopic dynamics. For a trivial tail, the reverse martingale theorem~\cite{Doob1953} further gives $\mE_t(f)\to\mu(f)1$ in $L^2$ for every $f\in L^2(X,\mu)$.

\subsection{Classical avatar of modular flow in Hamiltonian systems}
\label{subsec:classical-state-flow}

There are two established antecedents for this construction. Classical KMS theory characterizes equilibrium through an identity involving a state, a Poisson bracket, and a derivation. Its relation to Tomita--Takesaki theory was studied by Gallavotti and Pulvirenti~\cite{GallavottiVerboven1975,GallavottiPulvirenti1976}. Poisson geometry instead associates a modular vector field with a smooth density~\cite{Weinstein1997}. In the symplectic setting with Liouville reference measure, these viewpoints lead to the logarithmic state Hamiltonian used below. State-selected time also appears in the classical limit of the thermal-time proposal~\cite{ConnesRovelli1994}. None of these statements identifies the nontrivial Poisson flow with the modular automorphism group of the abelian multiplication algebra, which remains trivial.

So far, the state and the dynamics have been independent inputs. A probability density assigns weights to microstates, but measure theory alone does not turn these weights into a motion of the microstates. A Poisson bracket supplies the missing structure: it converts a function into a Hamiltonian vector field. Applied to the logarithm of a density, it gives a distinguished state-dependent dynamics. This construction will let us compare the dynamics associated with a state and with its coarse description.

\paragraph{Poisson brackets and Hamiltonian motion:}
A Poisson structure on a smooth manifold $X$ is an antisymmetric bivector $\Pi$ whose bracket
\begin{align}
    \{f,g\}=\Pi(df,dg),\qquad f,g\in C^\infty(X),
\end{align}
satisfies the Jacobi identity. The bracket obeys the Leibniz rule in each argument, so fixing a smooth real Hamiltonian $H$ defines a vector field. We use the convention
\begin{align}\label{eq:classical-Hamiltonian-convention}
    X_H(f)&=\{f,H\},\qquad
    \alpha_t^H(f)=f\circ\Phi_{-t}^H,\nn\\
    \delta_H(f)&:=\left.\partial_t\alpha_t^H(f)\right|_{t=0}=\{H,f\},
\end{align}
where $\Phi_t^H$ is the flow of $X_H$. Thus, the generator on observables has the opposite sign from the vector field transporting points. The Jacobi identity ensures that Hamiltonian evolution preserves the Poisson bracket. In canonical coordinates, $\{q_i,p_j\}=\delta_{ij}$ and the point trajectories satisfy the usual Hamilton equations.

\paragraph{Casimirs and symplectic leaves:}
A Casimir is a function $c$ with $\{c,f\}=0$ for every smooth $f$. It is conserved by every Hamiltonian, rather than by one chosen dynamics. The points accessible from a given microstate by successive Hamiltonian motions form a symplectic leaf. Every Casimir is constant on each leaf, although global Casimirs need not distinguish all leaves. A leaf is therefore a classical analogue of an irreducible sector for Hamiltonian motion. This analogy concerns the Poisson bracket: the pointwise algebra $L^\infty(X,\mu)$ remains abelian, and restricting it to a leaf does not turn it into a nonabelian factor.

\paragraph{Liouville measure and the state Hamiltonian:}
For the explicit constructions below, take $X$ to be a connected compact symplectic manifold without boundary, with symplectic form $\omega$ and normalized Liouville measure
\begin{align}
    d\mu=\frac{\omega^n/n!}{\int_X\omega^n/n!},\qquad \dim X=2n.
\end{align}
Hamiltonian flows preserve this measure. Equivalently,
\begin{align}\label{statepreserve}
    \int_X\{f,g\}\,d\mu=0
\end{align}
for smooth $f,g$. Compactness makes smooth Hamiltonian flows complete and avoids boundary terms. On a noncompact phase space these properties, together with the required integrability, must be checked separately.

Let $\rho$ be a smooth strictly positive probability density relative to $\mu$. Define its {\it state Hamiltonian} and the associated evolution by
\begin{align}\label{eq:classical-state-Hamiltonian}
    K_\rho=-\log\rho,\qquad
    \Phi_t^\rho:=\Phi_t^{K_\rho},\qquad
    \alpha_t^\rho:=\alpha_t^{K_\rho}.
\end{align}
The infinitesimal evolution of an observable is
\begin{align}\label{eq:classical-state-derivation}
    \delta_\rho(f)=\{K_\rho,f\}=\{f,\log\rho\}.
\end{align}
The flow moves along level sets of the density, since $X_{K_\rho}(\rho)=0$. It preserves both $\mu$ and the state $\omega_\rho(f)=\int_X\rho f\,d\mu$. For a Gibbs density $\rho=\mathcal Z_\beta^{-1}e^{-\beta H}$, we have $K_\rho=\beta H+\log\mathcal Z_\beta$ and $\Phi_t^\rho=\Phi_{\beta t}^H$. Thus the state recovers the Hamiltonian motion, with a rescaled parameter. For a general density, there is no prior identification of this parameter with physical time.

\paragraph{The classical KMS identity:}
The logarithm is distinguished by an integration-by-parts identity. Using \eqref{statepreserve},
\begin{align}
    0=\int_X\{g,f\rho\}\,d\mu
    =-\omega_\rho(\{f,g\})+
    \omega_\rho\bigl(f\{g,\log\rho\}\bigr).
\end{align}
Consequently,
\begin{align}\label{KMSclassical}
    \omega_\rho(\{f,g\})=\omega_\rho(f\delta_\rho(g)).
\end{align}
For a Gibbs state this reads $\omega_\beta(\{f,g\})=\beta\omega_\beta(f\delta_H(g))$. This is the classical, or static, KMS condition, written with the time orientation in \eqref{eq:classical-Hamiltonian-convention}~\cite{GallavottiVerboven1975,GallavottiPulvirenti1976,vanDeVen2022}. It is stronger than stationarity: for a given smooth positive density, testing against all smooth $f$ determines $\delta_\rho(g)$ for every $g$. Conversely, on a connected symplectic manifold, imposing this identity for a specified Hamiltonian dynamics determines a smooth positive density to be Gibbs, up to normalization. The identity also motivates formulations on algebras of local observables in infinite systems, where existence of the dynamics and equilibrium states requires additional assumptions. A global Gibbs density need not exist. For classical lattice spins with compact symplectic single-site phase space, Drago and van de Ven establish a DLR--KMS correspondence under their interaction assumptions~\cite{DragoVanDeVen2023}. That result supplies an infinite-system equilibrium framework, rather than a justification for an unrestricted formal use of a global density.

\paragraph{Poisson flow:}

We use Weinstein's density-dependent modular vector field~\cite{Weinstein1997}, with the Hamiltonian and inverse-pullback conventions fixed above. Its change under a rescaling of density is Hamiltonian. The corresponding modular class is independent of that choice. In the symplectic case, this class vanishes because the Liouville measure is invariant under all Hamiltonian flows, although the representative for a nonuniform density can be nonzero.
There is a geometric formulation of the same construction. For a smooth positive measure $\nu$, define its Poisson modular vector field by
\begin{align}
    \mathfrak m_\nu(f):=\operatorname{div}_\nu X_f.
\end{align}
This measures the failure of Hamiltonian motions to preserve $\nu$~\cite{Weinstein1997}. For $d\nu=\rho\,d\mu$ and Liouville $\mu$, the divergence formula gives
\begin{align}
    \mathfrak m_\nu(f)=X_f(\log\rho)
    =\{f,-\log\rho\}=X_{K_\rho}(f).
\end{align}
Thus $\Phi_t^\rho$ is precisely this Poisson modular flow. Its action on observables uses the inverse pullback, so $\delta_\rho=-\mathfrak m_\nu$. On a general Poisson manifold and for an arbitrary smooth reference measure, the formula instead reads $\mathfrak m_{\rho\mu}=\mathfrak m_\mu+X_{-\log\rho}$. A reference measure invariant under every Hamiltonian need not exist. Our symplectic setting removes this additional obstruction.

The construction uses the Poisson structure in an essential way. It is distinct from the multiplication group $U_t^\rho$ introduced above and from the trivial modular automorphism group of the abelian von Neumann algebra. It also supplies no automatic ergodic or entropy-production property. A nonconstant $\rho$ is itself a conserved observable, so the flow cannot be ergodic on the full probability space. Questions about ergodicity and mixing must be posed on invariant sectors with suitable invariant measures. The comparison of full and coarse state Hamiltonians, however, is meaningful without assuming ergodicity.

\subsection{Subsystem Poisson flows in classical many-body systems and field theory}
\label{subsec:classical-subalgebra-Hamiltonian}

A coarse description determines a reduced state and hence, when the coarse space is Poisson, its own state-generated dynamics. We can then ask whether the reduced dynamics agrees with what the full dynamics does to the observed variables. Comparing the two motions gives a classical model of the subsystem flows studied in this paper. The comparison separates evolution already visible within the retained algebra from evolution involving the discarded variables.

\paragraph{Poisson subsystems:} We say a subalgebra $\mN_Y\subset \mM$ of smooth observables is a Poisson subsystem if for all $b_1,b_2\in \mN_Y$, we have $\{b_1,b_2\}\in \mN_Y$. In the geometric setting below, $\pi:X\to Y$ is a smooth surjective Poisson submersion:
\begin{align}\label{eq:classical-Poisson-subsystem}
    \{\pi^*f,\pi^*g\}_X=\pi^*\{f,g\}_Y,
    \qquad f,g\in C^\infty(Y).
\end{align}
The smooth coarse observables $\mN=\pi^*C^\infty(Y)$ are closed under the bracket. This condition allows us to speak of Hamiltonian dynamics within the coarse description.

The simplest setting makes both the geometric and probabilistic structures explicit. Take a product $X=Y\times Z$ of connected compact symplectic manifolds, with product symplectic structure and normalized Liouville measure $\mu=\mu_Y\otimes\mu_Z$. The projection to $Y$ is Poisson, and the associated von Neumann subalgebra is $\mN_Y=L^\infty(Y,\mu_Y)\otimes1$. For a smooth positive density $\rho$, restriction of the state gives the marginal density
\begin{align}\label{eq:classical-marginal-Hamiltonian}
    \rho_Y(y)=\int_Z\rho(y,z)\,d\mu_Z(z),\qquad
    K_Y=-\log\rho_Y.
\end{align}
The expectation $\mE_Y$ from Appendix~\ref{subsec:classical-region-subsystem} extends this marginal back to $X$ as $\rho_0(y,z)=\rho_Y(y)$. Here, $\rho_Y=\pi_*\rho$ is the classical analog of the reduced state on $Y$, and $\rho_0=\pi^*\rho_Y=\pi^*\pi_* \rho$ is the pullback of $\rho_Y$ to $X$, which is the coarse-grained version of $\rho$. Its Hamiltonian is $K_0=\pi^*K_Y$, whose flow is
\begin{align}
    \Phi_t^0(y,z)=(\Phi_t^Y(y),z).
\end{align}
This provides a definite lift of the reduced motion to the full phase space. For a general Poisson submersion, the marginal measure need not be Liouville and a complementary subsystem need not exist, so we stay with the product setting.

\paragraph{Relation to the Hamiltonian of mean force:}
For a Gibbs density $\rho=\mathcal Z_\beta^{-1}e^{-\beta H}$ in the product setting, the reduced state Hamiltonian is
\begin{align}
 K_Y(y)=-\log\int_Z e^{-\beta H(y,z)}\,d\mu_Z(z)+\log\mathcal Z_\beta.
\end{align}
Thus $\beta^{-1}K_Y$ agrees, up to an additive normalization constant, with the classical Hamiltonian of mean force associated with integrating out the environment~\cite{TalknerHanggi2020}. This is an equilibrium reduction: it reproduces the marginal state, but need not reproduce the projected microscopic trajectories. The next criterion states precisely when that stronger dynamical agreement holds in our product symplectic model.

\paragraph{Agreement between reduced and full Poisson dynamics:}
Agreement means that the observed trajectory is determined by its initial coarse state and follows the flow generated by $K_Y$:
\begin{align}\label{eq:classical-state-intertwining}
    \pi\circ\Phi_t^\rho=\Phi_t^Y\circ\pi.
\end{align}
For our complete smooth flows, this is equivalent to the infinitesimal condition
\begin{align}\label{eq:classical-state-defect}
    \{\pi^*g,K_\rho\}_X-\pi^*\{g,K_Y\}_Y
    =\{\pi^*g,K_\rho-K_0\}_X=0,
    \qquad g\in C^\infty(Y).
\end{align}
The left side measures the part of the observed velocity missed by the reduced state Hamiltonian. A Hamiltonian invisible to all coarse observables belongs to the Poisson centralizer
\begin{align}
    \mN^{\mathrm c}_{\mathrm P}
    :=\{h\in C^\infty(X):\{h,b\}=0\text{ for every }b\in\mN\}.
\end{align}
Thus agreement is precisely the condition $K_\rho-K_0\in\mN^{\mathrm c}_{\mathrm P}$. This is an analogue of a relative commutant defined using Poisson brackets. The ordinary operator commutant of a classical subalgebra contains the whole abelian observable algebra and cannot express this condition.

In the product setting, nondegeneracy of the symplectic form and connectedness of $Y$ imply that this centralizer consists of functions of $z$ alone. Hence \eqref{eq:classical-state-intertwining} is equivalent to
\begin{align}\label{eq:classical-product-state-closure}
    K_\rho(y,z)=K_Y(y)+K_Z(z),
    \qquad
    \rho(y,z)=\rho_Y(y)\rho_Z(z).
\end{align}
Normalization of the marginal fixes $\int_Z\rho_Z\,d\mu_Z=1$. The condition \eqref{eq:classical-state-defect} implies that the coarse observables have the same Poisson evolution under $\rho_0$ and $\rho$. In other words:
\begin{align}
    \forall g\in C^\infty(Y):\qquad \alpha^\rho_t\circ\alpha_{-t}^0(\pi^*g)=\pi^*g\ .
\end{align}
This compares full and reduced observable evolution, as in Section~\ref{subsec:subsystem-flow-background}. The state-preserving expectation criterion itself differs in the classical and quantum theories.

\paragraph{Comparing the two flows:}

We now form the relative family directly from the two complete Hamiltonian flows. This is an interaction-picture comparison. The cocycle identity and the derivative at zero follow from composition and differentiation. We retain their derivation to specify exactly what is being compared, rather than attribute this particular subsystem construction to the references on classical KMS theory.
Define the relative transformation by first undoing the lifted coarse motion and then applying the full motion:
\begin{align}\label{eq:classical-relative-transformation}
    R_t=\Phi_t^\rho\circ\Phi_{-t}^0,
    \qquad
    \mathcal V_t(f):=f\circ R_t^{-1}
    =\alpha_t^\rho\circ\alpha_{-t}^0(f).
\end{align}
For each $t$, this is a symplectic transformation and a measure-preserving automorphism, respectively.

The relative family need not be a one-parameter group. Its cocycle relation is
\begin{equation}
 \mathcal V_{t+s}=\alpha_t^\rho\circ\mathcal V_s\circ\alpha_{-t}^\rho\circ\mathcal V_t.
\end{equation}
This is the classical counterpart of factoring out intrinsic motion, rather than a separate autonomous dynamics.

 Since the lifted coarse flow preserves $\mN_Y$,
\begin{align}\label{eq:classical-relative-algebra-flow}
    \mathcal V_t(\mN_Y)=\alpha_t^\rho(\mN_Y).
\end{align}
The relative transformation therefore follows the same moving subalgebra as the full dynamics, while canceling the chosen internal coarse evolution. It fixes every coarse observable exactly when \eqref{eq:classical-state-intertwining} holds. For a product state, it can still move hidden variables (the relative commutant): $R_t(y,z)=(y,\Phi_t^Z(z))$.

The infinitesimal generator of $R_t$ is the difference of the state Hamiltonians:
\begin{align}\label{eq:classical-relative-Hamiltonian}
    K_{\mathrm{rel}}&=K_\rho-K_0=-\log\frac{\rho}{\rho_0},\nn\\
    \left.\partial_tR_t\right|_{t=0}&=X_{K_{\mathrm{rel}}},\qquad
    \left.\partial_t\mathcal V_t(f)\right|_{t=0}=\{K_{\mathrm{rel}},f\}.
\end{align}

\paragraph{Example: a uniform marginal with nonautonomous observables:}
Take $Y$ and $Z$ to be two-dimensional tori with angular coordinates $(q,p)$ and $(Q,P)$, each of period $2\pi$, and the product symplectic form $dq\wedge dp+dQ\wedge dP$. Consider
\begin{align}
    \rho(q,p,Q,P)=\frac{e^{\epsilon\cos(q-Q)}}{\mathcal Z_\epsilon},
    \qquad
    \mathcal Z_\epsilon=\int_X e^{\epsilon\cos(q-Q)}\,d\mu,
    \qquad \epsilon\ne0.
\end{align}
Translation invariance of the angular integral gives $\rho_Y=1$, so $K_0=0$ and the reduced state predicts no motion. The full state Hamiltonian is $K_\rho=-\epsilon\cos(q-Q)+\log\mathcal Z_\epsilon$, giving
\begin{align}
    \dot q=\dot Q=0,\qquad
    \dot p=-\epsilon\sin(q-Q),\qquad
    \dot P=\epsilon\sin(q-Q).
\end{align}
For example, the bounded coarse observable $e^{ip}$ evolves as
\begin{align}
    \alpha_t^\rho(e^{ip})
    =\exp\bigl(i[p+\epsilon t\sin(q-Q)]\bigr).
\end{align}
Its future value depends on the hidden angle $Q$, so it leaves $\mN_Y$ for $t\ne0$. The full state and its uniform marginal are stationary, but the retained observables do not evolve autonomously. Here $R_t=\Phi_t^\rho$ directly displays the motion that the marginal state Hamiltonian misses. The example also exhibits the failure of monotonicity announced at the beginning of this appendix: the relative Hamiltonian $K_{\mathrm{rel}}=K_\rho-K_0=-\epsilon\cos(q-Q)+\log\mathcal Z_\epsilon$ is indefinite for every $\epsilon\ne0$, since $\mathcal Z_\epsilon=I_0(\epsilon)<e^{|\epsilon|}$. Its $\mu$-average $\log\mathcal Z_\epsilon$ is positive while its $\rho$-average $-S(\rho\|\rho_0)$ is negative. The quantum comparison is instead the form $V^\dagger K_\mM V-K_\mN$ on $\mH_\mN$, which is nonnegative under the assumptions of Proposition~\ref{prop:sufficiency}. The two modular operators need not act on the same Hilbert space.

\paragraph{Relative entropy and the information discarded:}
The same relative logarithm appeared in the phase construction at the beginning of this appendix. Here it generates a vector field, while its expectation gives
\begin{align}\label{eq:classical-relative-entropy-Hamiltonian}
    S(\rho\|\rho_0)=-\omega_\rho(K_{\mathrm{rel}})
    =H(\rho_0)-H(\rho).
\end{align}
It is the relative entropy lost by restricting to the coarse algebra, and it vanishes if and only if $\rho$ is measurable in the coarse $\sigma$-algebra, that is, if and only if that algebra is a sufficient statistic for the pair $(\rho,\mu)$~\cite{OhyaPetz1993}. The vector field alone does not determine this number: on a connected symplectic manifold it fixes its Hamiltonian only up to a constant, which the normalized densities supply. Information loss and failure of autonomous coarse dynamics are also distinct. For a product state the relative transformation acts only on $Z$, yet $S(\rho_Y\rho_Z\|\rho_Y\otimes1)=\int_Z\rho_Z\log\rho_Z\,d\mu_Z$ can be positive. Coarse-graining has discarded the hidden distribution even without correlations, and in general $S(\rho\|\rho_Y\otimes1)=S(\rho\|\rho_Y\rho_Z)+S(\rho_Z\|1)$ splits into the mutual information and the entropy of the hidden marginal relative to the reference. Neither quantity supplies a monotone entropy-production law without an additional coarse-graining or dynamical assumption.

\paragraph{Subregions, expectations, and the passage to field theory:}
The region algebras $\mM_O$ remain valid coarse descriptions. However, their coarse densities $\rho_0=\mE_O(\rho)=1_O\rho+c_O1_{O^c}$, with $c_O$ the average of $\rho$ over $O^c$, are discontinuous across the boundary of $O$, so the Poisson construction assigns them a Hamiltonian flow only after a regularization there. This is a limitation of the Hamiltonian construction, not of the measure-theoretic net. Every classical coarse algebra, by contrast, admits a state-preserving conditional expectation for a faithful probability state, in the product setting $\mE_{\rho,Y}(f)(y)=\int_Zf(y,z)\,\rho(y,z)/\rho_Y(y)\,d\mu_Z(z)$, whether or not its Poisson state flow preserves the algebra. Its existence does not imply \eqref{eq:classical-state-intertwining}. In quantum modular theory, the two questions merge: by Takesaki's theorem~\cite{Takesaki1972} (Section~\ref{subsec:QEC-takesaki} and Theorem~\ref{thm:correctable}), a reference-preserving expectation onto a subalgebra exists if and only if the subalgebra is invariant under the modular flow of the reference. In QFT, a spacetime region selects local field observables, whereas a region $O\subset X$ selects microstates. The connection to the subregion algebras of the companion paper~\cite{companionQFT} is through inclusions of observable algebras and their state-dependent flows, and does not identify these two notions of region.

\paragraph{A classical half-sided inclusion:}

The following explicit current-algebra calculation illustrates one-sided nesting and translation--dilation relations. It is not an application of a classical KMS reconstruction theorem, and the geometric relations alone do not establish a probabilistic $K$-system. The modular future/past-algebra comparison belongs to the quantum setting of~\cite{ouseph2024local}.
A simple example of strict nesting is a classical chiral current $j(x)$ on the line, with $\{j(x),j(y)\}=-\partial_x\delta(x-y)$ on rapidly decreasing configurations. The nonnegative Hamiltonian $P=\tfrac12\int_{\mathbb R}j(x)^2\,dx$ generates the translations $\tau_aj(x)=j(x+a)$ through $\{P,j(x)\}=\partial_xj(x)$, and for the algebra $\mN_b$ generated by $1$ and the currents smeared with $f\in C_c^\infty((b,\infty))$,
\begin{align}
    \tau_a(\mN_b)=\mN_{b+a}\subsetneq\mN_b\quad(a>0),
    \qquad
    \bigcup_{b\in\mathbb R}\mN_b=\mM_{\mathrm{loc}},
    \qquad
    \bigcap_{b\geq0}\mN_b=\mathbb R1\ ,
\end{align}
with $\mM_{\rm loc}$ the polynomial algebra of compactly smeared currents. The full translation group is reversible, but its positive half acts on $\mN_0$ by injective, nonsurjective Poisson endomorphisms that progressively discard field degrees of freedom. The dilation Hamiltonian $D=\tfrac12\int_{\mathbb R}xj(x)^2\,dx$ obeys $\{D,P\}=P$, preserves $\mN_0$ and acts half-sidedly on $\mN_b$ for $b>0$. This is a classical Hamiltonian realization of the translation--dilation relations of half-sided modular inclusions. It specifies neither an invariant probability measure nor a mixing or $K$-property, and identifies the dilation with a state-generated modular evolution only with additional input.

\bibliographystyle{unsrt}
\bibliography{subsystemflow}

\end{document}